\documentclass[a4paper,11pt]{article}

\usepackage{jheppub}
\usepackage{amsthm}
\theoremstyle{definition}

\newtheorem*{dfn*}{Definition}

\newcommand{\de}{\partial}
\newcommand{\be}{\begin{equation}}
\newcommand{\ba}{\begin{eqnarray}}
\newcommand{\ea}{\end{eqnarray}}
\newcommand{\ee}{\end{equation}}

\newcommand{\lr}{\leftrightarrow}
\newcommand{\f}{\frac}
\newcommand{\s}{\sqrt}
\newcommand{\vp}{\varphi}

\newcommand{\ti}{\tilde}
\newcommand{\ap}{\alpha}

\newcommand{\ddd}{\cdot\cdot\cdot}
\newcommand{\no}{\nonumber \\}
\newcommand{\la}{\langle}
\newcommand{\lb}{\rangle}
\newcommand{\bea}{\begin{eqnarray}}
\newcommand{\eea}{\end{eqnarray}}
\newcommand{\bes}{\begin{equation*}}
\newcommand{\beas}{\begin{eqnarray*}}
\newcommand{\eeas}{\end{eqnarray*}}
\newcommand{\bas}{\begin{array*}}
\newcommand{\eas}{\end{array*}}
\newcommand{\ees}{\end{equation*}}

\newcommand{\ep}{\epsilon}

\title{\boldmath  Free Fermion Cyclic/Symmetric Orbifold CFTs
and  Entanglement Entropy}
\author[a,b,c]{Tadashi Takayanagi}
\author[a]{and\ \ Takashi Tsuda}
\affiliation[a]{Center for Gravitational Physics and Quantum Information (CGPQI),\\
Yukawa Institute for Theoretical Physics,\\
Kyoto University,
Kitashirakawa Oiwakecho, Sakyo-ku, Kyoto 606-8501}
\affiliation[b]{Inamori Research Institute for Science,\\
620 Suiginya-cho, Shimogyo-ku,
Kyoto 600-8411 Japan}
\affiliation[c]{Kavli Institute for the Physics and Mathematics
 of the Universe (WPI),\\
University of Tokyo, Kashiwa, Chiba 277-8582, Japan}
\emailAdd{takayana@yukawa.kyoto-u.ac.jp, takashi.tsuda@yukawa.kyoto-u.ac.jp}

\abstract{
In this paper we study the properties of two-dimensional CFTs defined by
cyclic and symmetric orbifolds of free Dirac fermions, especially by focusing on the partition function and entanglement entropy. Via the bosonization, we construct the twist operators which glue two complex planes to calculate the partition function of $\mathbb{Z}_2$ orbifold CFT on a torus. We also find an expression of $\mathbb{Z}_N$ cyclic orbifold in terms of 
Hecke operators, which provides an explicit relation between the partition functions of cyclic orbifolds and those of symmetric ones.
We compute the entanglement entropy and Renyi entropy in cyclic orbifolds on a circle both for finite temperature states and for time-dependent states under quantum quenches. We find that the replica method calculation is highly non-trivial and new because of the contributions from replicas with different boundary conditions. We find the full expression for the $\mathbb{Z}_2$ orbifold and show that the periodicity gets doubled.
Finally, we discuss extensions of our results on entanglement entropy to symmetric orbifold CFTs and make a heuristic argument towards holographic CFTs.}

\begin{document} 
\begin{flushright}
YITP-22-91
\\
IPMU22-0042
\\
\vspace{-1.5cm}
\end{flushright}
\maketitle
\flushbottom

\section{Introduction}

Symmetric orbifold CFTs play important roles to explain black hole entropy microscopically, especially for D1-D5 systems \cite{Strominger:1996sh}.
Such a microscopic model is typically described by a two-dimensional sigma model, whose target space is an instanton moduli space \cite{Vafa:1995bm,Vafa:1995zh,Douglas:1995bn}. Although the geometry of this moduli space is complicated, the space is a singularity-resolved manifold of  a symmetric product of the moduli space of a single instanton \cite{Dijkgraaf:1998gf,Aspinwall:1995zi}. One specific type of the resolution is given by the symmetric orbifold CFTs. This can be regarded as a special example of AdS/CFT \cite{Maldacena:1997re}. In order to have a genuine dual of classical gravity (namely holographic CFTs), we need to introduce strong interactions in the symmetric orbifolds via a suitable marginal deformation \cite{Larsen:1999uk,Avery:2010er,Carson:2014ena}. Nevertheless, the free symmetric orbifold CFT before the deformation acquires the reputation of ``something similar to two-dimensional holographic CFTs''.  Indeed, the match of the BPS spectrum was pointed out \cite{deBoer:1998us} for example.

Generally, to define a symmetric orbifold CFT, we start with an arbitrary seed CFT $\mathcal{C}$ with a central charge $c$. The $N$-th symmetric orbifold CFT $\mathcal{C}_{N,S}$ is defined by orbifolding the product $\mathcal{C}^{\otimes N}$ by the symmetric group $S_N$:
\begin{align}
    \mathcal{C}_{N,S} =&~ \mathcal{C}^{\otimes N} / S_N ,
\end{align}
which has a central charge $Nc$. Every quantity in $\mathcal{C}_{N,S}$ is completely determined by the choice of seed theory $\mathcal{C}$ in principle.

Interestingly, symmetric orbifold CFTs are equipped with some important properties which are necessary conditions for 2-dimensional CFTs to be holographic \cite{Heemskerk:2009pn,El-Showk:2011yvt}. The most significant feature is the presence of the Hagedorn transition in the large $N$ limit regardless of the choice of the seed theory, as shown in \cite{Keller:2011xi,Hartman:2014oaa}. The reason for this is that each of them has a sparse low-energy spectrum in addition to that they obviously have a large central charge $Nc$. Moreover, some multi-point correlation functions in symmetric orbifold CFTs obey the large $N$ factorization \cite{Pakman:2009zz,Belin:2015hwa}.

However, it has been known that symmetric orbifold CFTs are not exactly holographic. This is almost clear because the CFTs do not include interactions other than the orbifold projections. Thus they are not chaotic, which can be explicitly seen from the Lyapunov exponent calculations  \cite{Perlmutter:2016pkf} and the evolution of entanglement entropy \cite{Caputa:2017tju}. Refer also to 
\cite{Giusto:2014aba,Balasubramanian:2014sra,Balasubramanian:2016ids,Balasubramanian:2018ajb} for other quantum information theoretic approaches to the symmetric product CFTs. Moreover, although symmetric orbifold CFTs have such sparse spectra that make Hagedorn transition possible, the spectra are still too dense to have classical gravity duals \cite{Apolo:2022fya}. 
In the ``bulk dual'' point of view, the spectrum of symmetric orbifold CFTs contains too many higher-spin modes and thus the bulk dual is not an ordinary classical (super)gravity even if bulk dual exists. 

Despite the above differences from what we expect for the holographic theory, investigating symmetric orbifold CFTs further is still very intriguing. This is partly because they provide tractable toy models of holographic CFTs. The mechanism of how holographic CFTs admit their AdS gravity duals is not yet fundamentally understood, and the emergence of bottom-up construction of holographic CFTs will aid in understanding. As we referred above, some specific marginal deformations are believed to remove unnecessary higher-spin modes and to deform the theory into an exact holographic CFT with a classical gravity dual. For the case of a symmetric orbifold CFT with an $\mathcal{N}=2$ minimal model as the seed theory, the marginal operator that removes such higher-spin modes, is constructed recently \cite{Belin:2020nmp,Apolo:2022fya}.
The study of symmetric orbifolds is also intriguing partly because they are an interesting class of CFTs whose dynamical properties have not been understood completely. For instance, recently, the complete set of boundary states in symmetric orbifold CFT was constructed and their typical behaviors were compared with those expected from gravity duals in an interesting way \cite{Belin:2020nmp,Belin:2021nck}.  In addition to these two motivations, specific symmetric orbifold theories alone, without any deformations, can admit higher-spin gravity dual: tensionless string theory on AdS geometry \cite{Gaberdiel:2014cha,Giribet:2018ada,Gaberdiel:2018rqv,Eberhardt:2018ouy,Eberhardt:2019ywk}. Research in this direction may provide a more clear understanding of string geometry and ensemble average \cite{Eberhardt:2021jvj}.

Every element of symmetric group $S_N$ can be decomposed into products of elements of cyclic group $\mathbb{Z}_{_i}$, where $1\leq i\leq N$.
We can find a strong connection between a cyclic orbifold CFT and a symmetric orbifold CFT. 
The $N$-th cyclic orbifold CFT $\mathcal{C}_{N,\mathbb{Z}}$ can be obtained by orbifolding the tensor product of seed theory $\mathcal{C}^{\otimes N}$ by the cyclic group $\mathbb{Z}_N$:
\begin{align}
    \mathcal{C}_{N,\mathbb{Z}} =&~ \mathcal{C}^{\otimes N} / \mathbb{Z}_N . 
\end{align}
The structure of cyclic orbifold CFT is much simpler than symmetric one, it is natural to start with a study of cyclic orbifold CFTs. 
We can assemble the results of cyclic orbifolds to obtain quantities in symmetric ones.

We must clarify here the strong connection we referred above.
As for one of the well-known results of symmetric orbifold CFTs, their partition functions on a torus can be expressed in terms of so-called Hecke operators.
We will show that the cyclic orbifold CFT partition function on the torus can also be expressed in terms of Hecke operators. The fact that Hecke operators can describe both orbifold theories offers a strong connection.

The main purpose of the present paper is to explore the dynamical properties of cyclic and symmetric orbifold CFTs, motivated by the above observations. A useful quantity that characterizes the dynamical aspects is the entanglement entropy and its cousin, called Renyi entropy. These entropies are defined in terms of the reduced density matrix obtained by tracing out a subsystem.
These quantities estimate the degrees of freedom in a given quantum many-body system for static states. In particular, for two-dimensional CFTs, these entropies follow universal rules for simple choices of subsystems \cite{Holzhey:1994we,Calabrese:2004eu}. However, for more non-trivial setups such as the entropies for double interval subsystems \cite{Furukawa_2009,Headrick:2010zt} and those for a single interval subsystem in a finite size system at finite temperature, the results highly depend on the details of a given CFT, i.e. its spectrum and its interaction, as is manifest in the free Dirac fermion case \cite{Azeyanagi:2007bj,Ogawa:2011bz,Mukhi:2017rex,Takayanagi:2010wp}.  Moreover, they are helpful to characterize how a quantum state gets thermalized, being sensible to whether the theory is chaotic or not, under non-equilibrium processes such as the quantum quenches \cite{Calabrese:2005in}.
If we apply the AdS/CFT correspondence \cite{Maldacena:1997re}, such a non-equilibrium process is typically dual to a black hole formation \cite{Abajo-Arrastia:2010ajo,Hartman:2013qma}, whose behavior of entanglement entropy can be computed from the holographic entanglement entropy \cite{Ryu:2006bv,Ryu:2006ef,Hubeny:2007xt} and from its generalization to boundary conformal field theories \cite{Takayanagi:2011zk,Fujita:2011fp}. In this paper, we will study the entanglement entropy and Renyi entropy for the above setups in cyclic and symmetric orbifold CFTs. 

This paper is organized as follows.
In section 2, we briefly review the symmetric/cyclic orbifold CFT partition functions on a torus and then introduce Hecke operators. A symmetric orbifold CFT partition function on a torus can be expressed in terms of Hecke operators.
In section 3, we review the replica methods in 2-dimensional CFTs, and develop the method on Dirac fermion cyclic orbifold CFTs. We explicitly construct replica twist operators for this cyclic orbifold.  We also discuss the relation between the second Renyi entropy for double intervals on a plane and the torus partition function. 
In section 4, we calculate the Renyi and entanglement entropy in a finite size system at a finite temperature. For the $\mathbb{Z}_N$ cyclic orbifold CFT, we evaluate them by limiting to the "diagonal sectors" in the CFT on replicated tori to Renyi and entanglement entropy analytically and show that they reproduce the expected part of the thermal entropy by taking the limit where the subsystem approaches the total system. However, the full contribution to the  Renyi entropy can be obtained only after summing over both the diagonal and non-diagonal sectors. Even though this is difficult for general $N$, we solve this problem by focusing on the simplest case of the second Renyi entropy at $N=2$ and obtain the full expression. We also find that a time-like 2-point function in $N$-th cyclic orbifold CFT can have  a periodicity $N$ times longer than that in the seed CFT.  
In section 5, we extend the analysis in section 4 for cyclic orbifolds to the dynamical process of quantum quenches described by the CFT on replicated cylinders. We find that the periodicity of time-dependent entanglement entropy  in the $N$-th cyclic orbifold CFT gets $N$ times larger than that in the seed CFT. 
In section 6, we introduce a new type of Hecke operator (we call square-free Hecke operator) by deforming the ordinary Hecke operator. This new operator is necessary for constructing cyclic orbifold CFT partition functions in terms of Hecke operators. Here, through Hecke operators, the connection between partition functions of symmetric orbifold CFTs and those of cyclic ones, becomes very clear.
By using this connection, we show the calculation of diagonal Renyi entropy in Dirac fermion symmetric orbifold theory.
In section 7, we discuss the connection between symmetric orbifold CFTs and holographic CFTs. We provide evidence for Dirac fermion symmetric orbifold theory to have $\exp N$ recurrence time, which is a significantly long time even for such an integrable CFT. 
In section 8, we summarize our conclusions and discuss future problems. 
In appendix A, we present the definitions and properties  of the theta functions. In appendix B, we present examples of Hecke operators. In appendix C, we explain the construction of orbifold partition functions. 

\section{Orbifold CFT preliminaries}
In this section, we briefly review orbifold CFTs, especially so-called symmetric orbifold CFTs and cyclic orbifold CFTs.
We mainly focus on the partition functions of these orbifold CFTs, on a torus with complex structure $\tau$ ($q=\exp [2\pi i \tau]$).
In subsection \ref{SymOrbPF}, we review partition functions of symmetric orbifold CFTs, and their generating function. We also review the well-organized expressions of symmetric orbifold CFT partition functions in terms of Hecke operators, which are mainly used in the context of number theory. Examples of Hecke operators and partition functions are displayed in appendix \ref{ExamplesHeckePartitionFunction}. We also review cyclic orbifold CFT partition functions in \ref{CycOrbPF} shortly. We will revisit this topic in \ref{CycOrbRevisit}.

We introduce a seed CFT $\mathcal{C}$ with a central charge $c$ which we can prepare arbitrarily.
We assume that the CFT $\mathcal{C}$ has a modular invariant torus partition function $Z(\tau)$, which has the $q$-expansion
\begin{align}
    Z(\tau)=\sum_{m,\bar{m}\in I}d(m,\bar{m})q^{m} \bar{q}^{\bar{m}}.
\end{align}
Cyclic and Symmetric orbifold CFTs can be obtained by orbifolding $\mathcal{C}^{\otimes N}$ by cyclic group $\mathbb{Z}_N$ or symmetric group $S_N$, respectively:
\begin{align}
    \mathcal{C}_{N,S} =&~ \mathcal{C}^{\otimes N} / S_N,  \\
    \mathcal{C}_{N,\mathbb{Z}} =&~ \mathcal{C}^{\otimes N} / \mathbb{Z}_N.
\end{align}
Their central charges are both $Nc$. From this construction, symmetric/cyclic orbifold CFTs are completely determined by the choice of seed theory $\mathcal{C}$ in principle. 

\subsection{General orbifold partition functions}
The partition function of CFT orbifolded by some discrete group $G$ is \cite{Dixon:1985jw}
\begin{align}
\label{GeneralOrbifoldPF}
    \frac{1}{|G|}\sum_{\substack{g,h\in G \\ gh=hg}} {\scriptstyle h} \, \underset{\scriptstyle g}{\Box} =&~
    \frac{1}{|G|}\sum_{\substack{g,h\in G \\ gh=hg}} \mathrm{Tr}_{(g)}\left[hq^{H_L}\Bar{q}^{H_R}\right],
\end{align}
where $H_L$ and $H_R$ are the Hamiltonian in the left and right moving sectors, respectively.
Here, $g$ labels the (un)twisted sectors, and running $h$ corresponds to the projection.

\subsection{Symmetric orbifold partition function}
\label{SymOrbPF}
The generating function of the $N$-th symmetric orbifold partition function $Z_N(\tau)$ is well-known \cite{Keller:2011xi,Dijkgraaf:1996xw,Hartman:2014oaa,Haehl:2014yla};
\begin{equation}
\label{GeneralSymPF}
    \sum_{N\geq0}p^N Z_{N,S}(\tau) = \prod_{n>0}
    \prod_{
    \substack{m,\bar{m}\in I \\
    m-\bar{m}\equiv0~\mathrm{mod}~n}
    }
    \frac{1}{(1-p^n q^{m/n} \bar{q}^{\bar{m}/n})^{d(m,\bar{m})}}.
\end{equation}
Here, product over $m,\bar{m}\in I$ $(m-\bar{m}\equiv0~\mathrm{mod}~n)$ can be interpreted as insertion of
\begin{equation}
    \delta^{(n)}(m-\bar{m})\equiv\frac{1}{n}\sum_{k=0}^{n-1} e^{2 \pi i \frac{k}{n} (m-\bar{m})}.
\end{equation}
This insertion can be interpreted as performing an orbifold projection. We can calculate the effect of this projection explicitly by replacing $\tau/n$ with $(\tau+1)/n$.

Alternatively, there is another expression of the generating function \cite{Haehl:2014yla,Keller:2011xi};
\begin{equation}
    Z_{N,S}(\tau) = 
    \sum_{\mathrm{partition~of}~N}
    \prod_{k=1}^{N}\frac{1}{(N_k)!} \left( T_k Z(\tau) \right)^{N_k},
\end{equation}
where the partition of $N$ runs over $\displaystyle (N_1,\dots,N_N) ~\mathrm{s.t.}~ \sum_{k=1}^{N} k N_k = N$. 
Here we introduced the Hecke operator $T_k$, 
\begin{equation}
    T_k Z\left(\tau\right) \equiv \frac{1}{k}\sum_{i|k} \sum_{j=0}^{i-1} Z\left(\frac{k \tau}{i^{2}}+\frac{j}{i}\right).
\end{equation}
We will discuss the Hecke operator in detail in subsection \ref{CycOrbRevisit}. Yet another expression of the generating function \cite{Haehl:2014yla} reads:
\begin{align}
    \sum_{N\geq0}p^N Z_{N,S}(\tau) = \exp\left[\sum_{k=1}^{\infty}p^k T_k Z(\tau)\right].
\end{align}

\subsection{Cyclic orbifold partition function}
\label{CycOrbPF}
The cyclic orbifold partition function is, $G=\mathbb{Z}_N$ version of (\ref{GeneralOrbifoldPF}),
\begin{align}
    Z_{N,\mathbb{Z}}(\tau) =&
    \sum_{l=0}^{N-1} \mathrm{Tr}_{(l)} \left(\frac{\sum_{l'=1}^{N}g^{l'}}{N}q^{H_L}\bar{q}^{H_R}\right) \nonumber\\
    =& \frac{1}{N} \sum_{l,l'}\mathrm{Tr}_{(l)} \left(g^{l'}q^{H_L}\bar{q}^{H_R}\right) \label{CyclicOrbifoldPF}
\end{align}
where $g$ is one of the generators of $\mathbb{Z}_N$ and $l$ labels the (un)twisted sectors. In the $l$-twisted sectors, operators in $j$-th CFT change to ones in $(j+l)$-th CFT when moving from $z$ to $z+1$.
In \cite{Haehl:2014yla}, the more apparent form of $N$-th cyclic orbifold partition function $Z_{N,\mathbb{Z}}(\tau)$ is proposed:
\begin{align}
    Z_{N,\mathbb{Z}}(\tau) =&~ \frac{1}{N}
    \sum_{r,s=1,\dots,N} Z\left( \frac{(N,r)}{N}\left(\frac{(N,r)}{(N,r,s)}\tau+\kappa (r,s)
    \right)\right)^{(N,r,s)} ,
\end{align}
where $\kappa(r,s)$ is defined to be the smallest integer in $\{0,1,\dots,\frac{N}{(N,r)}-1\}$ that satisfies $\kappa(r,s)r-\frac{(N,r)s}{(N,r,s)}\equiv0$ mod $N$. 
If $N$ is a prime number, this function becomes as
\begin{align}
Z_{N,\mathbb{Z}}(\tau) =&~ \frac{1}{N}Z(\tau)^N+(N-1)T_N Z(\tau) \\
=&~\frac{1}{N}\left(Z(\tau)^N+(N-1)\left( Z(N\tau)+\sum_{j=0}^{N-1}Z\left(\frac{\tau+j}{N}\right)\right)\right).
\end{align}
This prime $N$ case is indicated in \cite{Klemm:1990df}.

We will newly derive the representation of this partition function in terms of Hecke operators in subsection \ref{CycOrbRevisit}.

\section{Twist operators}
\begin{figure}[hhh]
  \centering
  \includegraphics[width=8cm]{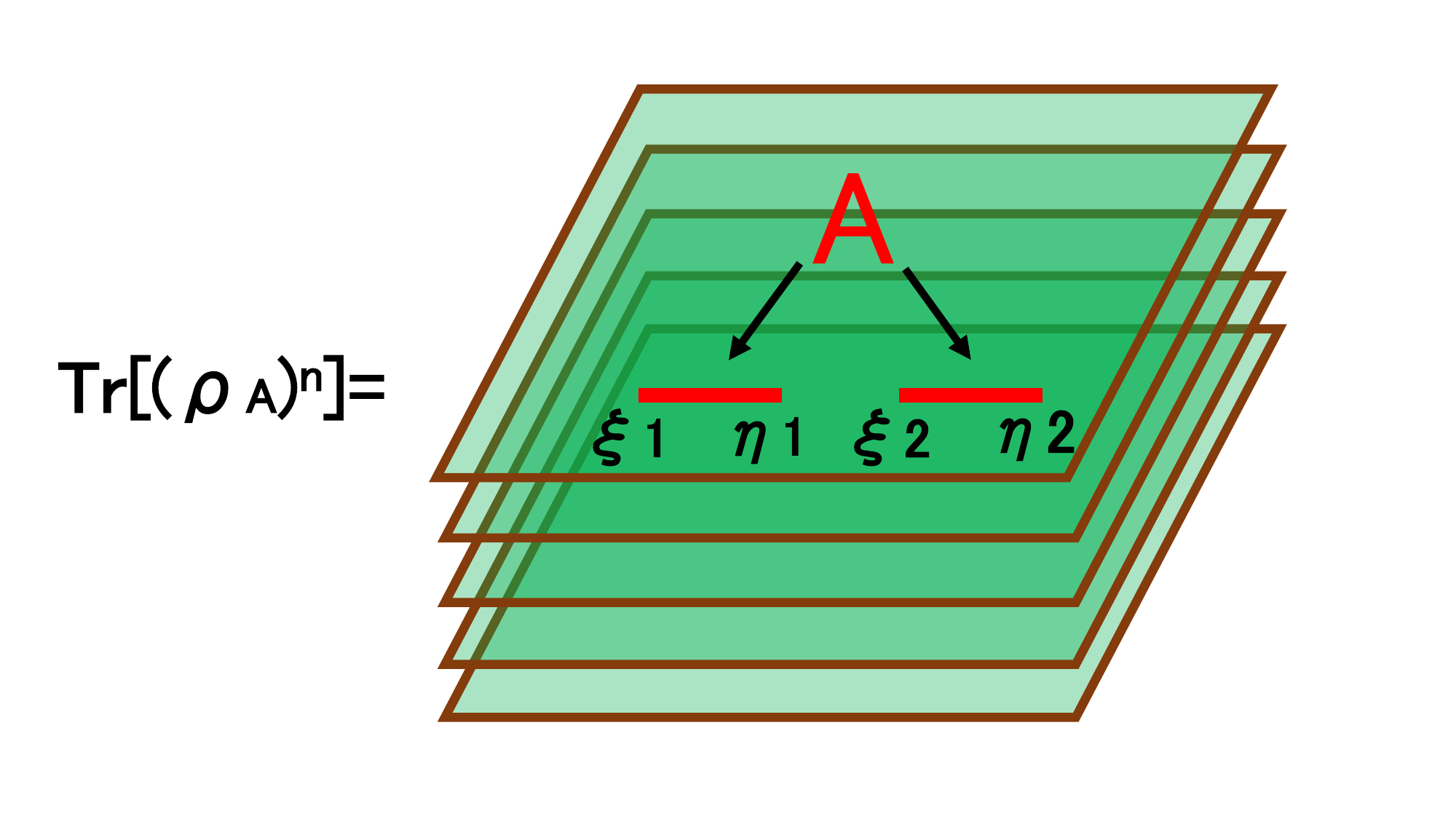}
  \caption{The replica method calculation of Tr$[(\rho_A)^n]$. We set $s=2$ and $n=5$.} 
\label{fig:replica}
\end{figure}

A basic and powerful way to calculate the entanglement entropy in two-dimensional CFTs is to employ the replica method \cite{Holzhey:1994we,Calabrese:2004eu}. The $n$-th Renyi entropy for subsystem $A$ is defined by 
\ba
S^{(n)}_A=\frac{1}{1-n}\log \mbox{Tr}[(\rho_A)^n],\label{nrenyi}
\ea
where $\rho_A$ is the reduced density matrix obtained by tracing out the complement of $A$ for a given quantum state.
In the Euclidean path-integral, the quantity Tr$[(\rho_A)^n]$ is equal to the partition function on an $n$-sheeted Riemann surface 
where the cuts are introduced on the subsystem $A$ as sketched in Fig.\ref{fig:replica}.
This partition function can be computed as a correlation function of twist operators, denoted by $\sigma_n$ and $\bar{\sigma}_n$, inserted on the boundaries $\de A$. When the subsystem $A$ is a union of $s$ intervals $[\xi_1,\eta_1]\cup [\xi_2,\eta_2]\ddd \cup [\xi_s,\eta_s]$, then we find that Tr$[(\rho_A)^n]$ is given by the $2s$ point function:
\ba
\mbox{Tr}[(\rho_A)^n]=\la \sigma_n(\xi_1)\bar{\sigma}_n(\eta_1)\ddd
\sigma_n(\xi_s)\bar{\sigma}_n(\eta_s)\lb.
\ea
Here the correlation function in the right-hand side is normalized such that $\la 1\lb=1$ and this corresponds to the normalization of density matrix Tr$\rho_A=1$.

The twist operator $\sigma_n(\xi)$ creates a cut at $z=\xi$ such that a $2\pi$ rotation around  $z=\xi$ reaches the next sheet. In other words, the periodicity around $z=\xi$ is $2\pi n$. The other twist operator $\bar{\sigma}_n(\xi)$ has the opposite winding such that it absorbs the cut created by $\sigma_n$. In this section below,  we would like to study the properties of the twist operators in the Dirac fermion CFT and its cyclic orbifolds.

\subsection{Twist operators in Dirac fermion CFT}

For the free massless Dirac fermion CFT with $c=1$, we can find an explicit formula of twist operators \cite{Casini:2005rm,Azeyanagi:2007bj}. This can be generalized to the cyclic orbifold CFTs as we explain below. Consider a free massless Dirac fermion on 
a complex plane whose coordinate is expressed as 
$(z,\bar{z})$. We write the Dirac fermion field as $(\psi_L(z),\bar{\psi}_L(z))$ for the left-moving mode
and $(\psi_R(\bar{z}),\bar{\psi}_R(\bar{z}))$ for the right-moving mode.  They satisfy the OPE relation 
\ba
\psi_{L}(z)\bar{\psi}_{L}(0)\sim \frac{1}{z},\ \ \ \psi_{R}(z)\bar{\psi}_{R}(0)\sim \frac{1}{\bar{z}}.
\ea
 In the replica calculation, we write $n$ replicas of the Dirac fermion as $\psi_{L,[a]}\ \ (a=1,2,\ddd,n)$ (and similarly in the right-moving sector). If the fermion goes around a point (simply we choose $z=0$) where the twist operator is inserted, the fermion will go back to the original sheet after $2\pi n$ rotation. However, in this $2\pi n$ rotation, the fermion will get the extra phase $-(-1)^{n}$. This can be found by conformally mapping from the $n$-sheeted surface described by $z$ coordinate into a single plane described by $w$ via $w=z^{1/n}$. The fermion transforms as $\psi_L(z)=z^{\frac{1-n}{2n}}\psi_L(w)$. Since in the $w$-plane there is no holonomy, we obtain the phase $e^{\pi i\frac{1-n}{n}}$ each time $\psi_L(z)$ goes around $z=0$. Therefore under the $2\pi$ rotation, the fermion transforms as follows:
\ba
\psi_{L,[a]}\to e^{\pi i\left(\frac{1-n}{n}\right)}\psi_{L,[a+1]}.
\ea

Now we perform the discrete Fourier transformation 
\ba
\ti{\psi}_{L,q}=\frac{1}{\s{n}}\sum_{a=1}^n e^{-2\pi i\frac{qa}{n}}\psi_{L,[a]}.\label{FTrep}
\ea
By introducing $q$ by shifting $p$ as $p=q+\frac{1-n}{2}$, under the $2\pi$ rotation $z\to z e^{2\pi i}$, the fermion transformation reads
\ba
\ti{\psi}_{L,p}\to e^{2\pi i\frac{p}{n}}\ti{\psi}_{L,p}.\label{monodro}
\ea
We can choose $p$ to be $p=-\frac{n-1}{2},-\frac{n-3}{2},\ddd,\frac{n-1}{2}$. Note that $p$ takes integer values (or half-integer values) when $n$ is odd (or even), respectively. Finally, the twist operator inserted at a point $z=z_*$ can be explicitly written in the form\footnote{Equivalently we can choose $\prod_{p=-\frac{n-1}{2}}^{\frac{n-1}{2}} e^{i\frac{p}{n}\left(\ti{\vp}_{L,p}(z_*)+\ti{\vp}_{R,p}(\bar{z}_*)\right)}$ by replacing $\psi_R$ with $\bar{\psi}_R$ and vice versa.}  \cite{Azeyanagi:2007bj}:
\ba
&& \sigma_n(z_*,\bar{z}_*)=\prod_{p=-\frac{n-1}{2}}^{\frac{n-1}{2}} e^{i\frac{p}{n}\left(\ti{\vp}_{L,p}(z_*)-\ti{\vp}_{R,p}(\bar{z}_*)\right)},  \no
&&  \bar{\sigma}_n(z_*,\bar{z}_*)=\prod_{p=-\frac{n-1}{2}}^{\frac{n-1}{2}} e^{-i\frac{p}{n}\left(\ti{\vp}_{L,p}(z_*)-\ti{\vp}_{R,p}(\bar{z}_*)\right)}, 
\label{twistopa}
\ea
where $\ti{\vp}_{L,R}$ is the massless free scalar field obtained from the bosonization 
\ba
&& \ti{\psi}_L(z)=e^{i\ti{\vp}_L(z)},\ \ \ \bar{\ti{\psi}}_L(z)=e^{-i\ti{\vp}_L(z)} ,\no
&& \ti{\psi}_R(\bar{z})=e^{i\ti{\vp}_R(\bar{z})},\ \ \ \bar{\ti{\psi}}_R(\bar{z})=e^{-i\ti{\vp}_R(\bar{z})} .
\ea
The OPE looks like
\ba
\ti{\vp}_{L}(z)\ti{\vp}_{L}(0)\sim -\log z,\ \ \ \  \ti{\vp}_{R}(\bar{z})\ti{\vp}_{R}(0)\sim -\log \bar{z}.
\ea
To see why the form of the twist operator (\ref{twistopa}) makes sense, note that the OPE 
\be
\ti{\psi}_L(z)e^{i\frac{p}{n}\ti{\vp}_L(z_*)}\sim (z-z_*)^{\frac{p}{n}}\cdot e^{i\left(1+\frac{p}{n}\right)\ti{\vp}_L(z_*)},
\ee
leads to the desired monodromy under the rotation (\ref{monodro}).

\subsection{Relation between second Renyi entropy and torus partition function}

\begin{figure}[hhh]
  \centering
  \includegraphics[width=8cm]{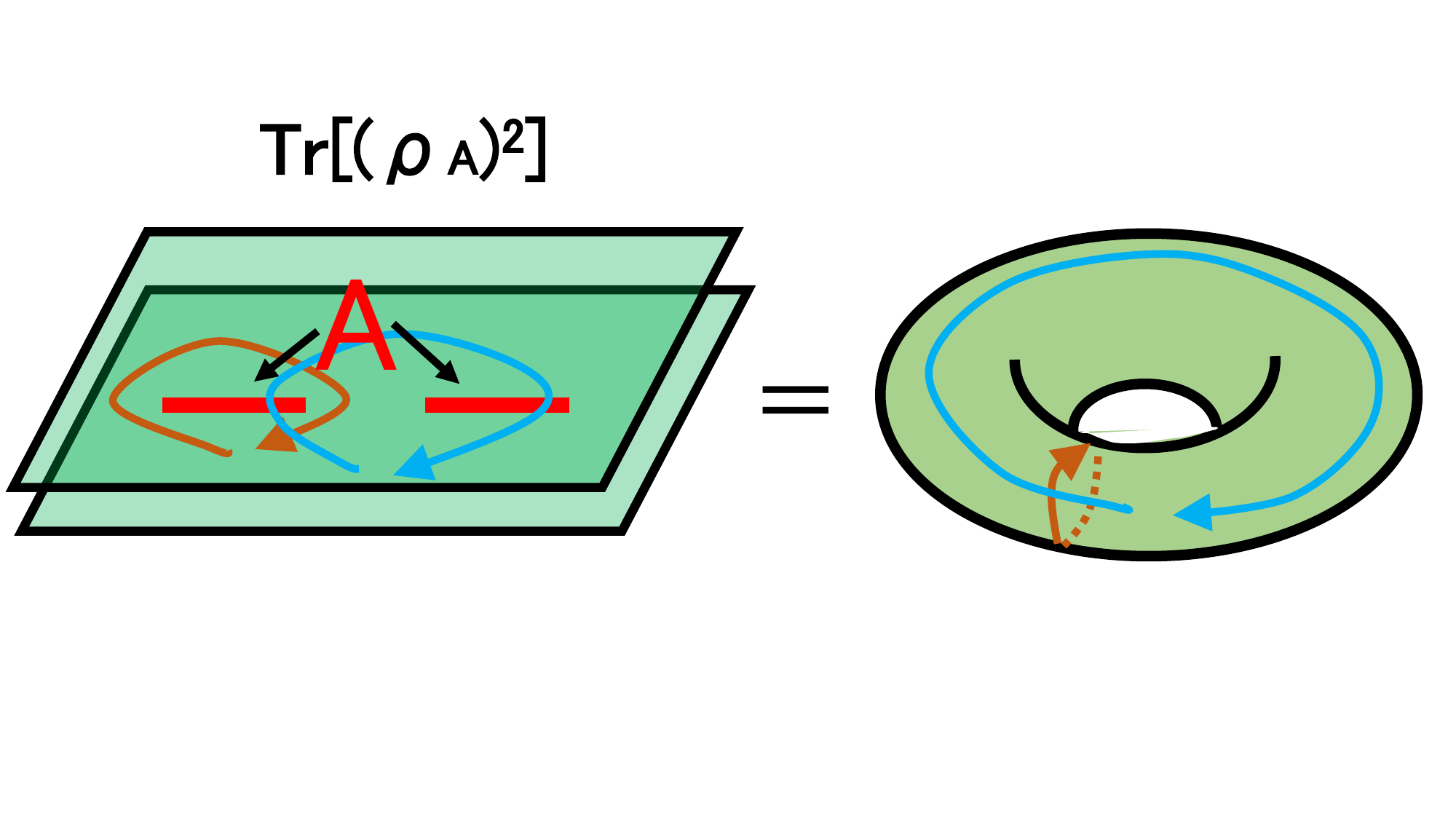}
  \caption{Equivalence between Tr$[(\rho_A)^2]$ and a torus partition function in two-dimensional CFTs.} 
\label{fig:replicab}
\end{figure}

For $n=2$, we know the following remarkable relation between the partition function of a general CFT with central charge $c$ defined on a 2-sheeted Riemann surface and that on a torus \cite{Lunin:2000yv,Headrick:2010zt} as depicted in Fig.\ref{fig:replicab}:
\ba
\mbox{Tr}[\rho_A^2]=\la \hat{\sigma}_2(0)\hat{\sigma}_2(x)\hat{\sigma}_2(1)\hat{\sigma}_2(\infty)\lb=(2^8x(1-x))^{-\frac{c}{12}}\cdot Z_{torus}(\tau),
\ea
where $\tau$ is the moduli of the torus and we assume it is rectangular i.e. $\tau=\bar{\tau}$. The twist operator $\hat{\sigma}_2$ at each location should be chosen appropriately among several candidates as we will explain below.
The moduli is related to $x$ via
\ba
x=\left[\frac{\theta_2(\tau)}{\theta_3(\tau)}\right]^4,\ \ \ 1-x=\left[\frac{\theta_4(\tau)}{\theta_3(\tau)}\right]^4.
\ea

In our Dirac fermion CFT we have $c=1$ and there are two types of twist operators:
\ba
&& \sigma_{2}=e^{\frac{i}{4}(\ti{\vp}_{L,\frac{1}{2}}-\ti{\vp}_{R,\frac{1}{2}})}\cdot e^{-\frac{i}{4}(\ti{\vp}_{L,-\frac{1}{2}}-\ti{\vp}_{R,-\frac{1}{2}})},\no
&& \bar{\sigma}_{2}=e^{-\frac{i}{4}(\ti{\vp}_{L,\frac{1}{2}}-\ti{\vp}_{R,\frac{1}{2}})}\cdot e^{\frac{i}{4}(\ti{\vp}_{L,-\frac{1}{2}}-\ti{\vp}_{R,-\frac{1}{2}})}.
\ea

We get the following three different partition functions depending on the different choices of the order of twist operators:
\ba
&&  \la \sigma_{2}(0)\bar{\sigma}_{2}(x)\sigma_{2}(1)\bar{\sigma}_{2}(\infty)\lb=\frac{\theta_3^2}{\theta_2\theta_4}=x^{-\frac{1}{4}}(1-x)^{-\frac{1}{4}},\no 
&& \ \ \ \ \ \ \ \ \ \ \ \ \ \ \ \ \ \ \ \  \lr \mbox{Tr}_{NS}(q^{L_0}\bar{q}^{\bar{L}_0})=\left|\frac{\theta_3(\tau)}{\eta(\tau)}\right|^2. \no
&& \la \sigma_{2}(0)\bar{\sigma}_{2}(x)\bar{\sigma}_{2}(1)\sigma_{2}(\infty)\lb=\frac{\theta_4}{\theta_2}=x^{-\frac{1}{4}}(1-x)^{\frac{1}{4}},\no 
&& \ \ \ \ \ \ \ \ \ \ \ \ \ \ \ \ \ \ \ \  \lr \mbox{Tr}_{NS}((-1)^F q^{L_0}\bar{q}^{\bar{L}_0})=\left|\frac{\theta_4(\tau)}{\eta(\tau)}\right|^2. \no
&& \la \sigma_{2}(0)\sigma_{2}(x)\bar{\sigma}_{2}(1)\bar{\sigma}_{2}(\infty)\lb=\frac{\theta_2}{\theta_4}=x^{\frac{1}{4}}(1-x)^{-\frac{1}{4}},\no 
&& \ \ \ \ \ \ \ \ \ \ \ \ \ \ \ \ \ \ \ \  \lr \mbox{Tr}_{R}(q^{L_0}\bar{q}^{\bar{L}_0})=\left|\frac{\theta_2(\tau)}{\eta(\tau)}\right|^2. \label{fermionto}
\ea
Note that the sector Tr$_R(-1)^F$  simply vanishes and thus is absent in the above. Our conventions of theta functions and eta function, used in the above, are summarized in Appendix \ref{apptheta}. 

The above identification (\ref{fermionto})  can be understood by examining the holonomy along the two cycles of the torus which correspond to the circle which surrounds $z=0$ and $z=x$ and the one which does $z=x$ and $z=1$. When the two twist operators surrounded by each circle have the same (or different) sign, then the circle follows the R (or NS) sector boundary condition. This analysis also reveals that our previous choice of the twist operator (\ref{twistopa}) corresponds to the NS sector. In this paper, we will focus on the NS sector contribution below.

\subsection{Twist operators in $\mathbb{Z}_N$ cyclic orbifold CFT}

Now we would like to turn to the cyclic orbifold CFT constructed from $N$ Dirac fermions, denoted by 
$\psi^j_{L,R}$ and $\bar{\psi}^j_{L,R}$ $(j=1,2,\ddd,N)$
by the $\mathbb{Z}_N$ orbifold action:
\ba
g: (\psi^{j}_{L,R},\bar{\psi}^j_{L,R})\to (\psi^{j+1}_{L,R},\bar{\psi}^{j+1}_{L,R}).
\ea
We can diagonalize again by taking the discrete Fourier transformation 
\ba
\ti{\psi}^{(k)}_{L,R}=\frac{1}{\s{N}}\sum_{j=1}^Ne^{-2\pi i\frac{kj}{N}}\psi^j_{L,R},\ \ \ \ (k=0,1,\ddd,N-1), \label{znactg}
\ea
so that the $\mathbb{Z}_N$ action simply multiplies the phase factor $g:\ti{\psi}^{(k)}_{L,R}\to e^{2\pi i\frac{k}{N}}\ti{\psi}^{(k)}_{L,R}$. For the calculation of Tr$(\rho_A)^n$ in the cyclic orbifold CFT, we consider the replicated fermions, denoted by 
\be
\psi^{j}_{L,[a]}(z), \ \  \ \bar{\psi}^{j}_{L,[a]}(z), \ \  \ \psi^{j}_{R,[a]}(\bar{z}), \ \  \ \bar{\psi}^{j}_{R,[a]}(\bar{z}),
\ee
where $j=1,2,\ddd,N$ and $a=1,2,\ddd,n$. After we perform the discrete Fourier transformations (\ref{FTrep}) and  (\ref{znactg}) with respect to $j$ and $a$, we obtain 
\be
\ti{\psi}^{(k)}_{L,p}(z), \ \  \ \bar{\ti{\psi}}^{(k)}_{L,p}(z), \ \  \ \ti{\psi}^{(k)}_{R,p}(\bar{z}), \ \  \ \bar{\ti{\psi}}^{(k)}_{R,p}(\bar{z}),
\ee
where $k=0,1,2,\ddd,N-1$ and $p=-\frac{n-1}{2},\ddd,\frac{n-1}{2}$.

The partition functions in NS sector look like (from (\ref{CyclicOrbifoldPF}))
\ba
Z^{NS}_{N,\mathbb{Z}}&=&\sum_{l=0}^{N-1}\mbox{Tr}^{(l)}_{NS}\left[\sum_{l'=0}^{N-1}\frac{g^{l'}}{N}q^{H_L}\bar{q}^{H_R}\right]\no
&& =\frac{1}{N}\sum_{l,l'=0}^{N-1}\prod_{k=0}^{N-1}
e^{i\frac{\pi k^2}{N^2}l^2(\tau-\bar{\tau})}\left|\frac{\theta_3\left(\frac{kl'}{N}+\frac{kl}{N}\tau|\tau\right)}{\eta(\tau)}\right|^2,
\label{partNS}
\ea
where $l=0,1,\ddd,N-1$ is the label of the twisted sectors such that $l=0$ is the untwisted sector and the summation over $l'$ realizes the $\mathbb{Z}_N$ projection. 
We can easily confirm the invariance under the $S$ transformation $\tau\to-1/\tau$.

In order to calculate the entanglement entropy, we again need to examine the twist operator $\sigma_n$ in the replica of the orbifold CFT. By bosonizing the $Nn$ fermions 
\ba
\ti{\psi}^{(k)}_{L,p}(z)=e^{i{\ti{\vp}}^{(k)}_{L,p}(z)},\ \ \ \ \ti{\psi}^{(k)}_{R,p}(\bar{z})=e^{i{\ti{\vp}}^{(k)}_{R,p}(\bar{z})},
\ea
the twist operator at $z=z_*$ can be explicitly written as in the $N=1$ case (\ref{twistopa}):
\ba
&& \sigma_n(z_*,\bar{z}_*)=\prod_{p=-\frac{n-1}{2}}^{\frac{n-1}{2}}\prod_{k=0}^{N-1}e^{i\frac{p}{n}({\ti{\vp}}^{(k)}_{L,p}(z_*)-{\ti{\vp}}^{(k)}_{R,p}(\bar{z}_*))},\no
&& \bar{\sigma}_n(z_*,\bar{z}_*)=\prod_{p=-\frac{n-1}{2}}^{\frac{n-1}{2}}\prod_{k=0}^{N-1}e^{-i\frac{p}{n}({\ti{\vp}}^{(k)}_{L,p}(z_*)-{\ti{\vp}}^{(k)}_{R,p}(\bar{z}_*))}.\label{twistcycl}
\ea
Note that the $\mathbb{Z}_N$ cyclic orbifold action shifts the scalar as follows: 
\ba
g:\ \ {\ti{\vp}}^{(k)}_{L,p}\to {\ti{\vp}}^{(k)}_{L,p}+\frac{2\pi k}{N},\ \ \ \ {\ti{\vp}}^{(k)}_{R,p}\to {\ti{\vp}}^{(k)}_{R,p}+\frac{2\pi k}{N}.  \label{ascshift}
\ea
It is obvious that the twist operator (\ref{twistcycl}) is invariant under this $\mathbb{Z}_N$ action and also that it provides the correct $\mathbb{Z}_N$ twist of the Dirac fermions. 

\subsection{Twist operator in $N=2$ cyclic/symmetric orbifold CFT}

As the simplest example among non-trivial cyclic orbifolds, we consider the $N=2$ cyclic orbifold. This can also be regarded as the $N=2$ symmetric orbifold CFT. 
In this case, the NS sector partition function takes the form:
\ba
Z^{NS}_{2,\mathbb{Z}}=\left|\frac{\theta_3(\tau)}{\eta(\tau)}\right|^2\cdot \left(\frac{|\theta_3(\tau)|^2+|\theta_4(\tau)|^2+|\theta_2(\tau)|^2}{2|\eta(\tau)|^2}\right).
\ea
If we compare this with
\ba
Z_{2,\mathbb{Z}}=\frac{1}{2}\left[Z(\tau)^2+Z(2\tau)+Z\left(\frac{\tau}{2}\right)
+Z\left(\frac{\tau+1}{2}\right)\right],
\ea
then we identify (we employ the formulas in appendix \ref{sec:doubleF})
\ba
&& Z(\tau)=\left|\frac{\theta_3(\tau)}{\eta(\tau)}\right|^2=2^{2/3}x^{-1/6}(1-x)^{-1/6},\no
&& Z(2\tau)=\left|\frac{\theta_4(2\tau)}{\eta(2\tau)}\right|^2=2^{4/3}x^{-1/3}(1-x)^{1/6},\no
&& Z\left(\frac{\tau}{2}\right)=\left|\frac{\theta_2\left(\frac{\tau}{2}\right)}{\eta\left(\frac{\tau}{2}\right)}\right|^2=2^{4/3}x^{1/6}(1-x)^{-1/3},\no
&& Z\left(\frac{\tau+1}{2}\right)=0.
\ea
In terms of the twist operator four-point functions, we find 
\ba
&& \mbox{Tr}[\rho_A^2]=2^{-\frac{4}{3}}x^{-\frac{1}{6}}(1-x)^{-\frac{1}{6}}Z^{NS}_{2,\mathbb{Z}}  \no
&& =2^{-\frac{4}{3}} \left(x^{-\frac{1}{2}}(1-x)^{-\frac{1}{2}}+(1-x)^{-\frac{1}{2}}+x^{-\frac{1}{2}}\right),\no
&&=\la\sigma_{2}(0)\bar{\sigma}_{2}(x)\sigma_{2}(1)\bar{\sigma}_{2}(\infty)\lb+\la\sigma_{2}(0)\bar{\ti{\sigma}}_{2}(x)\ti{\sigma}_{2}(1)\bar{\sigma}_{2}(\infty)\lb \no
&&\ \ \ \ +\la\sigma_{2}(0)\bar{\sigma}_{2}(x)\ti{\sigma}_{2}(1)\bar{\ti{\sigma}}_{2}(\infty)\lb,
\ea
where we defined the twist operators as follows:
\ba
&& \sigma_{2}=
e^{\frac{i}{4}{\ti{\vp}}^{(0)}_{L,\frac{1}{2}}(z_*)-\frac{i}{4}{\ti{\vp}}^{(0)}_{R,\frac{1}{2}}(\bar{z}_*)}\!\cdot\! 
e^{-\frac{i}{4}{\ti{\vp}}^{(0)}_{L,-\frac{1}{2}}(z_*)+\frac{i}{4}{\ti{\vp}}^{(0)}_{R,-\frac{1}{2}}(\bar{z}_*)}\cdot
e^{\frac{i}{4}{\ti{\vp}}^{(1)}_{L,\frac{1}{2}}(z_*)-\frac{i}{4}{\ti{\vp}}^{(1)}_{R,\frac{1}{2}}(\bar{z}_*)}\!\cdot \!
e^{-\frac{i}{4}{\ti{\vp}}^{(1)}_{L,-\frac{1}{2}}(z_*)+\frac{i}{4}{\ti{\vp}}^{(1)}_{R,-\frac{1}{2}}(\bar{z}_*)},\no
&& \ti{\sigma}_{2}=
e^{\frac{i}{4}{\ti{\vp}}^{(0)}_{L,\frac{1}{2}}(z_*)-\frac{i}{4}{\ti{\vp}}^{(0)}_{R,\frac{1}{2}}(\bar{z}_*)}\!\cdot \!
e^{-\frac{i}{4}{\ti{\vp}}^{(0)}_{L,-\frac{1}{2}}(z_*)+\frac{i}{4}{\ti{\vp}}^{(0)}_{R,-\frac{1}{2}}(\bar{z}_*)}\!\cdot\!
e^{-\frac{i}{4}{\ti{\vp}}^{(1)}_{L,\frac{1}{2}}(z_*)+\frac{i}{4}{\ti{\vp}}^{(1)}_{R,\frac{1}{2}}(\bar{z}_*)}\!\cdot \!
e^{\frac{i}{4}{\ti{\vp}}^{(1)}_{L,-\frac{1}{2}}(z_*)-\frac{i}{4}{\ti{\vp}}^{(1)}_{R,-\frac{1}{2}}(\bar{z}_*)}.\no
\label{nntwote}
\ea
Also, $\bar{\sigma}_{2}$ and $\bar{\ti{\sigma}}_{2}$ are simply the complex conjugate of  $\sigma_{2}$ and $\ti{\sigma}_{2}$, respectively. The reason why we need to introduce extra twist operators $\ti{\sigma}_{2}$ and $\bar{\ti{\sigma}}_{2}$ in addition to the original ones (\ref{twistcycl}) is because we need to twist the boundary conditions along the two cycles of the torus in a way that exchanges the two fermion fields in the $N=2$ cyclic orbifold. Indeed, if a fermion goes around the twist fields $\ti{\sigma}_{2}$ and $\bar{\ti{\sigma}}_{2}$, it is not only moved into the other sheet of replica but also replaced with the other one i.e. $\psi^{1}_{L,[1]}\to i\psi^{2}_{L,[2]}$.


\section{Entanglement entropy in cyclic orbifold CFTs on a finite size space at finite temperature}\label{sec:finitep}

Here we would like to study the entanglement entropy on a compactified space at finite temperature for the cyclic orbifolds. In general, we can calculate the quantity Tr$[(\rho_A)^n]$ as the two-point function of twist operators on a torus:
\ba
\mbox{Tr}[(\rho_A)^n]=\la \sigma_n(z_1,\bar{z}_1)\sigma_n(z_2,\bar{z}_2)\lb_{T^2}, \label{torusa}
\ea
where we introduce the torus coordinate $(z,\bar{z})$ such that it is compactified as $z\sim z+1$ and $z\sim z+\tau$, where we set $\tau=i\beta$
in terms of the inverse temperature $\beta$. For a single Dirac fermion CFT i.e. $N=1$ this was studied in 
\cite{Azeyanagi:2007bj}. Refer also to  \cite{Chen:2014ehg} for general CFT arguments.
Below we will generalize this previous calculation  to our orbifolds for $N>1$.

\subsection{Entanglement entropy in cyclic orbifolds}

Consider entanglement entropy in the cyclic orbifold CFTs.
We can bosonize the twist operators in terms of free scalars as we have seen in (\ref{twistcycl}). Thus, one may think the Renyi entropy and entanglement entropy can be analytically computed in the cyclic orbifold CFTs by a straight generalization of the Dirac fermion case. However, if we consider the summation over all twisted sectors of the cyclic orbifold CFTs, the $n$ replicated fermions can have different boundary conditions on each torus in general. This makes the analytical computation complicated. Therefore, in this subsection, we focus on the diagonal sectors where all of the fermions $\psi^{(k)}_{L,p}$ for all values of $k=0,1,2,\ddd, N-1$ satisfy the same boundary condition, described by 
$(l,l')$, on the torus. Later, we will provide a full analysis including all non-diagonal sectors by focusing on the simplest case $N=2$ and $n=2$, i.e. the second Renyi entropy in the $\mathbb{Z}_2$ cyclic / $S_2$ symmetric orbifold.

\subsubsection{Two-point functions in free scalar CFT on torus}

In the $\mathbb{Z}_N$ cyclic orbifold, the twist operators are expressed by (\ref{twistcycl}) in terms of scalar fields. Therefore to calculate the two-point function of twist operators (\ref{torusa}), we need to know the two-point functions of scalar fields on a torus.

\begin{figure}[hhh]
  \centering
  \includegraphics[width=8cm]{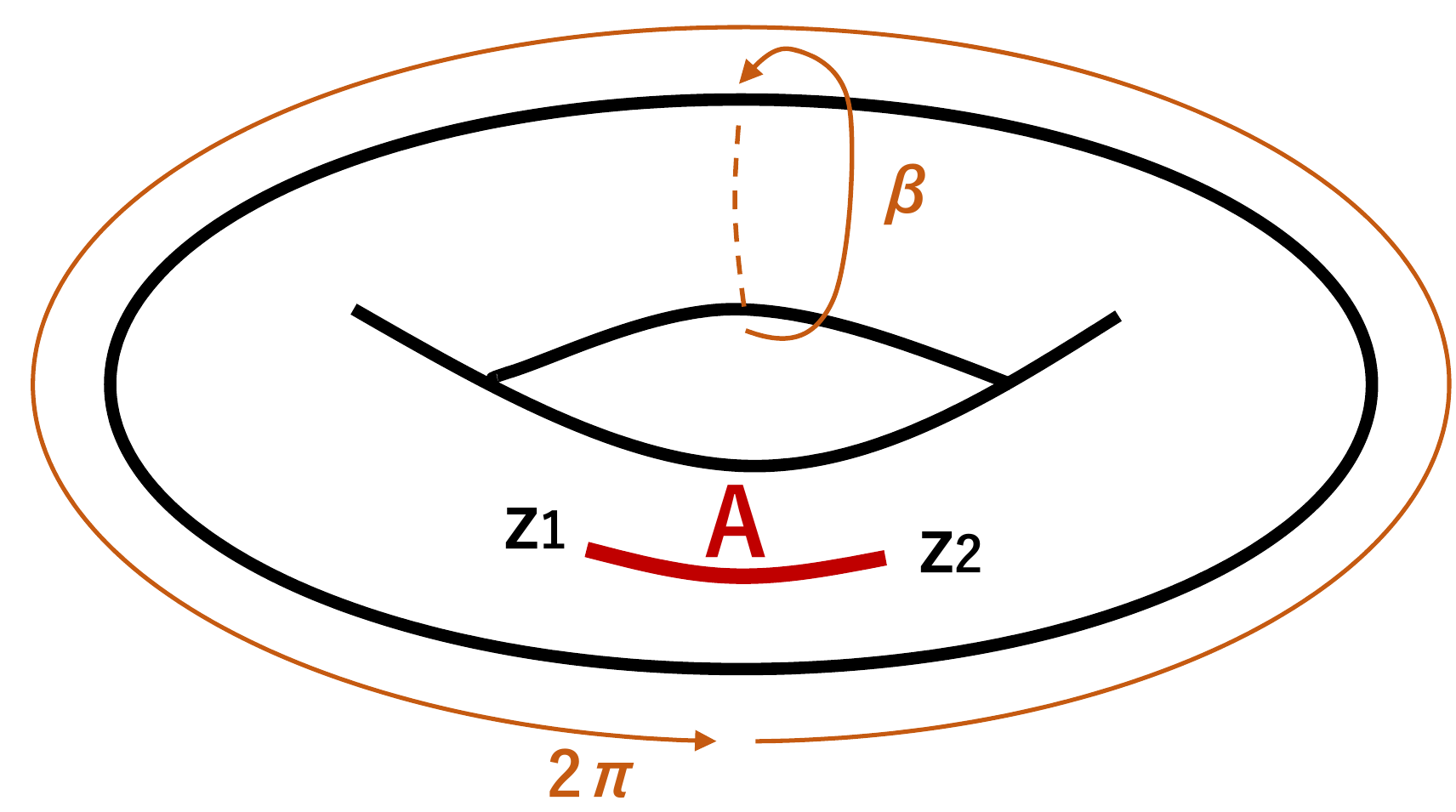}
  \caption{A sketch of calculation of entanglement entropy on a torus.} 
\label{fig:torus}
\end{figure}

Consider a massless free scalar field on a torus with the moduli $\tau=\tau_1+i\tau_2$. It obeys the twisted boundary condition compactified at radius $R$:
\ba
&& \vp(z+1,\bar{z}+1)=\vp(z,\bar{z})+2\pi R w+2\pi R\frac{k}{N}l,\no
&& \vp(z+\tau,\bar{z}+\bar{\tau})=\vp(z,\bar{z})+2\pi R w'+2\pi R\frac{k}{N}l',\label{twistbcgea}
\ea
where $w,w'\in \mathbb{Z}$ and $l,l'=0,1,2,\ddd,N-1$. Later we will set $R=2$. In this case, we find  $e^{i\vp_L}\to e^{2\pi i\frac{k}{N}l}e^{i\vp_L}$ and $e^{i\vp_R}\to e^{2\pi i\frac{k}{N}l}e^{i\vp_R}$, which fits nicely with the $\mathbb{Z_N}$ transformation of $\ti{\psi}^{(k)}_{L,R}$. In this interpretation, 
$l$ and $l'$ are the labels of twisted sectors and projections, respectively. Therefore, we will later fix the value of $l,l'$, and $k$ and sum over $w$ and $w'$ in order to focus on the contribution from a single scalar $\ti{\vp}^{(k)}_{p}$. 

Below we follow the computations in section 12.6 of the textbook \cite{Yellow} and generalize it to the twisted boundary condition shown above. We consider the two-point function $\la O_{e,m}O_{-e,-m}\lb$ on a torus, where 
\ba
&& O_{e,m}=e^{ip_L\vp_L+ip_R\vp_R}, \no
&& p_L=\frac{e}{R}+\frac{mR}{2},\ \ \ p_R=\frac{e}{R}-\frac{mR}{2}.
\ea
This has the conformal dimension 
\ba
h_{e,m}=\frac{1}{2}\left(\frac{e}{R}+\frac{mR}{2}\right)^2,\ \ \ 
\bar{h}_{e,m}=\frac{1}{2}\left(\frac{e}{R}-\frac{mR}{2}\right)^2.
\ea

We have
\ba
&& \mbox{Tr}_{(k,l)}\left[g^{l'}\cdot  O_{e,m}(z_1,\bar{z}_1)O_{-e,-m}(z_2,\bar{z}_2)q^{H}\bar{q}^{\bar{H}}\right]\no
&& =Z_{(l,l',k)}\cdot \la O_{e,m}(z_1,\bar{z}_1)O_{-e,-m}(z_2,\bar{z}_2)\lb_{(l,l',k)},
\ea
where $Z_{(l,l',k)}$ is the vacuum partition function with the twisted boundary condition (\ref{twistbcgea}). We also introduce $x_{12}+iy_{12}=z_{12}\equiv z_1-z_2$. We define 
$q=e^{2\pi i\tau}$ as usual.

This is evaluated in our setup as follows:
\ba
&& Z_{(l,l',k)}\cdot \la O_{e,m}(z_1,\bar{z}_1)O_{-e,-m}(z_2,\bar{z}_2)\lb_{(l,l',k)} \no
&& =\frac{R}{\s{2\tau_2}|\eta(\tau)|^2}\cdot \left[\frac{\de_z\theta_1(0|\tau)}{\theta_1(z_{12}|\tau)}\right]^{2h_{e,m}}
\left[\frac{\de_{\bar{z}}\theta_1(0|\bar{\tau})}{\theta_1(\bar{z}_{12}|\tau)}\right]^{2
\bar{h}_{e,m}}\no
&& \times \sum_{w,w'\in \mbox{Z}}
e^{-\frac{\pi R^2}{2\tau_2}\left[\left(w'-w\tau_1+\frac{k}{N}(l'-l\tau_1)\right)^2+\left(w+\frac{k}{N}l\right)^2\tau_2^2\right]}\cdot  e^{\frac{2\pi}{R^2\tau_2}\left(e y_{12}-im\frac{R^2}{2}x_{12}\right)^2}\no
&& \times e^{\left(\frac{2\pi i e}{\tau_2}y_{12}+\frac{\pi mR^2}{\tau_1}x_{12}\right)\left(w'+\frac{k}{N}l'\right)}
\cdot e^{\left[\frac{2\pi ie }{\tau_2}(x_{12}\tau_2-y_{12}\tau_1)-\frac{\pi m R^2}{\tau_2}(y_{12}\tau_2+x_{12}\tau_1)\right]\left(w+\frac{k}{N}l\right)}.\no
\label{parta}
\ea
In the above, the first factor $\frac{R}{\s{2\tau_2}|\eta(\tau)|^2}$ arises from the non-zero modes of the free scalar. 

By performing the Poisson resummation 
\ba
\sum_{w'}e^{-\pi a w'^2+2\pi
i b w'}=\frac{1}{\s{a}}\sum_{\ti{w}}e^{-\frac{\pi}{a}(\ti{w}-b)^2},
\ea
we can rewrite (\ref{parta}) as follows:
\ba
&& Z_{(l,l',k)}\cdot \la O_{e,m}(z_1,\bar{z}_1)O_{-e,-m}(z_2,\bar{z}_2)\lb_{(l,l',k)} \no
&& =\frac{1}{|\eta(\tau)|^2}\cdot \left[\frac{\de_z\theta_1(0|\tau)}{\theta_1(z_{12}|\tau)}\right]^{2h_{e,m}}
\left[\frac{\de_{\bar{z}}\theta_1(0|\bar{\tau})}{\theta_1(\bar{z}_{12}|\tau)}\right]^{2
\bar{h}_{e,m}}\no
&& \times \sum_{w,\ti{w}} e^{-\frac{2\pi\tau_2}{R^2}\ti{w}^2-\frac{\pi R^2\tau_2}{2}\left(w+\frac{k}{N}l\right)^2
+2\pi i\ti{w}\left(w\tau_1-\frac{k}{N}(l'-l\tau_1)\right)} \no
&&\times e^{2\pi i\left(\frac{2ie}{R^2}y_{12}+x_{12}m\right)\ti{w}+(2\pi ie x_{12}-\pi m R^2y_{12})\left(w+\frac{k}{N}l\right)}. \label{parfin}
\ea

In particular, if we set $e=m=0$ and $R=2$ (free fermion radius), we obtain\footnote{As we expect from the T-duality, we can get the same result for $R=1$. In this case, we need to rescale 
$k\to 2k$ in order to fit the definition of twisted boundary condition.}
\ba
&& Z_{(l,l',k)} \no
&& =\frac{1}{|\eta(\tau)|^2}\sum_{w,\ti{w}}e^{-\frac{\pi\tau_2}{2}\ti{w}^2+2\pi i\ti{w}\left(w\tau_1-\frac{k}{N}(l'-l\tau_1)\right)}
\cdot e^{-2\pi \left(w+\frac{k}{N}l\right)^2\tau_2}\no
&&  =\sum_{w,\ti{w}}q^{\frac{1}{2}\left(w+\ti{w}/2\right)^2}\bar{q}^{\frac{1}{2}\left(w-\ti{w}/2\right)^2}
\cdot e^{-2\pi i\ti{w}\frac{k}{N}(l'-l\tau_1)}\cdot e^{-4\pi \frac{k}{N}lw\tau_2-2\pi\frac{k^2 l^2}{N^2}\tau_2}.
\ea

When $\ti{w}$ is even, we introduce the integers $r=w+\ti{w}/2$ and $s=w-\ti{w}/2$. When  $\ti{w}$ is odd, we introduce the integers $r+\frac{1}{2}=w+\ti{w}/2$ and $s-\frac{1}{2}=w-\ti{w}/2$. In both cases, we find that $(r,s)$ are any integers that satisfy the constraint $r-s\equiv 0$ (mod $2$).
This leads to the decomposition into three sectors:
\ba
&& Z_{(l,l',k)} \no
&& = \frac{e^{-2\pi\frac{k^2 l^2}{N^2}\tau_2}}{2|\eta(\tau)^2|}\left[|Z_3|^2+|Z_4|^2+|Z_2|^2\right],
\ea
where 
\ba
&& Z_3=\sum_{r\in\mbox{Z}}q^{r^2/2}e^{-2\pi i\frac{k}{N}rl'}e^{2\pi i\frac{k}{N}lr\tau}=\theta_3\left(-\frac{k}{N}l'+\frac{k}{N}l\tau|\tau\right),\no
&& Z_4=\sum_{r\in\mbox{Z}}(-1)^r q^{r^2/2}e^{-2\pi i\frac{k}{N}rl'}e^{2\pi i\frac{k}{N}lr\tau}=\theta_4\left(-\frac{k}{N}l'+\frac{k}{N}l\tau|\tau\right),\no
&& Z_2=\sum_{r\in\mbox{Z}}q^{(r+1/2)^2/2}e^{-2\pi i\frac{k}{N}l'(r+1/2)}e^{2\pi i\frac{k}{N}l(r+1/2)\tau}=\theta_2\left(-\frac{k}{N}l'+\frac{k}{N}l\tau|\tau\right),\no
\ea
Since we can flip the sign $l'\to -l'$, the above result reproduces the vacuum partition function (\ref{partNS}) as
\begin{align}
Z_{N,\mathbb{Z}}^{NS}=&~ \frac{1}{N}\sum_{l,l'=0}^{N-1}\prod_{k=0}^{N-1} Z_{(l,l',k)}^{NS} \nonumber\\
=&~ \frac{1}{N}\sum_{l,l'=0}^{N-1}\prod_{k=0}^{N-1}e^{-2\pi\frac{k^2 l^2}{N^2}\tau_2}
\frac{|\theta_3\left(\frac{k}{N}l'+\frac{k}{N}l\tau|\tau\right)|^2}{|\eta(\tau)|^2},
\end{align}
where we restrict to the NS sector trace Tr$_{NS}[1]$.

\subsubsection{Two-point function of twist operators in diagonal sectors}

Now we would like to turn to the calculation of entanglement entropy.
For this purpose we need to calculate the two-point function of the twist operators (\ref{twistcycl}).  We can realize the twist operator $\ti{\sigma}_n$ in terms of $O_{e,m}$ as 
\ba
&& \sigma_n=\prod_{p=-(n-1)/2}^{(n-1)/2}\prod_{k=0}^{N-1}O^{(k)}_{0,\frac{p}{n}},\no
&&  O^{(k)}_{0,\frac{p}{n}}= e^{i\frac{p}{n}(\ti{\vp}^{(k)}_{L,p}-\ti{\vp}^{(k)}_{R,p})}.
\ea
for each sector $(l,l')$.  We chose $z_1=L$ and $z_2=0$ for the subsystem $A$. 

In this subsection, we focus on the diagonal sectors i.e. the fermions for any $k$ respect the same boundary condition $(l,l')$. This is simply because the non-diagonal sectors are too complicated, though we will work out the full expression for $N=n=2$ in subsection \ref{sec:Nn2}.

By using the result (\ref{parfin}), for fixed values of $(l,l')$ and the free fermion radius $R=2$, picking up the NS sector result, we obtain 
\ba
&&  \left(Z^{NS}_{(l,l')}\right)^n\cdot \la \sigma_n(L) \bar{\sigma}_n(0)\lb_{(l,l')}\ \ \no
&& =\prod_{p=-(n-1)/2}^{(n-1)/2}\prod_{k=0}^{N-1}\frac{e^{-2\pi
\frac{k^2 l^2}{N^2}\tau_2}}{|\eta(\tau)^2|}\left|\frac{2\pi\eta(\tau)^3}{\theta_1(L|\tau)}\right|^{4h_p}\cdot \left|\theta_3\left(-\frac{k}{N}l'+\frac{k}{N}l\tau+\frac{p}{n}L|\tau\right)\right|^2,\label{replicamethodtofermion}
\ea
where we introduced $h_p=\frac{p^2}{2n^2}$ and  $Z^{NS}_{(l,l')}=\prod_{k=0}^{N-1}Z_{(l,l',k)}^{NS}$.

\subsubsection{Renyi entropy in diagonal sectors}
\label{REiDS}

Now let us calculate the entanglement entropy for a fixed diagonal sector. The $n$-th Renyi entropy $S^{(n)}_{A(l,l')}$ is written as
\ba
S^{(n)}_{A(l,l')} &=& \frac{1}{1-n}\left[\log \la \sigma_n(L) \bar{\sigma}_n(0)\lb_{(l,l')}\right]\no
&\equiv& S^{(n)1}_{A(l,l')}+S^{(n)2}_{A(l,l')}.
\ea
Here, we defined the first term $S^{(n)1}_{A(l,l')}$ as the $\theta_1$ contribution:
\ba
S^{(n)1}_{A(l,l')}=\frac{N}{6}\left(1+\frac{1}{n}\right)\log\left|\frac{\theta_1(L|\tau)}{2\pi\delta\eta(\tau)^3}\right|,
\ea
where the central charge is given by $c=N$ and we inserted the UV cut-off $\delta$. 
The second contribution is from the other terms:
\begin{align}
    &~ S^{(n)2}_{A(l,l')}\nonumber\\
=&~ \frac{1}{1-n}\sum_{p=-(n-1)/2}^{(n-1)/2}\sum_{k=0}^{N-1}\log\left[\frac{e^{-2\pi
\frac{k^2 l^2}{N^2}\tau_2}}{|\eta(\tau)^2|}\cdot \left|\theta_3\left(-\frac{k}{N}l'+\frac{k}{N}l\tau+\frac{p}{n}L|\tau\right)\right|^2\right]-\frac{n}{1-n}\log\left[Z^{NS}_{(l,l')}\right].
\end{align}
We are interested in the difference between the Renyi entropy for the total system $L=1-\sigma$ and that for the infinitesimally small subsystem $L=\sigma$ with the limit 
$\sigma\to 0$:
\begin{align}
    &~\lim_{\sigma\to 0}\left[S^{(n)}_{A(l,l')}(1-\sigma)-S^{(n)}_{A(l,l')}(\sigma)\right] \nonumber\\
    =&~\lim_{\sigma\to 0}\left[S^{(n)2}_{A(l,l')}(1-\sigma)-S^{(n)2}_{A(l,l')}(\sigma)\right] \nonumber\\
    =&~\frac{1}{1-n}\sum_{p=-(n-1)/2}^{(n-1)/2}\sum_{k=0}^{N-1}\log\left[\frac{e^{-2\pi\frac{k^2 l^2}{N^2}\tau_2}}{|\eta(\tau)^2|}\cdot \left|\theta_3\left(-\frac{k}{N}l'+\frac{k}{N}l\tau+\frac{p}{n}|\tau\right)\right|^2\right]\nonumber\\
    &-\frac{1}{1-n}\sum_{p=-(n-1)/2}^{(n-1)/2}\sum_{k=0}^{N-1}\log\left[\frac{e^{-2\pi
\frac{k^2 l^2}{N^2}\tau_2}}{|\eta(\tau)^2|}\cdot \left|\theta_3\left(-\frac{k}{N}l'+\frac{k}{N}l\tau|\tau\right)\right|^2\right].
\end{align}
We would like to argue 
\begin{align}
&~ \lim_{\sigma\to 0}\left[S^{(n)}_{A(l,l')}(1-\sigma)-S^{(n)}_{A(l,l')}(\sigma)\right]\nonumber\\
=&~\frac{1}{1-n}\log\frac{Z_{(l,nl')}^{NS}(n\tau)}{Z_{(l,l')}^{NS}(\tau)^n},\label{relatuonfgdh}
\end{align}
where $Z_{(l,l')}^{NS}(\tau)$ is 
\begin{align}
    Z_{(l,l')}^{NS}(\tau)=&~\mbox{Tr}_{(l)}\left[g^{l'} q^{H_L}\bar{q}^{H_R}\right] \nonumber\\
    =&~\prod_{k=0}^{N-1}\frac{e^{-2\pi\frac{k^2 l^2}{N^2}\tau_2}}{|\eta(\tau)^2|}\cdot \left|\theta_3\left(-\frac{k}{N}l'+\frac{k}{N}l\tau|\tau\right)\right|^2, \\
    Z_{(l,nl')}^{NS}(n\tau)=&~\mbox{Tr}_{(l)}\left[g^{nl'} q^{nH_L}\bar{q}^{nH_R}\right] \nonumber\\
    =&~\prod_{k=0}^{N-1}\frac{e^{-2\pi\frac{k^2 l^2}{N^2}n\tau_2}}{|\eta(n\tau)^2|}\cdot \left|\theta_3\left(-\frac{k}{N}nl'+\frac{k}{N}ln\tau|n\tau\right)\right|^2.
\end{align}
We can show (\ref{relatuonfgdh}) by employing the formula (\ref{formulatthe}) to perform the sum over $p$.
Note that when considering $Z_{(l,nl')}^{NS}(n\tau)$, not only $\tau$ but also the $\mathbb{Z}_N$ action $g^{l'}$ get multiplied by $n$.

We can extend this procedure to the full diagonal Renyi entropy $S^{(n)}_{A,diag}$.
The first contribution is universal and thus is the same as the one in the fixed diagonal sector:
\begin{align}
    S^{(n)}_{A,diag}  =&~ \frac{1}{1-n}\left[\log\langle\sigma_n(L)\bar{\sigma }_n(0)\rangle\right] \nonumber\\
    \equiv&~ S^{(n)1}_{A,diag}+S^{(n)2}_{A,diag}, \\
    S^{(n)1}_{A,diag} =&~ S^{(n)1}_{A(l,l')}=\frac{N}{6}\left(1+\frac{1}{n}\right)\log\left|\frac{\theta_1(L|\tau)}{2\pi\delta\eta(\tau)^3}\right|.
\end{align}
The second contribution is from the other terms:
\begin{align}
    S^{(n)2}_{A,diag} =&~ \frac{1}{1-n}\log\left[\frac{1}{N}\sum_{l,l'=0}^{N-1}\prod_{p=-(n-1)/2}^{(n-1)/2}\prod_{k=0}^{N-1}\frac{e^{-2\pi
    \frac{k^2 l^2}{N^2}\tau_2}}{|\eta(\tau)^2|}\cdot \left|\theta_3\left(-\frac{k}{N}l'+\frac{k}{N}l\tau+\frac{p}{n}L|\tau\right)\right|^2\right] \nonumber\\
    &-\frac{n}{1-n}\log\left[\frac{1}{N}\sum_{l,l'=0}^{N-1}\prod_{k=0}^{N-1}\frac{e^{-2\pi\frac{k^2 l^2}{N^2}\tau_2}}{|\eta(\tau)^2|}\cdot \left|\theta_3\left(-\frac{k}{N}l'+\frac{k}{N}l\tau|\tau\right)\right|^2\right].
\end{align}
Although this equality is somewhat complicated, the following quantity is relatively understandable:
\begin{align}
    &~\lim_{\sigma\to 0}\left[S^{(n)}_{A,diag}(1-\sigma)-S^{(n)}_{A,diag}(\sigma)\right] \nonumber\\
    =&~ \frac{1}{1-n}\log\left[\frac{1}{N}\sum_{l,l'=0}^{N-1}\prod_{p=-(n-1)/2}^{(n-1)/2}\prod_{k=0}^{N-1}\frac{e^{-2\pi
    \frac{k^2 l^2}{N^2}\tau_2}}{|\eta(\tau)^2|}\cdot \left|\theta_3\left(-\frac{k}{N}l'+\frac{k}{N}l\tau+\frac{p}{n}|\tau\right)\right|^2\right] \nonumber\\
    &-\frac{1}{1-n}\log\left[\frac{1}{N}\sum_{l,l'=0}^{N-1}\prod_{p=-(n-1)/2}^{(n-1)/2}\prod_{k=0}^{N-1}\frac{e^{-2\pi
    \frac{k^2 l^2}{N^2}\tau_2}}{|\eta(\tau)^2|}\cdot \left|\theta_3\left(-\frac{k}{N}l'+\frac{k}{N}l\tau|\tau\right)\right|^2\right] \nonumber\\
    =&~ \frac{1}{1-n}\log\left[\frac{\frac{1}{N}\sum_{l,l'=0}^{N-1}Z_{(l,nl')}^{NS}(n\tau)}{\frac{1}{N}\sum_{l,l'=0}^{N-1}\left(Z_{(l,l')}^{NS}(\tau)\right)^n}\right]. \label{eq:diagonalRE}
\end{align}
We employ the formula (\ref{formulatthe}) again to perform the product over $p$. As we can see, only the first terms of $S^{(n)2}_{A,diag}(1-\sigma)$ and $S^{(n)2}_{A,diag}(\sigma)$ contribute. 

We should notice again that the above calculation is for the diagonal sectors. It is necessary to consider off-diagonal sectors---applying different twist-boundary conditions (labeled by $(l,l')$) for different replica sheets. Summing over all possible boundary condition will yield the complete Renyi entropy. However, naive construction of replica-twist operators cannot formulate the different boundary conditions for arbitrary $N$ and $n$.
If we could calculate the complete Renyi entropy $S^{(n)}_A$ including contributions from both diagonal and off-diagonal sectors, then we expect instead of (\ref{relatuonfgdh})
\ba
&& \lim_{\sigma\to 0}\left[S^{(n)}_{A}(1-\sigma)-S^{(n)}_{A}(\sigma)\right]
\no
&&=\frac{1}{1-n}\log\frac{Z(n\tau)}{Z(\tau)^n},\label{relatuonfgda}
\ea
where $Z(\tau)$ is the full partition function of any given two-dimensional CFT, including both the cyclic and symmetric orbifolds. Later we will explicitly confirm this relation (\ref{relatuonfgda}) for $N=n=2$. 

For reference, we perform here the high-temperature expansion of $S^{(n)}_{A(l,l')}$. 
By performing the modular S transformation to $S^{(n)1}_{A(l,l')}$, we obtain (we set $\tau=i\beta)$
\ba
S^{(n)1}_{A(l,l')}=\frac{N}{6}\left(1+\frac{1}{n}\right)\log\left[\frac{\beta}{\pi\delta}
e^{-\frac{\pi L^2}{\beta}}\sinh\left(\frac{\pi L}{\beta}\right)\prod_{m=1}^{\infty}
\frac{(1-e^{\frac{2\pi L}{\beta}}e^{-\frac{2\pi}{\beta} m})(1-e^{-\frac{2\pi L}{\beta}}e^{-\frac{2\pi}{\beta} m})}{(1-e^{-\frac{2\pi}{\beta}m})^2}\right].\no
\ea
To evaluate $S^{(n)2}_{A(l,l')}$, we employ the following formula:
\begin{align}
&~ \left|\theta_3\left(-\frac{k}{N}l'+\frac{k}{N}l\tau+\frac{p}{n}L\biggl|\tau\right)\right|^2
\no
=&~ \beta\cdot e^{-\frac{2\pi}{\beta}\left(\frac{kl'}{N}+\frac{pL}{n}\right)^2}
\cdot e^{2\pi\beta\frac{k^2l^2}{N^2}}\left|\theta_3\left(-\frac{k}{N\tau}l'+\frac{k}{N}l+\frac{p}{n\tau}L\biggl|-\frac{1}{\tau}\right)\right|^2.
\end{align}
This leads to
\ba
S^{(n)2}_{A(l,l')}&=&\frac{1}{1-n}\sum_{p=-(n-1)/2}^{(n-1)/2}\left(-\frac{2\pi N}{\beta}\right)\frac{p^2L^2}{n^2}+\ti{S}^{(n)2}_{A(l,l')}\no
&=&\frac{\pi N}{6\beta}\left(1+\frac{1}{n}\right)L^2+\ti{S}^{(n)2}_{A(l,l')},
\ea
where the contribution $\ti{S}^{(n)2}_{A(l,l')}$ is evaluated as follows
\ba
&&\ti{S}^{(n)2}_{A(l,l')}\no
&&=\frac{1}{1-n}\sum_{p=-(n-1)/2}^{(n-1)/2}\sum_{k=0}^{N-1}\log\left[
\frac{e^{-\frac{2\pi k^2l'^2}{N^2\beta }}e^{-\frac{4\pi kpl'L}{Nn\beta}}}{|\eta(\tau)|^2}
\left|\theta_3\left(\frac{k}{N}l+i\frac{k}{N\beta}l'-i\frac{pL}{n\beta}\biggl|\frac{i}{\beta}\right)\right|^2\right]\no
&& \ \ \ \ -\frac{n}{1-n}\sum_{k=0}^{N-1}\log\left[
\frac{e^{-\frac{2\pi k^2l'^2}{N^2\beta }}}{|\eta(\tau)|^2}
\left|\theta_3\left(\frac{k}{N}l+i\frac{k}{N\beta}l'\biggl|\frac{i}{\beta}\right)\right|^2\right]\no
&&=\frac{1}{1-n}\sum_{p=-(n-1)/2}^{(n-1)/2}\sum_{k=0}^{N-1}\log\left[
\left|\theta_3\left(\frac{k}{N}l+i\frac{k}{N\beta}l'-i\frac{pL}{n\beta}\biggl|\frac{i}{\beta}\right)\right|^2\cdot \left|\theta_3\left(\frac{k}{N}l+i\frac{k}{N\beta}l'\biggl|\frac{i}{\beta}\right)\right|^{-2}\right].\no
\ea

\subsubsection{Von-Neumann entanglement entropy in diagonal sectors}
Next, we compute the entanglement entropy by taking the von-Neumann limit $n=1$ in the replica method. We again focus on the diagonal sectors.
The part of the Renyi entropy 
$\ti{S}^{(n)2}_{A(l,l')}$ can be explicitly written as 
\ba
&& \ti{S}^{(n)2}_{A(l,l')}=\frac{1}{1-n}\sum_{p=-(n-1)/2}^{(n-1)/2}\sum_{k=0}^{N-1}
\log|F_{k,p}|^2,\no
&& F_{k,p}\equiv\prod_{m=1}^\infty\frac{(1+e^{\frac{2\pi ikl}{N}}e^{-2\pi\frac{kl'}{N\beta}}e^{2\pi\frac{pL}{n\beta}}e^{-\frac{2\pi}{\beta}(m-1/2)})(1+e^{\frac{-2\pi ikl}{N}}e^{2\pi\frac{kl'}{N\beta}}e^{-2\pi\frac{pL}{n\beta}}e^{\frac{2\pi}{\beta}(m-1/2)})}{(1+e^{\frac{2\pi ikl}{N}}e^{-2\pi\frac{kl'}{N\beta}}e^{-\frac{2\pi}{\beta}(m-1/2)})(1+e^{-\frac{2\pi ikl}{N}}e^{2\pi\frac{kl'}{N\beta}}e^{\frac{2\pi}{\beta}(m-1/2)})}.\no
\ea
Then we can expand the logarithm as follows
\ba
 \ti{S}^{(n)2}_{A(l,l')}
&=&\frac{1}{1-n}\sum_{p=-(n-1)/2}^{(n-1)/2}\sum_{k=0}^{N-1}\sum_{m=1}^\infty\sum_{s=1}^\infty\frac{(-1)^{s-1}}{s}e^{-2\pi\left(m-\frac{1}{2}\right)\frac{s}{\beta}}\left[G_{s,m,p,k}-G_{s,m,0,k}\right],\no
G_{s,m,p,k}&\equiv& e^{2\pi\left(i\frac{lks}{N}-\frac{l'ks}{N\beta}+\frac{Lps}{n\beta}\right)}+e^{2\pi\left(-i\frac{lks}{N}-\frac{l'ks}{N\beta}+\frac{Lps}{n\beta}\right)}+e^{2\pi\left(i\frac{lks}{N}+\frac{l'ks}{N\beta}-\frac{Lps}{n\beta}\right)}\no
&& +e^{2\pi\left(-i\frac{lks}{N}+\frac{l'ks}{N\beta}-\frac{Lps}{n\beta}\right)}.
\ea
By performing the summation over $m$, we obtain
\ba
 \ti{S}^{(n)2}_{A(l,l')}
&=&\frac{1}{1-n}\sum_{p=-(n-1)/2}^{(n-1)/2}\sum_{s=1}^{\infty}
\frac{(-1)^{s-1}}{s\cdot \sinh\left(\frac{\pi s}{\beta}\right)}\cdot \left[\frac{\sinh\left(\frac{\pi Ls}{\beta}\right)}{\sinh\left(\frac{\pi Ls}{n\beta}\right)}-n\right]\no
&&\ \ \ \ \times \ \left[\frac{\sinh\left[N\left(\pi i\frac{ls}{N}-\frac{\pi l's}{N\beta}\right)\right]}{\sinh\left[\pi i\frac{ls}{N}-\frac{\pi l's}{N\beta}\right]}+\frac{\sinh\left[N\left(\pi i\frac{ls}{N}+\frac{\pi l's}{N\beta}\right)\right]}{\sinh\left[\pi i\frac{ls}{N}+\frac{\pi l's}{N\beta}\right]}\right].\nonumber
\ea
By taking the von-Neumann limit $n\to 1$, we find
\ba
 \ti{S}^{(1)2}_{A(l,l')}&=&\sum_{s=1}^\infty\frac{(-1)^{s}}{s\cdot \sinh\left(\frac{\pi s}{\beta}\right)}\cdot \left[\frac{\pi Ls}{\beta}\coth\left(\frac{\pi sL}{\beta}\right)-1\right]\no
 &&\ \ \ \times \left[\frac{\sinh\left[N\left(\pi i\frac{ls}{N}-\frac{\pi l's}{N\beta}\right)\right]}{\sinh\left[\pi i\frac{ls}{N}-\frac{\pi l's}{N\beta}\right]}+\frac{\sinh\left[N\left(\pi i\frac{ls}{N}+\frac{\pi l's}{N\beta}\right)\right]}{\sinh\left[\pi i\frac{ls}{N}+\frac{\pi l's}{N\beta}\right]}\right].\nonumber
\ea
The final expression of the entanglement entropy for the $(l,l')$ sector is given by 
\ba
&& S^{(1)}_{A(l,l')}\no
&& =\frac{N}{3}\!\log\! \left[\frac{\beta}{\pi\delta}\sinh\left(\frac{\pi L}{\beta}\right)\right]\!+\!\frac{N}{3}\!\sum_{m=1}^\infty\log\!\left[\frac{\left(1-e^{\frac{2\pi L}{\beta}}e^{-\frac{2\pi}{\beta}m}\right)\left(1-e^{-\frac{2\pi L}{\beta}}e^{-\frac{2\pi}{\beta}m}\right)}{\left(1-e^{-\frac{2\pi}{\beta}m}\right)^2}
\right]\!+\!\ti{S}^{(1)2}_{A(l,l')}.\no
\ea

\subsection{Time-like correlation function in cyclic orbifolds}
\label{sec:time-like C}

Before we proceed to study entanglement entropy further, we would like to stop here to explore the periodicity under the time evolution in cyclic orbifolds.  First note that in the $k$-th fermion, $l$-th twisted sector is defined by
the twisted boundary condition 
\ba
\psi^{(k)}_{L,R}(x+1)=e^{\frac{2\pi ikl}{N}}\psi(x)^{(k)}_{L,R}
\ea
for both the left $(L)$ and right $(R)$ moving sectors. Then the quantized fermionic modes look like
\ba
\left(\psi^{(k)}_L\right)_{-n+\frac{kl}{N}},\ \ \ \left(\psi^{(k)}_R\right)_{-n-\frac{kl}{N}},\ \ \ \ \ (n\in \mathbb{Z}).
\ea

Now we set
\ba
O(z,\bar{z})=\psi^{(k)}_L(z)\bar{\psi}^{(k)}_R(\bar{z})=e^{i\vp^{(k)}_L(z)-i\vp^{(k)}_R(\bar{z})},
\ea
which has a non-trivial phase under the time translation as 
\ba
e^{i2\pi Ht}\left[\left(\psi^{(k)}_L\right)_{-n+\frac{kl}{N}} \left(\bar{\psi}^{(k)}_R\right)_{-m+\frac{kl}{N}}\right]e^{-2\pi iHt}
=e^{2\pi i(n+m-\frac{2kl}{N})t}\left[\left(\psi^{(k)}_L\right)_{-n+\frac{kl}{N}} \left(\bar{\psi}^{(k)}_R\right)_{-m+\frac{kl}{N}}\right].\no\label{timeev}
\ea

We would like to evaluate the two-point function $\la O(z,\bar{z})O(0,0)\lb$, by choosing the time-like separation: $(z,\bar{z})=(T,-T)$. Or if we set $z_{12}=x_{12}+iy_{12}$ we set $y_{12}=iT$. By plugging this and $(e,m)=(0,1)$ into (\ref{parta}), we obtain 
\begin{align}
&~ Z_{(l,l')}\cdot \la O(T,-T)O(0,0)\lb_{(l,l')} \nonumber\\
=&~ Z_{(l,l')}~
e^{-4\pi i \frac{klT}{N}}\left|\frac{2\pi\eta^3}{\theta_1(T|\tau)}\right|^2
\frac{\theta_3(\frac{kl\tau-kl'}{N}-T|\tau)\overline{\theta_3(\frac{kl\tau-kl'}{N}+T|\tau)}}
{\left|{\theta_3(\frac{kl\tau-kl'}{N}|\tau)}\right|^2}.
\end{align}
The phase factor $e^{-4\pi iT\frac{kl}{N}}$ is consistent with (\ref{timeev}). Summation over $(l,l')$ gives the two-point function $Z\cdot \langle O(T,-T)O(0,0)\rangle$. This shows that the periodicity in the time direction is $T\sim T+N$ for odd $N$ and $T\sim T+N/2$ for even $N$.

\subsection{Second Renyi entropy at $N=2$}
\label{sec:Nn2}

\begin{figure}[hhh]
  \centering
  \includegraphics[width=8cm]{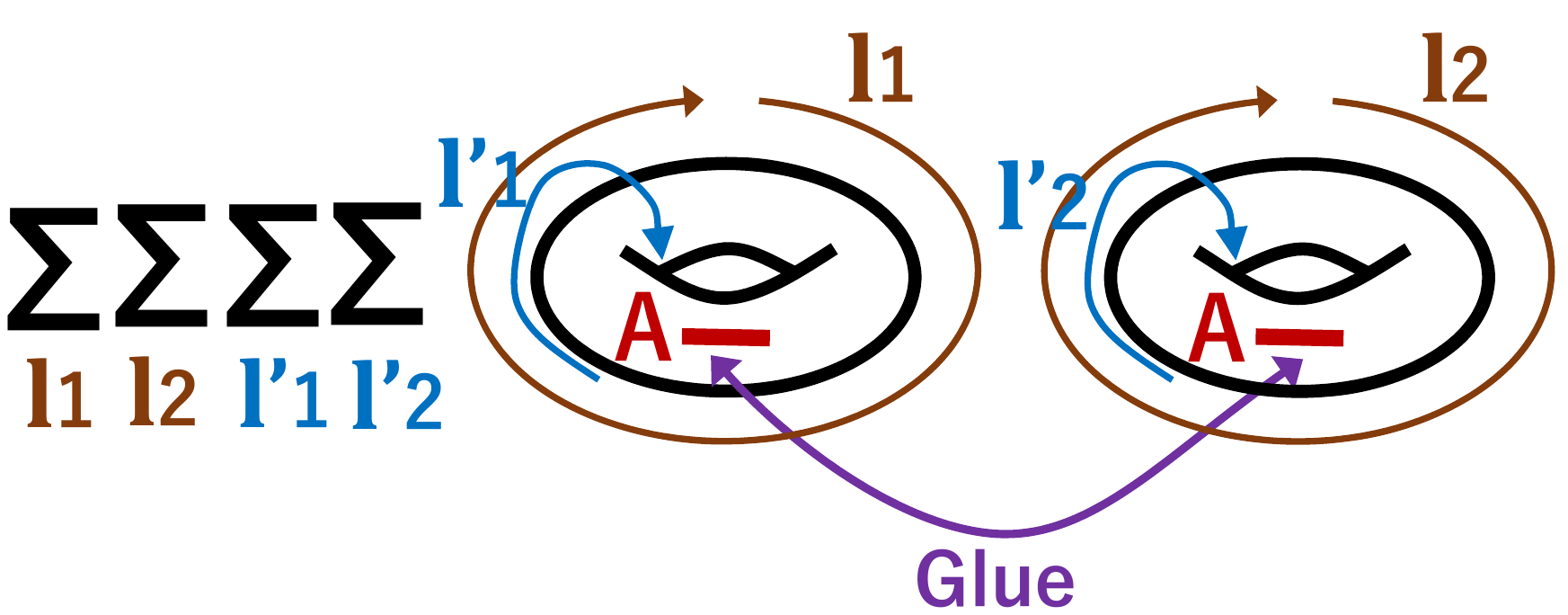}
  \caption{The replica method calculation of second Renyi entropy on a torus for a cyclic orbifold CFT.} 
\label{fig:torusum}
\end{figure}

To fully evaluate the Renyi entropy including both the diagonal and non-diagonal sectors, here we analyze the second Renyi entropy in the $\mathbb{Z}_2$ cyclic orbifold or equally the $S_2$ symmetric orbifold, i.e. $N=n=2$. In this case, the twisted operators are found as (\ref{nntwote}). We again write the full partition function at finite temperature $\beta$ to be $Z(\beta)$.
When we fully sum over all sectors,  the Tr$\rho_A^2$ is written in the form
\ba
 Z(\beta)\cdot \la \sigma_2(L)\bar{\sigma}_2(0)\lb
 =Z_{k=0}(L,\tau)\cdot Z_{k=1}(L,\tau),
\ea
where we find the $k=0$ contributions from our previous calculation as follows:
\ba
Z_{k=0}(L,\tau)=\frac{1}
 {|\eta(\tau)|^4}\cdot \left|\frac{2\pi \eta(\tau)^3}{\theta_1(L|\tau)}\right|^{1/4}\cdot \left|\theta_3\left(\frac{L}{4}|\tau\right)\right|^4.
\ea

We can decompose the final part as follows
\ba
Z_{k=1}(L,\tau)=\frac{1}{4}\sum_{l_1,l'_1=0}^1 \sum_{l_2,l'_2=0}^1
Z^{(l_2,l'_2)}_{(l_1,l'_1)}(L,\tau),
\ea
where $(l_1,l'_1)$ and $(l_2,l'_2)$ describe the first and second sheets of replicated partition functions. The sectors with $(l_1,l_2)\neq (l_2,l'_2)$ are off-diagonal.

It is useful to extract the factor related to 
$\theta_1(L|\tau)$ as 
\ba
 Z(\beta)\cdot \la \sigma_2(L)\bar{\sigma}_2(0)\lb
 =\frac{1}
 {|\eta(\tau)|^4}\cdot \left|\frac{2\pi \eta(\tau)^3}{\theta_1(L|\tau)}\right|^{1/2}\cdot \left|\theta_3\left(\frac{L}{4}|\tau\right)\right|^4\cdot \ti{Z}_{k=1}(L,\tau),
\ea
in order to remove the singular behavior in the $L\to 0$ limit.

We can again decompose the final part as follows
\ba
\ti{Z}_{k=1}(L,\tau)=\frac{1}{4}\sum_{l_1,l'_1=0}^1 \sum_{l_2,l'_2=0}^1
\ti{Z}^{(l_2,l'_2)}_{(l_1,l'_1)}(L,\tau).
\ea

The second Renyi entropy is computed from the two-point function 
\ba
S^{(2)}_A=-\log  \la \sigma_2(L)\bar{\sigma}_2(0)\lb.
\ea
Thus we find the difference $S_A-S_{A^c}$ in the limit where $A$ approaches the total space reads
\ba
S^{(2)}_A(L=1)-S^{(2)}_A(L=0)=-\log\left[\frac{Z^{(0)}(2\tau)}{Z^{(0)}(\tau)^2}\right]
-\log\left[\frac{\ti{Z}_{k=1}(1,\tau)}{\ti{Z}_{k=1}(0,\tau)}\right],
\label{entdift}
\ea
where $Z^{(0)}(\tau)$ is the vacuum partition function of $k=0$ part
(i.e. a single Dirac fermion)
\be
Z^{(0)}(\tau)=\frac{|\theta_3(\tau)|^2}{|\eta(\tau)|^2}.
\ee
Indeed, the first term in the right-hand side of (\ref{entdift}) is equal to the 2nd Renyi thermal entropy of the $k=0$ part of
the $N=2$ cyclic orbifold CFT, which is identical to the single Dirac fermion CFT. We would like to argue that the second term coincides with the 2nd Renyi thermal entropy of the $k=1$ part as follows:
\ba
\frac{\ti{Z}_{k=1}(1,\tau)}{\ti{Z}_{k=1}(0,\tau)}=\frac{Z^{(1)}(2\tau)}{Z^{(1)}(\tau)^2},\label{entropref}
\ea
where $Z^{(1)}(\tau)$ is the vacuum partition function of $k=1$ part
(i.e. a $\mathbb{Z}_2$ orbifolded fermion)
\ba
Z^{(1)}(\tau)=\frac{|\theta_3(\tau)|^2+|\theta_2(\tau)|^2+|\theta_4(\tau)|^2}{2|\eta(\tau)|^2}.
\ea
This relation between the difference of the Renyi entropy and the thermal entropy is physically obvious as $L=1$ means that the subsystem $A$ coincides with the total system. However, its explicit derivation is non-trivial as we will present below.

\subsubsection{Diagonal parts}
We already computed the diagonal case i.e. $l_1=l_2(=l)$ and $l'_1=l'_2(=l')$, which leads to 
\ba
\ti{Z}^{(l,l')}_{(l,l')}(L,\tau)=e^{-\pi l^2\tau_2}\frac{\left|\theta_3\left(-\frac{l'}{2}+\frac{\tau l}{2}+\frac{L}{4}\right)\right|^2\left|\theta_3\left(-\frac{l'}{2}+\frac{\tau l}{2}-\frac{L}{4}\right)\right|^2}{|\eta(\tau)|^4}.
\label{lzerobb}
\ea
In the limit $L\to 0$,  we  obtain  
\ba
&& \ti{Z}^{(0,0)}_{(0,0)}(0,\tau)=\frac{|\theta_3(\tau)|^4}{|\eta(\tau)|^4},\ \ \ \
 \ti{Z}^{(0,1)}_{(0,1)}(0,\tau)=\frac{|\theta_4(\tau)|^4}{|\eta(\tau)|^4},\no
&& \ti{Z}^{(1,0)}_{(1,0)}(0,\tau)=\frac{|\theta_2(\tau)|^4}{|\eta(\tau)|^4}, \ \ \ \
 \ti{Z}^{(1,1)}_{(1,1)}(0,\tau)=0.\label{lzeroa}
\ea
In the limit $L\to 1$,  we obtain  
\ba
&& \ti{Z}^{(0,0)}_{(0,0)}(1,\tau)=\ti{Z}^{(0,1)}_{(0,1)}(1,\tau)=\frac{|\theta_3(2\tau)|^2}{|\eta(2\tau)|^2},\no
&& \ti{Z}^{(1,0)}_{(1,0)}(1,\tau)=\frac{|\theta_2(2\tau)|^2}{|\eta(2\tau)|^2},\ \ \ \ 
 \ti{Z}^{(1,1)}_{(1,1)}(1,\tau)=0.\label{lonea}
\ea

\subsubsection{Off-diagonal parts}

The off-diagonal partition functions are new. First, note the symmetry 
\ba
Z^{(a,b)}_{(c,d)}(L,\tau)=Z^{(c,d)}_{(a,b)}(L,\tau).
\ea

Let us start with 
$(l_1,l'_1)=(0,0)$ and $(l_2,l'_2)=(1,0)$. In this case, the 
fermion follows the boundary condition in the real $z$-direction:
\ba
&& \ti{\psi}^{(0)}_{\pm}(z+1)=\ti{\psi}^{(0)}_{\pm}(z),\no
&& \ti{\psi}^{(1)}_{\pm}(z+1)=\ti{\psi}^{(1)}_{\mp}(z).
\ea
Therefore, in the $k=1$ sector, the periodicity in the Re$z$ direction gets doubled as $\ti{\psi}^{(1)}_{\pm}(z+2)=\ti{\psi}^{(1)}_{\pm}(z)$, which makes the moduli of the torus halved i.e. $\frac{\tau}{2}$.
Therefore $ \ti{Z}^{(0,0)}_{(1,0)}(L,\tau)$ is computed by the four-point function rather than a square of two-point functions:
\ba
Z^{(1,0)}_{(0,0)}(L,\tau)&=&\la e^{\frac{i}{4}\ti{\vp}^{(-)}_+(L)}e^{-\frac{i}{4}\ti{\vp}^{(-)}_-(L)} e^{\frac{i}{4}\ti{\vp}^{(-)}_+(0)}e^{-\frac{i}{4}\ti{\vp}^{(-)}_-(0)}\lb\no
&=&\la e^{\frac{i}{4}\ti{\vp}^{(-)}_+(L)}e^{-\frac{i}{4}\ti{\vp}^{(-)}_+(L+1)} e^{\frac{i}{4}\ti{\vp}^{(-)}_+(0)}e^{-\frac{i}{4}\ti{\vp}^{(-)}_+(1)}\lb. \label{fourpta}
\ea
This is evaluated as follows 
\ba
&& Z^{(1,0)}_{(0,0)}(L,\tau)\no
&&=\frac{1}{|\eta\left(\frac{\tau}{2}\right)|^2}\left|\frac{2\pi
\eta\left(\frac{\tau}{2}\right)^3}{
\theta_1\left(\frac{L}{2}|\frac{\tau}{2}\right)}\right|^{\frac{1}{4}}
\left|\frac{2\pi
\eta\left(\frac{\tau}{2}\right)^3}{
\theta_1\left(\frac{L+1}{2}|\frac{\tau}{2}\right)}\right|^{-\frac{1}{4}}
\left|\frac{2\pi
\eta\left(\frac{\tau}{2}\right)^3}{
\theta_1\left(\frac{1}{2}|\frac{\tau}{2}\right)}\right|^{\frac{1}{4}}\left|\theta_2\left(\frac{\tau}{2}\right)\right|^2.\label{calcuthe}
\ea
Here the final factor $\left|\theta_2\left(\frac{\tau}{2}\right)\right|^2$ comes from the winding mode summation.  

This leads to 
\ba
\ti{Z}^{(1,0)}_{(0,0)}(L,\tau)
=\frac{1}{|\eta(\tau)|^4}\left|\theta_2\left(\frac{\tau}{2}\right)\right|^{\frac{7}{2}}\left|\theta_2\left(\frac{L}{2}|\frac{\tau}{2}\right)\right|^{\frac{1}{2}}.
\ea
By adding the diagonal projection $l'_1=l'_2=1$ to the above result, we obtain
\ba
\ti{Z}^{(1,1)}_{(0,1)}(L,\tau)
=\frac{1}{|\eta(\tau)|^4}\left|\theta_1\left(\frac{\tau}{2}\right)\right|^{\frac{7}{2}}\left|\theta_1\left(\frac{L}{2}|\frac{\tau}{2}\right)\right|^{\frac{1}{2}}=0.
\ea

Next consider the asymmetric projection case 
$(l_1,l'_1)=(0,0)$ and $(l_2,l'_2)=(0,1)$.
In this case, the imaginary direction Im$z$ gets doubled due to the twisted boundary condition 
\ba
&& \ti{\psi}^{(0)}_{\pm}(z+\tau)=\ti{\psi}^{(0)}_{\pm}(z),\no
&& \ti{\psi}^{(1)}_{\pm}(z+\tau)=\ti{\psi}^{(1)}_{\mp}(z).
\ea
Therefore the moduli of the torus is given by $2\tau$. 
\ba
&& Z^{(0,1)}_{(0,0)}(L,\tau)\no
&&=\frac{1}{|\eta\left(2\tau\right)|^2}\left|\frac{2\pi
\eta\left(2\tau\right)^3}{
\theta_1\left(L|2\tau\right)}\right|^{\frac{1}{4}}
\left|\frac{2\pi
\eta\left(2\tau\right)^3}{
\theta_1\left(L+\tau|2\tau\right)}\right|^{-\frac{1}{4}}
\left|\frac{2\pi
\eta\left(2\tau\right)^3}{
\theta_1\left(\tau|2\tau\right)}\right|^{\frac{1}{4}}\left|\theta_4\left(2\tau\right)\right|^2.\no\label{calcuthea}
\ea
This leads to 
\ba
&& \ti{Z}^{(0,1)}_{(0,0)}(L,\tau)
=\frac{1}{|\eta(\tau)|^4}\left|\theta_4\left(2\tau\right)\right|^{\frac{7}{2}}\left|\theta_4\left(\frac{L}{2}|2\tau\right)\right|^{\frac{1}{2}},\no
&& \ti{Z}^{(1,1)}_{(1,0)}(L,\tau)
=\frac{1}{|\eta(\tau)|^4}\left|\theta_2\left(2\tau\right)\right|^{\frac{7}{2}}\left|\theta_2\left(\frac{L}{2}|2\tau\right)\right|^{\frac{1}{2}},
\ea
where in the latter we added a diagonal twist in Re$z$ direction.

Finally we consider the case $(l_1,l'_1)=(1,0)$ and $(l_2,l'_2)=(0,1)$.
In this case, the both Re$z$ and Im$z$ for $k=1$ are twisted as 
\ba
&& \ti{\psi}^{(1)}_{\pm}(z+1)=\ti{\psi}^{(0)}_{\mp}(z),\no
&& \ti{\psi}^{(1)}_{\pm}(z+\tau)=\ti{\psi}^{(1)}_{\mp}(z).
\ea
Since $\ti{\psi}^{(1)}_{\pm}(z+\tau+1)=\ti{\psi}^{(1)}_{\pm}(z)$, we can regard the torus as defined by the identification $z\sim z+2$ and 
$z\sim z+\tau+1$. Therefore the moduli is given by $\frac{\tau+1}{2}$.
This is evaluated as follows 
\ba
&& Z^{(0,1)}_{(1,0)}(L,\tau)\no
&&=\frac{1}{|\eta\left(\frac{\tau+1}{2}\right)|^2}\left|\frac{2\pi
\eta\left(\frac{\tau+1}{2}\right)^3}{
\theta_1\left(\frac{L}{2}|\frac{\tau+1}{2}\right)}\right|^{\frac{1}{4}}
\left|\frac{2\pi
\eta\left(\frac{\tau+1}{2}\right)^3}{
\theta_1\left(\frac{L+1}{2}|\frac{\tau+1}{2}\right)}\right|^{-\frac{1}{4}}
\left|\frac{2\pi
\eta\left(\frac{\tau+1}{2}\right)^3}{
\theta_1\left(\frac{1}{2}|\frac{\tau+1}{2}\right)}\right|^{\frac{1}{4}}\left|\theta_2\left(\frac{\tau+1}{2}\right)\right|^2.\no
\label{calcuthee}
\ea
This leads to
\ba
&& \ti{Z}^{(1,1)}_{(0,0)}(L,\tau)
=\frac{1}{|\eta(\tau)|^4}\left|\theta_1\left(\frac{\tau+1}{2}\right)\right|^{\frac{7}{2}}\left|\theta_1\left(\frac{L}{2}|\f{\tau+1}{2}\right)\right|^{\frac{1}{2}}=0,\no
&& \ti{Z}^{(0,1)}_{(1,0)}(L,\tau)
=\frac{1}{|\eta(\tau)|^4}\left|\theta_2\left(\frac{\tau+1}{2}\right)\right|^{\frac{7}{2}}\left|\theta_2\left(\frac{L}{2}|\f{\tau+1}{2}\right)\right|^{\frac{1}{2}}.
\ea

If we take the limit $L\to 0$, we obtain
\ba
&& \ti{Z}^{(1,0)}_{(0,0)}(0,\tau)=\frac{|\theta_2(\tau)\theta_3(\tau)|^2}{|\eta(\tau)|^4},
\ \ \  \ti{Z}^{(0,1)}_{(0,0)}(0,\tau)=\frac{|\theta_3(\tau)\theta_4(\tau)|^2}{|\eta(\tau)|^4},\no
&& \ti{Z}^{(1,1)}_{(0,1)}(0,\tau)=\ti{Z}^{(1,1)}_{(1,0)}(0,\tau)=\ti{Z}^{(1,1)}_{(0,0)}(0,\tau)=0, \ \ \ \ 
\ti{Z}^{(0,1)}_{(1,0)}(0,\tau)=\frac{|\theta_2(\tau)\theta_4(\tau)|^2}{|\eta(\tau)|^4}.
\label{lzerob}
\ea

In the limit $L\to 1$, we obtain
\ba
&& \ti{Z}^{(1,0)}_{(0,0)}(1,\tau)=\ti{Z}^{(1,1)}_{(0,1)}(1,\tau)=0,\ \ \ 
 \ti{Z}^{(0,1)}_{(0,0)}(1,\tau)=\frac{|\theta_4(2\tau)|^2}{|\eta(2\tau)|^2},\no
&& \ti{Z}^{(1,1)}_{(1,0)}(1,\tau)=\ti{Z}^{(1,1)}_{(0,0)}(1,\tau)=\ti{Z}^{(0,1)}_{(1,0)}(1,\tau)=0.\label{loneb}
\ea

These results (\ref{lzeroa}),  (\ref{lzerob}),  (\ref{lonea}), and  (\ref{loneb}) confirm the identity (\ref{entropref}) we wanted to show.

\section{Entanglement entropy in cyclic orbifold CFTs under quantum quenches}

\begin{figure}[hhh]
  \centering
  \includegraphics[width=8cm]{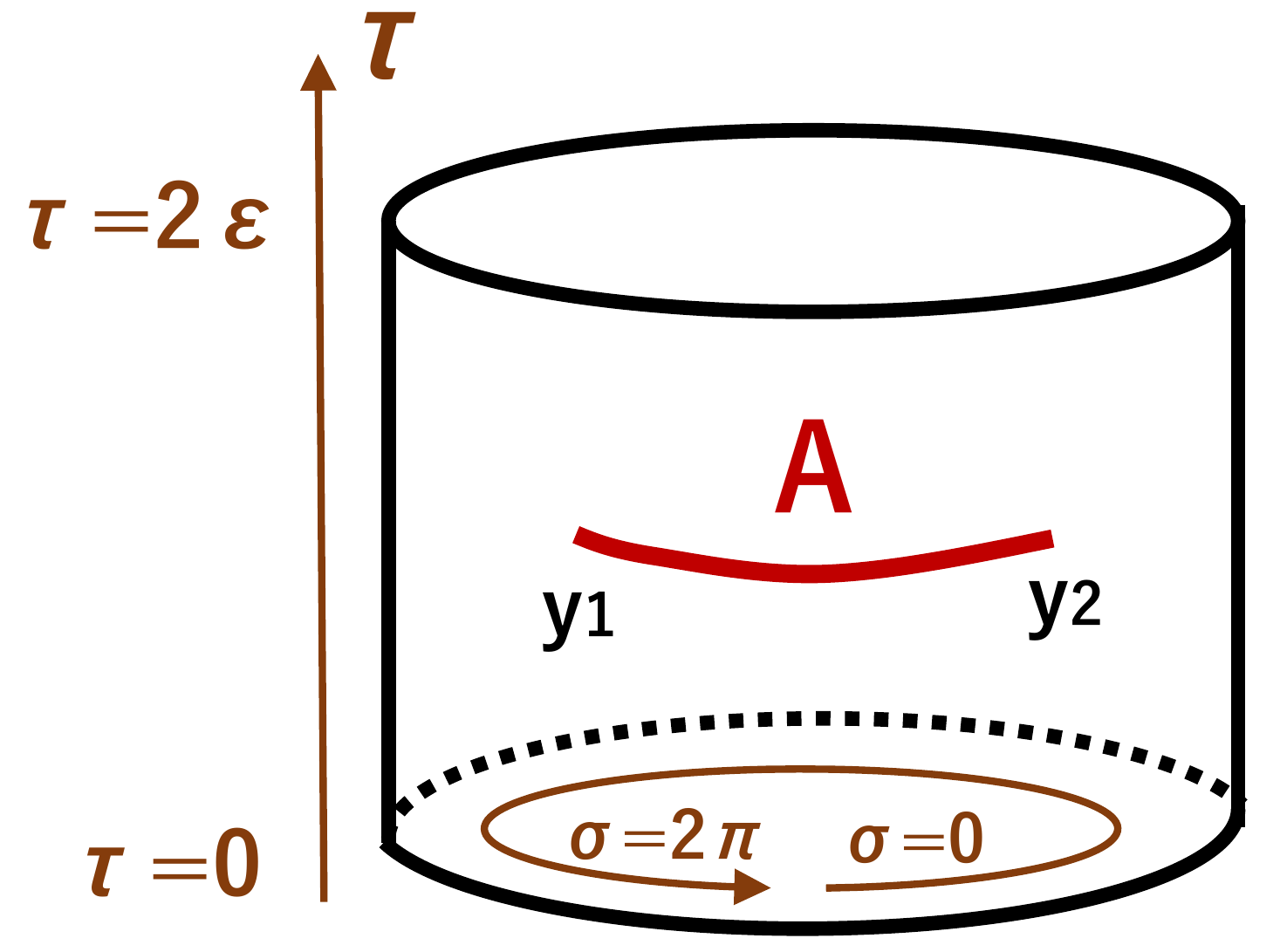}
  \caption{A sketch of the calculation of entanglement entropy on a cylinder.} 
\label{fig:cylinder}
\end{figure}

An excited state produced by a global quantum quench can be modeled by a regularized boundary state \cite{Calabrese:2005in}:
\be
|\Psi(t)\lb={\cal N} e^{-iHt}e^{-\ep H}|B\lb,
\ee
where $|B\lb$ is a boundary state (or Cardy state) in a given CFT. In this section, we would like to study the evolution of entanglement entropy for such excited states in the cyclic orbifold CFT on a circle. At $N=1$ i.e. the $c=1$ Dirac fermion CFT,  this was studied in \cite{Takayanagi:2010wp} and we will closely follow the convention in that paper below. We write the complex coordinate of the cylinder by $(y,\bar{y})=(\tau+i\sigma,\tau-i\sigma)$, where the spacial coordinate $\sigma$ is compactified as $\sigma\sim \sigma+2\pi$.
We choose subsystem $A$ to be an interval with length $\sigma$ at the real-time $t$ as depicted in Fig.\ref{fig:cylinder}. The endpoints of $A$, denoted by $(y_1,\bar{y}_1)$ and $(y_2,\bar{y}_2)$, are given by 
\be
(y_1,\bar{y}_1)=(\ep+it,\ep+it),\ \ \ (y_2,\bar{y}_2)=(\ep+it+i\sigma,\ep+it-i\sigma).
\ee

The entanglement entropy can be computed from the partition function of $n$ sheeted cylinder
\ba
Z^{cyl}_{n}=\la B_n|\sigma_n(y_1,\bar{y}_1)\bar{\sigma}_n(y_2,\bar{y}_2)|B_n\lb,
\label{partcyl}
\ea
where $|B_n\lb$ is the boundary state of the $n$ folded CFT.

\subsection{Boundary state of a single fermion CFT}

Consider the $N=1$ case, i.e. the $c=1$ free Dirac fermion. We can bosonize the fermion into a scalar and this leads to the following expression of boundary state
\ba
|B\lb_{N=1}={\cal N}_1\cdot e^{-\sum_{n=1}^\infty\frac{1}{n}\ap_{L,-n}\ap_{R,-n}}\sum_{w\in Z}|w\lb,
\ea
where $\ap_{L,n}$ and $\ap_{R,n}$ are the left and right-moving parts of the oscillators in the scalar field $\ti{\vp}=\ti{\vp}_L(y)+\ti{\vp}_R(\bar{y})$.
They satisfy the familiar commutation relation
\ba
[\ap_{L,n},\ap_{L,m}]=[\ap_{R,n},\ap_{R,m}]=n\delta_{n+m,0}.
\ea

The two-point functions of typical primary fields can be computed as follows 
(note that we set $\tau=\frac{2i\ep}{\pi}$ and that we have $k_R=-k_L$ for winding modes) \cite{Takayanagi:2010wp}:
\ba
&&\la B|e^{-2\ep H}e^{ik_L(\ti{\vp}_L(y_1)-\ti{\vp}_{R}(\bar{y}_1))}
e^{-ik_L(\ti{\vp}_L(y_2)-\ti{\vp}_{R}(\bar{y}_2))}|B\lb \no
&& =\sum_w e^{-\frac{R^2}{2}\ep w^2}\cdot \frac{1}{\eta(\tau)}\cdot 
\left[\frac{\eta(\tau)^6\theta_1\left(\frac{y_1+\bar{y}_2}{2\pi i}\right)\theta_1\left(\frac{y_2+\bar{y}_1}{2\pi i}\right)}{\theta_1\left(\frac{y_2-y_1}{2\pi i}\right)
\theta_1\left(\frac{\bar{y}_2-\bar{y}_1}{2\pi i}\right)
\theta_1\left(\frac{y_1+\bar{y}_1}{2\pi i}\right)\theta_1\left(\frac{y_2+\bar{y}_2}{2\pi i}\right)}\right]^{k_L^2}. \label{cylnamp}
\ea
We choose the free fermion radius i.e. $R=2$ as before. Then the summation over $w$ leads to the NS sector partition function
\ba
\sum_w e^{-\frac{R^2}{2}\ep w^2}=\theta_3(\tau).
\ea

\begin{figure}[hhh]
  \centering
  \includegraphics[width=8cm]{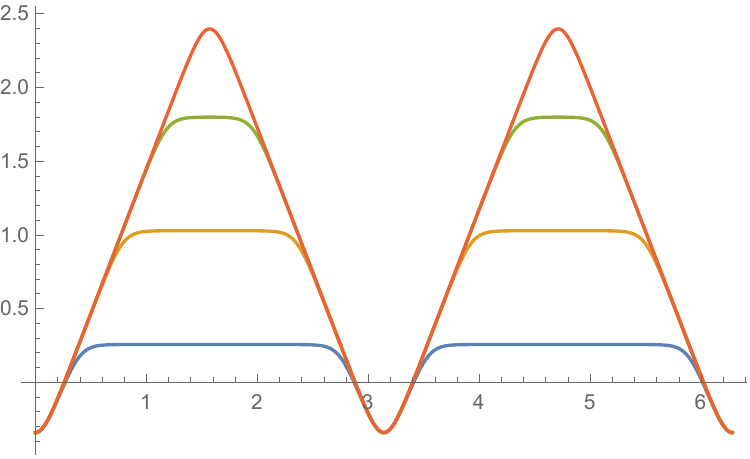}
  \caption{The plot of entanglement entropy as a function of time $t$ for the $c=1$ Dirac fermion CFT under quantum quenches.
  We set $n=2$ (2nd Renyi entropy), $\ep=0.2$ and $\delta=1$. The blue (bottom), orange, green, and red (top) curves describe the entanglement entropy for  $\sigma=\frac{\pi}{4},\frac{\pi}{2},\frac{3\pi}{4}$ and 
  $\pi$, respectively.} 
\label{fig:DiracEET}
\end{figure}

To calculate Tr$(\rho_A)^n$, we choose $k_L=\frac{p}{n}$ and take the product as follows
\ba
\mbox{Tr}[(\rho_A)^n]=\prod_{p=-\frac{n-1}{2}}^{\frac{n-1}{2}}
\la B_n|e^{-2\ep H}e^{i\frac{p}{n}(\ti{\vp}_L(y_1)-\ti{\vp}_{R}(\bar{y}_1))}
e^{-i\frac{p}{n}(\ti{\vp}_L(y_2)-\ti{\vp}_{R}(\bar{y}_2))}|B_n\lb.
\ea
Since we have $\sum_{p=-\frac{n-1}{2}}^{\frac{n-1}{2}}\frac{p^2}{n^2}=\frac{1}{12}\left(n-\frac{1}{n}\right)$, we find the $n$-th Renyi entropy is given by \cite{Takayanagi:2011zk}:
\ba
&& S^{(n)}_A=-\frac{\de}{\de n}\log \mbox{Tr}[(\rho_A)^n] \no
&&=\frac{1}{12}\left(1+\frac{1}{n}\right)
\log\frac{\left|\theta_1\left(\frac{\sigma}{2\pi}|\tau\right)\right|^2\left|\theta_1\left(\frac{\ep+it}{\pi i}|\tau\right)\right|^2}
{\left|\eta\left(\frac{2i\ep}{\pi}\right)\right|^6\left|\theta_1\left(\frac{\ep+it}{\pi i}+\frac{\sigma}{2\pi}|\tau\right)\right|\left|\theta_1\left(\frac{\ep+it}{\pi i}-\frac{\sigma}{2\pi}|\tau\right)\right|\delta^2},\label{singlecyl}
\ea
where we inserted the infinitesimally small parameter $\delta$ as the UV cut-off, which is a standard prescription in the CFT calculation of entanglement entropy (e.g. see \cite{Calabrese:2004eu}). This time evolution of entanglement entropy has the periodicity by $\pi$ as  
\be
S^{(n)}_A(t+\pi,\sigma)=S^{(n)}_A(t,\sigma).  \label{snperf}
\ee
Refer to the numerical plot shown in Fig.\ref{fig:DiracEET}. We find the linear growth 
until $t=\frac{\sigma}{2}$ and then the plateau later until $t=\pi-\frac{\sigma}{2}$ as the usual evolution of entanglement entropy in quantum quenches \cite{Calabrese:2005in}, which can be explained by the propagation of entangled pairs created by the quench at $t=0$.

\subsection{Entanglement entropy in cyclic orbifold CFTs}

Next, we calculate entanglement entropy in the cyclic orbifolds using the boundary state formalism. 
The boundary state in the $\mathbb{Z}_N$ cyclic orbifold can be written in the form
\ba
|B(\ap)\lb_N={\cal N'} \sum_{l=0}^{N-1}\prod_{k=0}^{N-1}e^{\frac{2\pi i\ap l}{N}}|B^{(k)}(l)\lb,
\ea
where $\ap=0,1,\ddd,N-1$ labels the type of the boundary state, so-called the fractional branes \cite{Diaconescu:1997br,Billo:1999nf,Diaconescu:1999dt,Billo:2000yb,Takayanagi:2001aj}. Also, the state $|B^{(k)}(l)\lb$ is the boundary state for the fermion $\ti{\psi}^{(k)}_{L,R}$ and  $\ti{\ti{\psi}}^{(k)}_{L,R}$ in the $l$-th twisted sector
in terms of their bosonized scalar fields. 

\begin{figure}[hhh]
  \centering
  \includegraphics[width=8cm]{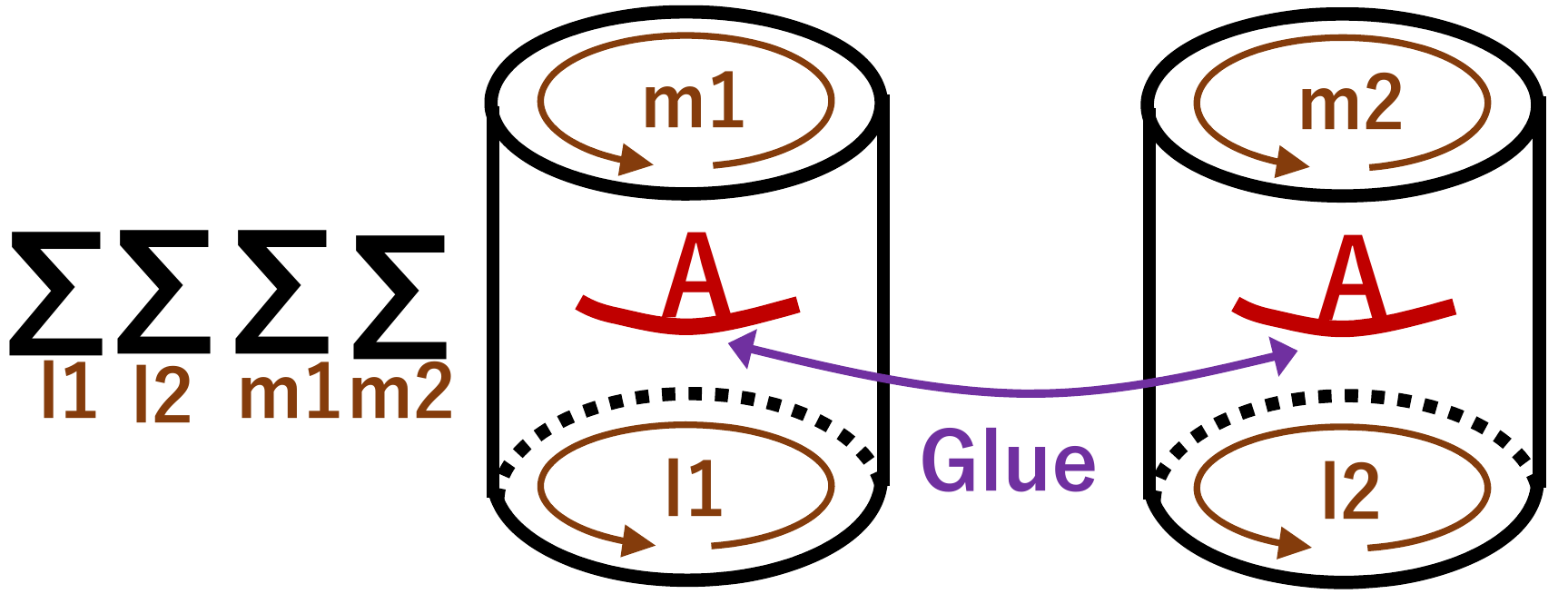}
  \caption{The replica method calculation of second Renyi entropy on a cylinder for the $Z_N$ orbifold CFT.} 
\label{fig:cylindersum}
\end{figure}

To find the entanglement entropy we need to compute the replicated partition function which is written in terms of a two-point function of twist operators on a cylinder (\ref{partcyl}). For example, when we consider the second Renyi entropy $n=2$, this is given by a summation over the twisted sectors of the boundary states as depicted in Fig.\ref{fig:cylindersum}. For general $N$, it looks quite difficult to calculate each contribution and perform the summation. However, it is straightforward to calculate the entanglement entropy for the diagonal sectors $l_1=m_1=l_2=m_2=\ddd=l_n=m_n(\equiv l)$ as we did in the finite temperature case. In this case, the zero mode of the winding mode is shifted by $w\to w+\frac{lk}{N}$.
Since the winding mode summation just gives an overall factor that does not depend on $t$ and $\sigma$, the final entanglement entropy is simply $N$ times that of the single fermion result (\ref{singlecyl}) i.e.
\ba
S^{(n)}_{A}|_{diagonal}=\frac{N}{12}\left(1+\frac{1}{n}\right)
\log\frac{\left|\theta_1\left(\frac{\sigma}{2\pi}|\tau\right)\right|^2\left|\theta_1\left(\frac{\ep+it}{\pi i}|\tau\right)\right|^2}
{\left|\eta\left(\frac{2i\ep}{\pi}\right)\right|^6\left|\theta_1\left(\frac{\ep+it}{\pi i}+\frac{\sigma}{2\pi}|\tau\right)\right|\left|\theta_1\left(\frac{\ep+it}{\pi i}-\frac{\sigma}{2\pi}|\tau\right)\right|\delta^2}.\label{NEEcyl}
\ea
Note that this is again periodic under the time translation with the same periodicity, given by (\ref{snperf}). Its profile is identical to Fig.\ref{fig:DiracEET}.

However, if we add all sectors including off-diagonal ones, we expect that the periodicity gets $N$ times longer such that $S_A(t+N\pi)=S_A(t)$. This is because the twisted sector, which describes long strings, is equivalent to an $N$ folded string i.e. the spatial coordinate has a larger periodicity: $\sigma\sim \sigma+ 2\pi N$.  In the next subsection, we will confirm this in the $N=2$ case by an explicit calculation.

\subsection{Full analysis in $N=n=2$ case}\label{sec:fullQ}

We can explicitly evaluate the replicated partition function (\ref{partcyl}) for the second Renyi entropy of $N=2$ cyclic orbifold. In this $N=n=2$ case, the two-point function of twist operators
is written as follows
\ba
\la B(\ap)|e^{-2\ep H}\sigma_2(y_1,\bar{y}_1)\bar{\sigma}_2(y_2,\bar{y}_2)|B(\ap)\lb=\frac{1}{4}\sum_{l_1,l_2,m_1,m_2=0}^1 C^{(l_2,m_2)}_{(l_1,m_1)}(\sigma,t).
\ea
First, we note the symmetry which exchanges the two replicas:
\ba
 C^{(l_2,m_2)}_{(l_1,m_1)}(\sigma,t)= C^{(l_1,m_1)}_{(l_2,m_2)}(\sigma,t).
\ea

The $k=0$ fermion $\ti{\psi}^{(0)}_{p}$ follows the untwisted boundary condition. Thus for any values of $(l_1,m_1)$ and $(l_2,m_2)$ of the summation depicted in Fig.\ref{fig:cylindersum}, the $k=0$ partition function is simply given by $\left|\frac{\theta_3(\tau)}{\eta(\tau)}\right|^2$. The $k=1$ fermion  $\ti{\psi}^{(1)}_{p}$ follows the $\mathbb{Z}_2$ twisted boundary condition in the $l=1$ sector, while it is untwisted in the $l=0$ sector. 

Let us start with the diagonal cases. When $(l_1,m_1)=(l_2,m_2)=(0,0)$, the partition function can be computed as the square of (\ref{cylnamp}):
\ba
C^{(0,0)}_{(0,0)}(\sigma,t)=\left|\frac{\theta_3(\tau)}{\eta(\tau)}\right|^4\cdot \left[\frac
{\left|\eta\left(\frac{2i\ep}{\pi}\right)\right|^6\left|\theta_1\left(\frac{\ep+it}{\pi i}+\frac{\sigma}{2\pi}|\tau\right)\right|\left|\theta_1\left(\frac{\ep+it}{\pi i}-\frac{\sigma}{2\pi}|\tau\right)\right|}{\left|\theta_1\left(\frac{\sigma}{2\pi}|\tau\right)\right|^2\left|\theta_1\left(\frac{\ep+it}{\pi i}|\tau\right)\right|^2}\right]^{\f{1}{4}}.
\ea
When $(l_1,m_1)=(l_2,m_2)=(1,1)$, the $k=1$ sector is twisted leading to the $k=1$ partition function $\left|\frac{\theta_2(\tau)}{\eta(\tau)}\right|^2$. By combining with the $k=0$ part we obtain
\ba
C^{(1,1)}_{(1,1)}(\sigma,t)=\left|\frac{\theta_3(\tau)}{\eta(\tau)}\right|^2\cdot \left|\frac{\theta_2(\tau)}{\eta(\tau)}\right|^2\cdot \left[\frac
{\left|\eta\left(\frac{2i\ep}{\pi}\right)\right|^6\left|\theta_1\left(\frac{\ep+it}{\pi i}+\frac{\sigma}{2\pi}|\tau\right)\right|\left|\theta_1\left(\frac{\ep+it}{\pi i}-\frac{\sigma}{2\pi}|\tau\right)\right|}{\left|\theta_1\left(\frac{\sigma}{2\pi}|\tau\right)\right|^2\left|\theta_1\left(\frac{\ep+it}{\pi i}|\tau\right)\right|^2}\right]^{\f{1}{4}}.
\ea

In the off-diagonal case $(l_1,m_1)=(0,0)$ and $(l_2,m_2)=(1,1)$, the periodicity of the Im$y$ direction gets doubled $y\sim y+4\pi i$ due to the twisted boundary condition 
$\ti{\psi}^{(1)}_{\pm}(y+2\pi i)=\ti{\psi}^{(1)}_{\mp}(y)$. Therefore the two-point function should be regarded as a four-point function on a cylinder as we have already seen  in (\ref{fourpta}). In the end, we obtain the following expression
\ba
&& C^{(1,1)}_{(0,0)}(\sigma,t)\no
&& =\left|\frac{\theta_3(\tau)^3\theta_2(\tau)}{\eta(\tau)^4}\right|\cdot 
\left[\frac
{\left|\eta\left(\tau\right)\right|^6\left|\theta_1\left(\frac{\ep+it}{\pi i}+\frac{\sigma}{2\pi}|\tau\right)\right|\left|\theta_1\left(\frac{\ep+it}{\pi i}-\frac{\sigma}{2\pi}|\tau\right)\right|}{\left|\theta_1\left(\frac{\sigma}{2\pi}|\tau\right)\right|^2\left|\theta_1\left(\frac{\ep+it}{\pi i}|\tau\right)\right|^2}\right]^{\f{1}{8}}\no
&&\ \ \ \ \times
\left|\frac{\eta(\frac{\tau}{2})^6~\theta_2\left(\frac{\sigma}{4\pi}|\frac{\tau}{2}\right)^2
\theta_2\left(\frac{\ep+it}{2\pi i}|\frac{\tau}{2}\right)^2
\theta_1\left(\frac{\ep+it}{2\pi i}+\frac{\sigma}{4\pi}|\frac{\tau}{2}\right)\theta_1\left(\frac{\ep+it}{2\pi i}-\frac{\sigma}{4\pi}|\frac{\tau}{2}\right)}
{4\theta_1\left(\frac{\sigma}{4\pi}|\frac{\tau}{2}\right)^2\theta_2\left(0|\frac{\tau}{2}\right)^2
\theta_1\left(\frac{\ep+it}{2\pi i}|\frac{\tau}{2}\right)^2\theta_2\left(\frac{\ep+it}{2\pi i}+\frac{\sigma}{4\pi}|\frac{\tau}{2}\right)\theta_2\left(\frac{\ep+it}{2\pi i}-\frac{\sigma}{4\pi}|\frac{\tau}{2}\right)}\right|^{\frac{1}{8}}.
\ea

We can also obtain the result for $(l_1,m_1)=(1,0)$ and $(l_2,m_2)=(0,1)$
 by performing transformation $\sigma\to 2\pi-\sigma$
\ba
C^{(0,1)}_{(1,0)}(\sigma,t)=C^{(1,1)}_{(0,0)}(2\pi-\sigma,t).
\ea
All other components of  $C^{(l_2,m_2)}_{(l_1,m_1)}(\sigma,t)$ are vanishing.

The total partition function takes the form
\ba
&&\la B(\ap)|e^{-2\ep H}\sigma_2(y_1,\bar{y}_1)\bar{\sigma}_2(y_2,\bar{y}_2)|B(\ap)\lb
\no
&&=
\frac{1}{4}\left[C^{(0,0)}_{(0,0)}(\sigma,t)+C^{(1,1)}_{(1,1)}(\sigma,t)
+2C^{(1,1)}_{(0,0)}(\sigma,t)+2C^{(0,1)}_{(1,0)}(\sigma,t)\right].
\ea
It is easy to confirm 
\ba
&& \lim_{\sigma\to 0} C^{(0,0)}_{(0,0)}(\sigma,t)=\lim_{\sigma\to 2\pi} C^{(0,0)}_{(0,0)}(\sigma,t)=\left|\frac{\theta_3(\tau)^4}{\eta(\tau)^4}\right|\cdot\left|\frac{\eta(\tau)^3}{\theta_1\left(\frac{\sigma}{2\pi}|\tau\right)}\right|^{\frac{1}{2}}, \no
&& \lim_{\sigma\to 0} C^{(1,1)}_{(1,1)}(\sigma,t)=\lim_{\sigma\to 2\pi} C^{(1,1)}_{(1,1)}(\sigma,t)=\left|\frac{\theta_3(\tau)^2\theta_2(\tau)^2}{\eta(\tau)^4}\right|\cdot\left|\frac{\eta(\tau)^3}{\theta_1\left(\frac{\sigma}{2\pi}|\tau\right)}\right|^{\frac{1}{2}}, \no
&& \lim_{\sigma\to 0} C^{(1,1)}_{(0,0)}(\sigma,t)=\left|\frac{\theta_3(\tau)^3\theta_2(\tau)}{\eta(\tau)^4}\right|\cdot\left|\frac{\eta(\tau)^3}{\theta_1\left(\frac{\sigma}{2\pi}|\tau\right)}\right|^{\frac{1}{2}}, \no
&& \lim_{\sigma\to 2\pi} C^{(1,1)}_{(0,0)}(\sigma,t)=\lim_{\sigma\to 0} C^{(0,1)}_{(1,0)}(\sigma,t)=0,\no
&& \lim_{\sigma\to 2\pi} C^{(0,1)}_{(1,0)}(\sigma,t)=\left|\frac{\theta_3(\tau)^2\theta_2(\tau)}{\eta(\tau)^4}\right|\cdot\left|\frac{\eta(\tau)^3}{\theta_1\left(\frac{\sigma}{2\pi}|\tau\right)}\right|^{\frac{1}{2}}, \no
\ea
where we employed the identities in appendix \ref{apptheta}.
Thus we obtain
\ba
&& \lim_{\sigma\to 0}\la B(\ap)|e^{-2\ep H}\sigma_2(y_1,\bar{y}_1)\bar{\sigma}_2(y_2,\bar{y}_2)|B(\ap)\lb \no
&&= \lim_{\sigma\to 2\pi}\la B(\ap)|e^{-2\ep H}\sigma_2(y_1,\bar{y}_1)\bar{\sigma}_2(y_2,\bar{y}_2)|B(\ap)\lb \no
&&=\left|\frac{\eta(\tau)^3}{\theta_1\left(\frac{\sigma}{2\pi}|\tau\right)}\right|^{\frac{1}{2}}\cdot 
\left|\frac{\theta_3(\tau)^2}{\eta(\tau)^2}\right|\cdot 
\left|\frac{\theta_3(\tau)+\theta_2(\tau)}{2\eta(\tau)}\right|^2.
\ea
This result is consistent with the fact that the partition function should be factorized into the product of two cylinder partition functions in the limit $\sigma\to 0$ and
 $\sigma\to 2\pi$. Indeed, the final term is a square of cylinder amplitudes of the $N=2$ cyclic orbifold CFT in the NS sector:
  \ba
Z^{N=2}_{cyl}(\tau)=\left|\frac{\theta_3(\tau)}{\eta(\tau)}\right|\cdot 
\left|\frac{\theta_3(\tau)+\theta_2(\tau)}{2\eta(\tau)}\right|.
\ea

\begin{figure}[hhh]
  \centering
  \includegraphics[width=8cm]{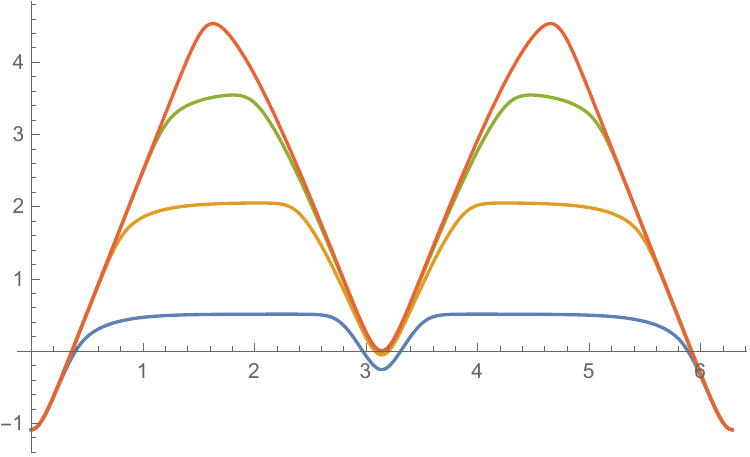}
  \caption{The plot of entanglement entropy as a function of time $t$ for the $N=2$ orbifolded Dirac fermion CFT under quantum quenches.  We set $n=2$ (2nd Renyi entropy), $\ep=0.2$ and $\delta=1$. The blue (bottom), orange, green, and red (top) curves describe the entanglement entropy for  $\sigma=\frac{\pi}{4},\frac{\pi}{2},\frac{3\pi}{4}$ and 
  $\pi$, respectively.} 
\label{fig:OrbEET}
\end{figure}

Thus the final expression of the second Renyi entropy reads
\ba
S^{(2)}_A(\sigma,t)=-\log\frac{\la B(\ap)|e^{-2\ep H}\sigma_2(y_1,\bar{y}_1)\bar{\sigma}_2(y_2,\bar{y}_2)|B(\ap)\lb\cdot \delta^{\frac{1}{2}}}{Z^{N=2}_{cyl}(\tau)^2},
\ea
where we again inserted the UV cut-off $\delta$ dependence.
It is clear that the periodicity in the time direction is now doubled:
\ba
S^{(2)}_A(\sigma,t+2\pi)=S^{(2)}_A(\sigma,t),
\ea
as we argued in the last subsection. We can also confirm the pure state identity 
\ba
S^{(2)}_A(\sigma,t)=S^{(2)}_A(2\pi-\sigma,t).
\ea
In the small $\sigma$ limit, we reproduced the well-known behavior \cite{Calabrese:2004eu} (at the central charge $c=N=2$)
\ba
S^{(2)}_A(\sigma,t)\simeq \frac{1}{2}\log\frac{\sigma}{\delta}.
\ea
Refer to Fig.\ref{fig:OrbEET} for numerical plots.

For general $N$, we cannot explicitly evaluate the entanglement entropy because the detailed form the twist operator $\sigma_2$ is not available for $N>2$.
However, it is quite natural to expect that the periodicity of the time evolution of the entanglement entropy becomes $N\pi$ i.e.
\ba
S^{(2)}_A(\sigma,t+N\pi)=S^{(2)}_A(\sigma,t),
\ea
for the $\mathbb{Z}_N$ cyclic orbifold CFT, due to the non-diagonal contributions.

\section{Connection to symmetric orbifold CFTs}
In this section, we examine the extension of results in cyclic orbifolds to symmetric orbifolds.
We start with re-analyzing their partition functions, in terms of Hecke operators.
By introducing deformed Hecke operators (we named square-free Hecke operators), we can relate these partition functions.
In addition to the connection in partition functions, by studying the structure of twisted sectors in detail, we can calculate quantities in symmetric orbifold CFTs through cyclic ones.

\subsection{Cyclic orbifold CFT revisited: connection to symmetric orbifold}
\label{CycOrbRevisit}
Recall that the $N$-th symmetric orbifold CFT $\mathcal{C}_{N,S}$ has a partition function of the following form
\begin{equation}
    Z_{N,S}(\tau) = 
    \sum_{\mathrm{partition~of}~N}
    \prod_{k=1}^{N}\frac{1}{(N_k)!} \left( T_k Z(\tau) \right)^{N_k},
\end{equation}
where the partition of $N$ runs over $\displaystyle (N_1,\dots,N_N) ~\mathrm{s.t.}~ \sum_{k=1}^{N} k N_k = N$ and $T_k$ is the Hecke operator, 
\begin{equation}
    T_k Z\left(\tau\right) = \frac{1}{k}\sum_{i|k} \sum_{j=0}^{i-1} Z\left(\frac{k \tau}{i^{2}}+\frac{j}{i}\right).
\end{equation}
In this subsection, we focus on these Hecke operators in detail. 
By introducing deformed Hecke operators (we named square-free Hecke operators), we can relate these partition functions.

\subsubsection{Hecke operators and square-free Hecke operators}
\label{sec:squarefreeheckeopIntro}
Firstly we study Hecke operators.
Hecke operators, which map modular forms into themselves\footnote{In this paper we focus on only modular forms with weight zero.}, are used in mathematics, especially in number theories.
For some modular invariant function $Z\left(\tau\right)$,
the $k$-th Hecke operator $T_k$ is defined to be ($i$ runs over the divisors of $k$),
\begin{equation}
\label{HeckeDef}
    T_k Z\left(\tau\right) \equiv \frac{1}{k}\sum_{i|k} \sum_{j=0}^{i-1} Z\left(\frac{k \tau}{i^{2}}+\frac{j}{i}\right).
\end{equation}
For the case $k$ is prime ($k=p$), this definition can be simplified into
\begin{equation}
    T_p Z\left(\tau\right) = \frac{1}{p}
    \left(
    Z(p\tau)
    +
    \sum_{i=0}^{p-1} 
    Z\left(\frac{\tau+i}{p}\right)
    \right).
\end{equation}

In order to write down the cyclic orbifold partition function in terms of Hecke operators, we need to introduce a new type of Hecke operator, which we call square-free Hecke operators.
As one can observe in appendix \ref{ExamplesHeckePartitionFunction}, if $k$ has some square divisors (we take $a^2$ for example), the $k$-th Hecke operator includes terms that constitute $\frac{1}{a^2}$ times $\frac{k}{a^2}$-th Hecke operator\footnote{
Actually we can check this fact explicitly:
in the definition of the $k$-th Hecke operator, if we restrict $i$ in the summation to be $a$ times divisors of $k/a^2$ and $j$ to be multiple of $a$, we get
\begin{align*}
    T_k Z\left(\tau\right) =&~ \frac{1}{k}\sum_{i|k} \sum_{j=0}^{i-1} Z\left(\frac{k}{i^{2}}\tau+\frac{j}{i}\right) \\
    \supset&~\frac{1}{k}\sum_{i'|\frac{k}{a^2}} \sum_{j'=0}^{i'-1} Z\left(\frac{k}{(ai')^{2}}\tau+\frac{aj'}{ai'}\right) \\
    =&~\frac{1}{a^2}\cdot\frac{1}{k/a^2}\sum_{i'|\frac{k}{a^2}} \sum_{j'=0}^{i'-1} Z\left(\frac{k/a^2}{i'^{2}}\tau+\frac{j'}{i'}\right)
\end{align*}
where we set $(i,j)=(ai',aj')$. The last line is exactly the $k/a^2$-th Hecke operator multiplied by $1/a^2$. 
}.
This suggests that a generic Hecke operator is not {\it minimally} modular invariant, in other words, we can divide a Hecke operator into multiple parts which are independently modular invariant. Since it is useful for our purpose to specify the minimally modular invariant part(s) in each Hecke operator,
we introduce the $k$-th square-free Hecke operator $T_k^{sf}$ as the minimally modular invariant part of $T_k$. We can write down the definition recursively:
\begin{align}
T_k^{sf}Z(\tau) = T_k Z(\tau) - \sum_{
    \substack{a\in \mathbb{Z}>1 \\
    a^2 | k}
    }
\frac{1}{a^2}T_{\frac{k}{a^2}}^{sf}Z(\tau).
\end{align}
This definition coincides with the ordinary Hecke operator (\ref{HeckeDef}) if $k$ has no square factors.
One can explicitly check that some specific term, $\frac{1}{k}Z(k\tau)$ for example, runs every term in $T_k^{sf}Z(\tau)$ by acting modular-$S$ and/or $T$ transformations and accordingly $T_k^{sf}Z(\tau)$ is minimally modular invariant.

Hecke operators are used in mathematics, especially in number theories. These operators are helpful for us to organize symmetric/cyclic orbifold CFT partition functions, and to reveal the connection between them.
We prepare some examples in appendix \ref{ExamplesHeckePartitionFunction}.

\subsubsection{Cyclic orbifold partition function with Hecke operators}
Recall that the $N$-th cyclic orbifold partition function is, $G=\mathbb{Z}_N$ version of (\ref{GeneralOrbifoldPF}),
\begin{align*}
    Z_{N,\mathbb{Z}}(\tau) =&
    \sum_{l=0}^{N-1} \mathrm{Tr}_{(l)} \left(\frac{\sum_{l'=1}^{N}g^{l'}}{N}q^{H_L}\bar{q}^{H_R}\right) \\
    =& \frac{1}{N} \sum_{l,l'}\mathrm{Tr}_{(l)} \left(g^{l'}q^{H_L}\bar{q}^{H_R}\right),
\end{align*}
where $g$ is one of the generators of $\mathbb{Z}_N$ and $l$ labels the (un)twisted sectors as we have seen in \ref{CycOrbPF}. 
We newly derived the representation of this partition function in terms of (square-free) Hecke operators. 
The $N$-th cyclic orbifold partition function $Z_{N,\mathbb{Z}}(\tau)$ can be written in terms of square-free Hecke operators\footnote{We can compare this equation (\ref{GeneralCyclicPF}) to the result discussed in \cite{Haehl:2014yla} (we already referred to in \ref{CycOrbPF}). By equating each expression of $Z_{N,\mathbb{Z}}$, fixing $d$, and restricting the range of $r$ and $s$ to be $(N,r,s)=d$, we get
\begin{align}
    \frac{(\mathrm{\#~of}~k\leq N~\mathrm{s.t.}~\gcd(N,k)=d)}{d} T_{\frac{N}{d}}^{sf}\left(Z(\tau)^{d}\right)
    =& \frac{1}{N}
    \sum_{\substack{r,s=1,\dots,N \\
    (N,r,s)=d}} Z\left( \frac{(N,r)}{N}\left(\frac{(N,r)}{d}\tau+\kappa (r,s)
    \right)\right)^d .
\end{align}
One can explicitly verify this equation for a small $N$ case.},
\begin{align}
    Z_{N,\mathbb{Z}}(\tau) =& \sum_{k=1}^{N} \frac{1}{\gcd(N,k)} T_{\frac{N}{\gcd(N,k)}}^{sf}\left(Z(\tau)^{\gcd(N,k)}\right) \nonumber\\
    =& \sum_{d|N} \frac{(\mathrm{\#}k~(1\leq k\leq N)~\mathrm{s.t.}~\gcd(N,k)=d)}{d} T_{\frac{N}{d}}^{sf}\left(Z(\tau)^{d}\right) \nonumber\\
    =& \sum_{d|N} \frac{\phi(N/d)}{d} T_{\frac{N}{d}}^{sf}\left(Z(\tau)^{d}\right).\label{GeneralCyclicPF}
\end{align}
where $\phi$ is the Euler function.
We provide the proof of this construction in appendix \ref{COPproof}.
$N=8$ case for example, the number of $k$ $(k=1,\dots,8)$ which satisfies
\begin{itemize}
    \item $\gcd(8,k)=8$ is 1 $(k=8)$,
    \item $\gcd(8,k)=4$ is 1 $(k=4)$,
    \item $\gcd(8,k)=2$ is 2 $(k=2,6)$,
    \item $\gcd(8,k)=1$ is 4 $(k=1,3,5,7)$.
\end{itemize}
Thus we get
\begin{align}
Z_{8,\mathbb{Z}}(\tau) 
=& \frac{1}{8} T_{1}^{sf}\left(Z(\tau)^8\right)+ \frac{1}{4} T_{2}^{sf}\left(Z(\tau)^4\right)+ \frac{2}{2} T_{4}^{sf}\left(Z(\tau)^2\right)+ \frac{4}{1} T_{8}^{sf}\left(Z(\tau)^1\right).
\end{align}
More examples are provided in Appendix \ref{ExamplesHeckePartitionFunction}.
If $N$ is a prime number, this function becomes as follows (for prime $N$ the $N$-th square-free Hecke operator is the same as the normal Hecke operator);
\begin{align}
Z_{N,\mathbb{Z}}(\tau) = \frac{1}{N}Z(\tau)^N+(N-1)T_N Z(\tau).
\end{align}
This prime $N$ case is indicated in \cite{Klemm:1990df}.

\subsection{Detailed investigation of the Dirac fermion symmetric orbifold CFT}
We consider Dirac fermion symmetric orbifold CFT in this section.
Let us analyze the partition function and proceed to calculate the diagonal Renyi entropy. (Full Renyi entropy is too complicated to determine: $N=n=2$ case is already considered in \ref{sec:Nn2}. Note that for $N=2$ the cyclic orbifold is identical to the symmetric orbifold.)

\subsubsection{Partition function}
For the case of Dirac fermion, the $i$-th cyclic orbifold partition function is of the form
\begin{align}
    Z_{i,\mathbb{Z}}(\tau) =&~ \frac{1}{i}\sum_{l,l'=0}^{i-1}\prod_{k=0}^{i-1}\sum_{a=2,3,4}\frac{e^{i\pi\frac{k^2l^2}{i^2}(\tau-\bar{\tau})}}{2}\left|\frac{\theta_a\left(\frac{kl'}{i}+\frac{kl}{i}\tau|\tau\right)}{\eta(\tau)}\right|^2 \\
    =&~ \frac{1}{i}\sum_{l,l'=1}^{i}\prod_{k=0}^{i-1}\sum_{a=2,3,4}\frac{e^{i\pi\frac{k^2l^2}{i^2}(\tau-\bar{\tau})}}{2}\left|\frac{\theta_a\left(\frac{kl'}{i}+\frac{kl}{i}\tau|\tau\right)}{\eta(\tau)}\right|^2.
\end{align}
Especially its NS sector partition function is
\begin{align}
    Z_{i,\mathbb{Z}}^{NS}(\tau) =&~ \frac{1}{i}\sum_{l,l'=1}^{i}\prod_{k=0}^{i-1}
e^{i\pi\frac{ k^2l^2}{i^2}(\tau-\bar{\tau})}
\left|\frac{\theta_3\left(\frac{kl'}{i}+\frac{kl}{i}\tau|\tau\right)}{\eta(\tau)}\right|^2,
\end{align}
which was explained in section \ref{sec:finitep}. By comparing these two equations, we can identify each sector:
\begin{align}
\label{eq:squarefreeHeckeidentification}
    \frac{\phi(i/d)}{d} T_{\frac{i}{d}}^{sf}\left(Z(\tau)^{d}\right) =&~ \frac{1}{i}\sum_{\substack{
    l,l'=1 \\
    (i,l,l')=d
    }    }^{i}\prod_{k=0}^{i-1}\sum_{a=2,3,4}\frac{e^{i\frac{\pi k^2}{i^2}l^2(\tau-\bar{\tau})}}{2}\left|\frac{\theta_a\left(\frac{kl'}{i}+\frac{kl}{i}\tau|\tau\right)}{\eta(\tau)}\right|^2.
\end{align}

In this subsection, what we want to do is to write down the NS partition function of the Dirac fermion symmetric orbifold CFT in terms of theta functions.
Recall that the partition function of the $N$-th symmetric theory can be  written in terms of the normal Hecke operators,
\begin{equation}
    Z_{N,S}(\tau) = 
    \sum_{\mathrm{partition~of}~N}
    \prod_{i=1}^{N}\frac{1}{(N_i)!} \left( T_i Z(\tau) \right)^{N_i}.
\end{equation}
Here, by definition, we can reconstruct the normal Hecke operator by square-free ones, 
\begin{align}
T_i Z(\tau) =&~ T_i^{sf} Z(\tau) + \sum_{
    \substack{a\in \mathbb{Z}>1 \\
    a^2 | i}
    }
\frac{1}{a^2}T_{i/a^2}^{sf}Z(\tau) \\
 =&~ \sum_{
    \substack{a\in \mathbb{Z}>0 \\
    a^2 | i}
    }
\frac{1}{a^2}T_{i/a^2}^{sf}Z(\tau).
\end{align}

\paragraph{$a=1$ case}
As we see above, we can identify the NS sector of 
$T_i^{sf} Z(\tau)$, $\left(T_i^{sf} Z(\tau)\right)^{NS}$ to be
\begin{align}
\label{eq:aeq1Tsfid}
    \left(T_i^{sf} Z(\tau)\right)^{NS} =&~ \frac{1}{i\phi(i)}\sum_{\substack{
    l,l' = 1,\dots,i \\
    (l,l',i) = 1
    }}
    \prod_{k=0}^{i-1}
    e^{i\pi\frac{k^2l^2}{i^2}(\tau-\bar{\tau})}
    \left|\frac{\theta_3\left(\frac{kl'}{i}+\frac{kl}{i}\tau|\tau\right)}{\eta(\tau)}\right|^2.
\end{align}
This identification is from the $i$-th cyclic orbifold, $d=1$ case of (\ref{eq:squarefreeHeckeidentification}).

\paragraph{$a>1$ case}
If $i$ has some square-factors, we must add $T_{i/a^2}^{sf} Z(\tau)$ to reconstruct $T_i Z(\tau)$. 
Considering the $i/a^2$-th cyclic orbifold, $d=a^2$ case of (\ref{eq:squarefreeHeckeidentification}), we can evaluate $T_{i/a^2}^{sf} Z(\tau)$ to be 
\begin{align}
\label{eq:ageq1TsfidPrev}
    \left(T_{i/a^2}^{sf} Z(\tau)\right)^{NS} =&~ \frac{a^2}{i\phi(i/a^2)}\sum_{\substack{
    l,l' = 1,\dots,{i/a^2} \\
    (l,l',i) = 1
    }}
    \prod_{k=0}^{{i/a^2}-1}
    e^{i\pi\frac{a^2k^2l^2}{i^2}(\tau-\bar{\tau})}
    \left|\frac{\theta_3\left(\frac{a^2kl'}{i}+\frac{a^2kl}{i}\tau|\tau\right)}{\eta(\tau)}\right|^2.
\end{align}
This form seems to include only $i/a^2$ fermions. However, by applying the identity (\ref{fermionsquareidentity}), this product can be interpreted as the product of $i$ twisted Dirac fermions in NS sector\footnote{Obviously, if we want to calculate only the partition function, the equation (\ref{eq:ageq1TsfidPrev}) is sufficient and there is no need to apply the identity \ref{fermionsquareidentity} and interpret each $\theta$ to be $a^2$ fermions. We will use this interpretation in the next subsubsection.}. Substituting the following equation
\begin{align}
\label{eq:ageq1Tsfid}
    &~e^{i\pi\frac{a^2k^2l^2}{i^2}(\tau-\bar{\tau})}
    \left|\frac{\theta_3\left(\frac{a^2kl'}{i}+\frac{a^2kl}{i}\tau|\tau\right)}{\eta(\tau)}\right|^2 \nonumber\\
    =&~\prod_{\kappa = -\frac{a-1}{2}}^{\frac{a-1}{2}}\prod_{\lambda = -\frac{a-1}{2}}^{\frac{a-1}{2}}
    e^{i\pi\left(\frac{\lambda}{a}+\frac{akl}{i}\right)^2(\tau-\bar{\tau})}
    \left|\frac{\theta_3\left(\frac{\kappa}{a}+\frac{akl'}{i}+\left(\frac{\lambda}{a}+\frac{akl}{i}\right)\tau|\tau\right)}{\eta(\tau)}\right|^2,\nonumber
\end{align}
we get
\begin{align}
    &~\left(T_{i/a^2}^{sf} Z(\tau)\right)^{NS}  \nonumber\\
    =&~ \frac{a^2}{i\phi(i/a^2)}\sum_{\substack{
    l,l' = 1,\dots,{i/a^2} \\
    (l,l',i/a^2) = 1
    }}
    \prod_{k=0}^{{i/a^2}-1}
    \prod_{\kappa = -\frac{a-1}{2}}^{\frac{a-1}{2}}\prod_{\lambda = -\frac{a-1}{2}}^{\frac{a-1}{2}}
    e^{i\pi\left(\frac{\lambda}{a}+\frac{akl}{i}\right)^2(\tau-\bar{\tau})}
    \left|\frac{\theta_3\left(\frac{\kappa}{a}+\frac{akl'}{i}+\left(\frac{\lambda}{a}+\frac{akl}{i}\right)\tau|\tau\right)}{\eta(\tau)}\right|^2.
\end{align}
If we set $a=1$, this equation reproduces the result in the previous paragraph.

Combining (\ref{eq:aeq1Tsfid}) and (\ref{eq:ageq1Tsfid}), we obtain
\begin{align}
    \left(T_i Z(\tau)\right)^{NS}  =&~ \frac{1}{i}\sum_{
    \substack{a\in \mathbb{Z}>0 \\
    a^2 | i}
    }
    \frac{1}{\phi(i/a^2)}\sum_{\substack{
    l,l' = 1,\dots,{i/a^2} \\
    (l,l',i/a^2) = 1
    }}
    \prod_{k=0}^{{i/a^2}-1}
    \prod_{\kappa = -\frac{a-1}{2}}^{\frac{a-1}{2}}\prod_{\lambda = -\frac{a-1}{2}}^{\frac{a-1}{2}} \nonumber\\
    &~~~~~~~~~~~e^{i\pi\left(\frac{\lambda}{a}+\frac{akl}{i}\right)^2(\tau-\bar{\tau})}
    \left|\frac{\theta_3\left(\frac{\kappa}{a}+\frac{akl'}{i}+\left(\frac{\lambda}{a}+\frac{akl}{i}\right)\tau|\tau\right)}{\eta(\tau)}\right|^2.
\end{align}
This contains $i$ twisted Dirac fermions in each twisted sector.
By summing this $T_i^{NS}$ over the partition of $N$, we obtain the NS partition function of the symmetric orbifold.

\subsubsection{Diagonal Renyi entropy}

Now we can extend our calculation to the diagonal Renyi entropy in the NS sector of Dirac fermion symmetric orbifold CFT.

As we have seen in (\ref{replicamethodtofermion}), the bosonization and replica method changes each twisted fermion as
\begin{align}
    &~e^{i\pi\beta^2(\tau-\bar{\tau})} 
    \left|\frac{\theta_3\left(\alpha+\beta\tau|\tau\right)}{\eta(\tau)}\right|^2 \nonumber\\
    \to &~\prod_{p=-(n-1)/2}^{(n-1)/2}e^{i\pi\beta^2(\tau-\bar{\tau})} \left|\frac{2\pi\eta(\tau)^3}{\theta_1(L|\tau)}\right|^{4h_p}\cdot \left|\frac{\theta_3\left(\alpha+\beta\tau+\frac{p}{n}L|\tau\right)}{|\eta(\tau)^2|}\right|^2
\end{align}
where $h_p=\frac{p^2}{2n^2}$.
Now let us calculate the entanglement entropy for a fixed diagonal sector. 
As we expected, the universal $\theta_1$ part factorizes:
\begin{align}
    Z_{N,S}^{NS}\langle \sigma_n(L)\bar{\sigma}_n(0) \rangle =&~ \left(\prod_{p=-(n-1)/2}^{(n-1)/2}\left|\frac{2\pi\eta(\tau)^3}{\theta_1(L|\tau)}\right|^{4h_p}\right)^N \nonumber\\
    &\times \left(\sum_{\mathrm{partition~of}~N}
    \prod_{i=1}^{N}\frac{1}{(N_i)!} \langle\langle\sigma_n(L)\bar{\sigma}_n(0)\rangle\rangle_i ^{N_i}\right)
\end{align}
where
\begin{align}
    \langle\langle\sigma_n(L)\bar{\sigma}_n(0)\rangle\rangle_i =&~\frac{1}{i}\sum_{
    \substack{a\in \mathbb{Z}>0 \\
    a^2 | i}
    }
    \frac{1}{\phi(i/a^2)}\sum_{\substack{
    l,l' = 1,\dots,{i/a^2} \\
    (l,l',i) = 1
    }}
    \prod_{k=0}^{{i/a^2}-1}
    \prod_{\kappa = -\frac{a-1}{2}}^{\frac{a-1}{2}}\prod_{\lambda = -\frac{a-1}{2}}^{\frac{a-1}{2}} 
    \prod_{p=-\frac{n-1}{2}}^{\frac{n-1}{2}}\nonumber\\
    &~~~~~~~~~~~e^{i\pi\left(\frac{\lambda}{a}+\frac{akl}{i}\right)^2(\tau-\bar{\tau})}
    \left|\frac{\theta_3\left(\frac{\kappa}{a}+\frac{akl'}{i}+\left(\frac{\lambda}{a}+\frac{akl}{i}\right)\tau+\frac{p}{n}L|\tau\right)}{\eta(\tau)}\right|^2.
\end{align}

By iterating the procedure in \ref{REiDS}, we can calculate the diagonal Renyi entropy $S^{(n)}_{A,diag}$. For example, we can derive an equation similar to (\ref{eq:diagonalRE}), thermal diagonal Renyi entropy:
\begin{align}
    &~\lim_{\delta\to 0}\left[S^{(n)}_{A,diag}(1-\delta)-S^{(n)}_{A,diag}(\delta)\right] \nonumber\\
    =&~\frac{1}{1-n}\log \left[\frac{
        \sum_{\mathrm{partition~of}~N}\prod_{i=1}^{N}\frac{1}{(N_i)!} \left(\frac{1}{i}\sum_{\substack{a\in \mathbb{Z}>0 \\a^2 | i}}\frac{1}{\phi(i/a^2)}\sum_{\substack{l,l' = 1,\dots,{i/a^2} \\(l,l',i) = 1}}
        Z_{(i,a,l,nl')}^{NS}(n\tau)
        \right)^{N_i}
    }{
        \sum_{\mathrm{partition~of}~N}\prod_{i=1}^{N}\frac{1}{(N_i)!} \left(\frac{1}{i}\sum_{\substack{a\in \mathbb{Z}>0 \\a^2 | i}}\frac{1}{\phi(i/a^2)}\sum_{\substack{l,l' = 1,\dots,{i/a^2} \\(l,l',i) = 1}}
        Z_{(i,a,l,l')}^{NS}(\tau)^n
        \right)^{N_i}
    }\right].
\end{align}
where we introduced $Z_{(i,a,l,l')}^{NS}$ for shorthand:
\begin{align}
    Z_{(i,a,l,l')}^{NS}(\tau)=&~\prod_{k=0}^{{i/a^2}-1}
    \prod_{\kappa = -\frac{a-1}{2}}^{\frac{a-1}{2}}\prod_{\lambda = -\frac{a-1}{2}}^{\frac{a-1}{2}}
    e^{i\pi\left(\frac{\lambda}{a}+\frac{akl}{i}\right)^2(\tau-\bar{\tau})}
    \left|\frac{\theta_3\left(\frac{\kappa}{a}+\frac{akl'}{i}+\left(\frac{\lambda}{a}+\frac{akl}{i}\right)\tau|\tau\right)}{\eta(\tau)}\right|^2 \nonumber\\
    Z_{(i,a,l,nl')}^{NS}(n\tau)=&~\prod_{k=0}^{{i/a^2}-1}
    \prod_{\kappa = -\frac{a-1}{2}}^{\frac{a-1}{2}}\prod_{\lambda = -\frac{a-1}{2}}^{\frac{a-1}{2}}
    e^{i\pi\left(\frac{\lambda}{a}+\frac{akl}{i}\right)^2(n\tau-n\bar{\tau})}
    \left|\frac{\theta_3\left(\frac{n\kappa}{a}+\frac{aknl'}{i}+\left(\frac{\lambda}{a}+\frac{akl}{i}\right)n\tau|n\tau\right)}{\eta(n\tau)}\right|^2
\end{align}
Note that when considering $Z_{(i,a,l,nl')}^{NS}(n\tau)$, not only $\tau$ but also $l'$ and $\kappa$
get multiplied by $n$.
Calculating the full Renyi entropy is one of the future problems.


\section{Towards holographic CFTs: periodicity and BTZ black hole}

Our results of two-point functions in the $\mathbb{Z}_i$ cyclic orbifold presented in section \ref{sec:time-like C} and those of the Renyi entropy for $\mathbb{Z}_2$ cyclic orbifold in section \ref{sec:fullQ}, strongly imply that the real-time periodicity of the $\mathbb{Z}_i$ orbifold CFT based on the free Dirac fermions is $i$ times that of the seed CFT i.e. $c=1$ free Dirac fermion CFT. Recall that the $S_N$ symmetric orbifold theory includes the $\mathbb{Z}_i$ orbifolded sector (or $T_{i}$ sector) for all $1\leq i \leq N$. One way to understand this behavior is the fact that the twisted sector describes the long string where the spacial coordinate gets folded by $i$ times \cite{Maldacena:1996ds,Maldacena:2001kr}.
This suggests that the real-time periodicity in symmetric orbifold CFT under the time evolution is $\mathrm{LCM}[1,\dots ,N]$ times that of the seed CFT.
Here, $\mathrm{LCM}[a,b,\ddd]$ denotes the least common multiple of $a,b,\ddd$.

It is well-known in number theory that $\mathrm{LCM}[1,\dots ,N]$ grows as $\exp N$ in the limit $N\to \infty$. This result can be found in several references, e.g. theorem 3.8 of \cite{tenenbaum2015introduction}\footnote{In this reference $\psi(n)$ is set to be $\ln [\mathrm{LCM}[1,\dots ,n]]$. Theorem 3.8 of \cite{tenenbaum2015introduction} says that $\psi(n)\sim n$ at large $n$.}. This fact mainly follows from the celebrated Prime number theorem, which says that the number of primes less than or equal to $n$, is approximately given by $\pi(n)\sim\frac{n}{\log n}$.

Thus, the recurrence time for the symmetric orbifold CFT 
is estimated as $e^{S_{orb}}\sim e^{c_{orb}}$, where $S_{orb}$ is the thermal entropy.
Note that the central charge $c_{orb}$ of the $S_N$ orbifold of the free Dirac fermion is $N$. Interestingly, this 
qualitatively agrees with the Poincare recurrence time of holographic two-dimensional CFTs at finite temperature.

\section{Conlusions and discussions}
In this paper, we developed the methods of calculations in cyclic/symmetric orbifold CFTs with the Dirac fermion as a seed theory. 
We started with a bosonization of the Dirac fermion cyclic theory and constructed replica-twist operators. For the twist operators for the second Renyi entropy, we confirmed that their four-point functions correctly reproduce the torus partition function for 
the $\mathbb{Z}_2$ cyclic orbifold CFT, which can also be regarded as the S$_2$ symmetric orbifold CFT. In this analysis, we found that we need to modify the replica-twist operator to calculate the contributions from twisted sectors of the orbifold.

Using these twisted operators, we evaluated the entanglement entropy in the orbifold CFTs on a torus which describes CFT states at finite temperature in a finite size system. For general $\mathbb{Z}_N$ cyclic orbifolds, we calculated the diagonal contributions to the entanglement entropy and confirmed that their difference (\ref{relatuonfgdh}) reproduces a part of thermal entropy.
However, a naive extension of the standard replica method for un-orbifolded fermions does not yield the complete Renyi entropy: we must take into account different twist-boundary conditions for different replica sheets. Even though the replica method is difficult for arbitrary $N$ and $n$, we succeeded to construct the method to calculate full second Renyi entropy in the $\mathbb{Z}_2$ cyclic / $S_2$ symmetric orbifold theory on torus and cylinder. As a result, we found that the difference of the second thermal Renyi entropy on a torus perfectly agrees with the expected form $-\log[Z(2\tau)/Z(\tau)^2]$, which is the thermal Renyi entropy.

We also studied the time evolution of entanglement entropy under quantum quenches. This can be computed by evaluating the two-point functions of twist operators on a cylinder. As in the torus case, we evaluated this for
$\mathbb{Z}_N$ cyclic orbifolds with respect to the diagonal sectors. Moreover, we found a full expression for $N=2$ case, summing over both the diagonal and non-diagonal sectors. This shows that the real-time periodicity gets doubled, compared with the $c=1$ Dirac fermion CFT. 
For general $N$, we expect that the periodicity will be $N$ times longer, though we did not explicitly show this due to a technical difficulty.

Apart from the calculations in Dirac fermion, we explicitly constructed the expression of cyclic orbifold partition functions in terms of Hecke operators. Although the identification of $T^{sf}_k$ and theta function is specific to Dirac fermion orbifold theory, Hecke operators will enable us to deal with the problems in more general symmetric orbifold theory from a cyclic orbifold point of view\footnote{The connection between symmetric group and cyclic group is obviously well-understood, but connection between their orbifold theory has been unclear.}.

In the S$_N$ symmetric orbifold, by combining our results, we argued that the real-time periodicity of generic correlation functions and entanglement entropy is exponential as $\sim\exp N$ in the large $N$ limit. This result looks qualitatively consistent with the expected Poincare recurrence of holographic CFTs with the central charge $c=N$.\\

To conclude this paper, we clarify some future tasks and open problems.
\begin{description}
   \item[$\bullet$ Developing the replica method]\mbox{}\\
        As we indicated iteratively, the naive replica method excludes contributions from off-diagonal parts.
        To be precise, the off-diagonal contributions can be evaluated in principle by applying different twist-boundary conditions for different replica sheets.
        This problem can be rephrased: orbifold-twist (not replica-twist) operators should also be replicated as ordinary excitation e.g. $\psi(z)$.
        Hence, the key to developing the methods of calculating the off-diagonal parts might be to consider orbifold-twist operator correlations for arbitrary $n$. 
   \item[$\bullet$ Extension to symmetric orbifolds on cylinder]\mbox{}\\
	    By using the result of the recent research on boundary states in symmetric orbifolds \cite{Belin:2021nck}, we can extend our calculations to arbitrary $N$ symmetric orbifolds on a cylinder. This extension will be one of the concrete approaches to AdS/BCFT.
   \item[$\bullet$ Application to general symmetric orbifold CFTs]\mbox{}\\
	    We hope to extend our observation to the case with other seed CFTs. The most straightforward and beneficial extension will be the orbifold theory with $T^4$ (rectangular, $R=2$) SCFT as a seed CFT.
        This $T^4$ orbifold theory is much closer to real holographic CFT than the Dirac fermion orbifold theory.
        Marginal deformations to eliminate higher-spin modes and matching the spectrum to classical supergravity dual should be revisited.
    \item[$\bullet$ Heavy operators that contribute to the BTZ black hole]\mbox{}\\
        In the Dirac fermion cyclic orbifolds, we found examples of correlation functions that have a long periodicity. Although we conjectured that symmetric orbifold CFT has the periodicity  $\exp N$, we did not explicitly specify operators/correlators that show this periodicity, analogous to the BTZ black hole entropy. Specifying such excitation will be a novel approach to investigating the structure of black holes.
\end{description}

\vspace{5mm}
{\bf Acknowledgements} 

Takashi Tsuda thanks Masaki Okada for useful discussions about section \ref{COPproof}.
This work is supported by the Simons Foundation through the ``It from Qubit'' collaboration
and by MEXT KAKENHI Grant-in-Aid for Transformative Research Areas (A) through the ``Extreme Universe'' collaboration: Grant Number 21H05187.
This work is also supported by Inamori Research Institute for Science and World Premier International Research Center Initiative (WPI Initiative)
from the Japan Ministry of Education, Culture, Sports, Science and Technology (MEXT), and JSPS Grant-in-Aid for Scientific Research (A) No.~21H04469.

\appendix

\section{Theta function identities}\label{apptheta}
When to consider the Dirac fermion orbifold theory partition functions, theta function identities is useful.
To share our convention, we have described this section more carefully than necessary.

\subsection{Definitions of theta and eta functions, fundamental formulas}
\subsubsection{Definitions}
We set $q\equiv e^{2\pi i \tau}$ and $y\equiv e^{2\pi i z}$.
The Dedekind eta function is 
\begin{align}
    \eta(\tau) = q^{\frac{1}{24}}\prod_{m=1}^{\infty} (1-q^m).
\end{align}
The Jacobi theta functions have two different (but equivalent) representations.
In the sum representation,
\begin{align}
\begin{alignedat}{2}
    &\theta_{1}(z \mid \tau) = &~\sum_{r\in \mathbb{Z}+1/2} q^{\frac{r^2}{2}}y^r&e^{-\pi i r}, \\
    &\theta_{2}(z \mid \tau) = &~\sum_{r\in \mathbb{Z}+1/2} q^{\frac{r^2}{2}}y^r&, \\
    &\theta_{3}(z \mid \tau) = &~\sum_{r\in \mathbb{Z}} q^{\frac{r^2}{2}}y^r&, \\
    &\theta_{4}(z \mid \tau) = &~\sum_{r\in \mathbb{Z}} q^{\frac{r^2}{2}}y^r&e^{-\pi i r}.
\end{alignedat}
\end{align}
In the product representation,
\begin{align}
\begin{aligned}
&\theta_{1}(z \mid \tau)=2 \sin (\pi z) q^{\frac{1}{8}} \prod_{m=1}^{\infty}\left(1-q^{m}\right)\left(1-y q^{m} \right)\left( 1-y^{-1} q^{m}\right), \\
&\theta_{2}(z \mid \tau)=2 \cos (\pi z) q^{\frac{1}{8}} \prod_{m=1}^{\infty}\left(1-q^{m}\right)\left(1+y q^{m} \right)\left( 1+y^{-1} q^{m}\right), \\
&\theta_{3}(z \mid \tau)=\prod_{m=1}^{\infty}\left(1-q^{m}\right)\left(1+y q^{m-\frac{1}{2}}\right)\left(1+y^{-1} q^{m-\frac{1}{2}}\right), \\
&\theta_{4}(z \mid \tau)=\prod_{m=1}^{\infty}\left(1-q^{m}\right)\left(1-y q^{m-\frac{1}{2}}\right)\left(1-y^{-1} q^{m-\frac{1}{2}}\right).
\end{aligned}
\end{align}
We write $\theta_a(\tau)$ shorthand for $\theta_a(0|\tau)$.

\subsubsection{Fundamental formulas}
$\theta_1$ satisfies
\begin{align}
    \theta_1 (0|\tau) = \theta_1(\tau) = 0, \\
    \left.\partial_z\theta_1 (z|\tau)\right|_{z=0} = 2\pi \eta (\tau)^3.  
\end{align}
$\eta$ satisfies
\begin{align}
    2 \eta (\tau)^3 =&~ \theta_2(\tau)\theta_3(\tau)\theta_4(\tau) .
\end{align}
The famous Jacobi identity is 
\begin{align}
    \theta_1(z|\tau)^4-\theta_2(z|\tau)^4+\theta_3(z|\tau)^4-\theta_4(z|\tau)^4 = 0, 
\end{align}
or $z=0$ version
\begin{align}
    \theta_3(\tau)^4-\theta_2(\tau)^4-\theta_4(\tau)^4 = 0. 
\end{align}
The modular-$S$ and modular-$T$ transformation is as follows:
\begin{align}
\begin{aligned}
    \eta(\tau+1)=&~e^{\frac{i\pi}{12}}\eta(\tau), &\eta\left(-\frac{1}{\tau}\right) =&~ \sqrt{-i\tau}\eta(\tau), \\
    \theta_1(z|\tau+1)=&~e^{\frac{i\pi}{4}}\theta_1(z|\tau), & \theta_1\left(\frac{z}{\tau}|-\frac{1}{\tau}\right)=&~ -i\sqrt{-i\tau}e^{\frac{i\pi z^2}{\tau}}\theta_1(z|\tau),\\
    \theta_2(z|\tau+1)=&~e^{\frac{i\pi}{4}}\theta_2(z|\tau), & \theta_2\left(\frac{z}{\tau}|-\frac{1}{\tau}\right)=&~ \sqrt{-i\tau}e^{\frac{i\pi z^2}{\tau}}\theta_4(z|\tau),\\
    \theta_3(z|\tau+1)=&~\theta_4(z|\tau), & \theta_3\left(\frac{z}{\tau}|-\frac{1}{\tau}\right)=&~ \sqrt{-i\tau}e^{\frac{i\pi z^2}{\tau}}\theta_3(z|\tau),\\
    \theta_4(z|\tau+1)=&~\theta_3(z|\tau), & \theta_4\left(\frac{z}{\tau}|-\frac{1}{\tau}\right)=&~ \sqrt{-i\tau}e^{\frac{i\pi z^2}{\tau}}\theta_2(z|\tau).
\end{aligned}
\end{align}
Theta functions satisfy the following quasi-double periodicity:
\begin{align}
    \begin{alignedat}{2}
    \theta_1(z+m\tau+n|\tau)=&~& (-1)^{m+n} &~q^{-\frac{m^2}{2}}y^{-m} \theta_1(z|\tau), \\
    \theta_2(z+m\tau+n|\tau)=&~& (-1)^{n} &~q^{-\frac{m^2}{2}}y^{-m} \theta_2(z|\tau), \\
    \theta_3(z+m\tau+n|\tau)=&~&  &~q^{-\frac{m^2}{2}}y^{-m} \theta_3(z|\tau), \\
    \theta_4(z+m\tau+n|\tau)=&~& (-1)^{m} &~q^{-\frac{m^2}{2}}y^{-m} \theta_4(z|\tau),
    \end{alignedat}
\end{align}
where $m$ and $n$ are integers. Actually, we can consider the case $m$ and $n$ are half-integers by using the following relations,
\begin{align}
    \begin{aligned}
    \theta_1(z\pm1/2|\tau)=&~\pm\theta_2(z|\tau),   \\
    \theta_3(z\pm1/2|\tau)=&~\theta_4(z|\tau),
    \end{aligned}
\end{align}
\begin{align}
\begin{aligned}
    \theta_1(z\pm\tau/2|\tau)=&~&\pm iq^{-\frac{1}{8}}y^{\mp\frac{1}{2}}&\theta_4(z|\tau),\\
    \theta_2(z\pm\tau/2|\tau)=&~& q^{-\frac{1}{8}}y^{\mp\frac{1}{2}}&\theta_3(z|\tau), \\
    \theta_3(z\pm\tau/2|\tau)=&~& q^{-\frac{1}{8}}y^{\mp\frac{1}{2}}&\theta_2(z|\tau), \\
    \theta_4(z\pm\tau/2|\tau)=&~&\pm iq^{-\frac{1}{8}}y^{\mp\frac{1}{2}}&\theta_1(z|\tau).
\end{aligned}
\end{align}

\subsection{``Double-angle'' formulas}\label{sec:doubleF}
Since we use the ``double-angle\footnote{The reason we write with quotation marks is that ordinary double-angle ($n$-tuple angle) formulas in the context of theta functions refer to $2z$ ($nz$). In this subsection and the next subsection, we mainly focus on the case of multiple $\tau$.}'' formulas repeatedly in this paper, we present this special case here. There are specific relations in the ``double-angle'' case, namely
\begin{align}
\begin{aligned}
    \theta_2(2\tau) =&~ \left(\frac{\theta_3(\tau)^2-\theta_4(\tau)^2}{2}\right)^{1/2}, \\
    \theta_3(2\tau) =&~ \left(\frac{\theta_3(\tau)^2+\theta_4(\tau)^2}{2}\right)^{1/2}, \\
    \theta_4(2\tau) =&~ \left(\theta_3(\tau)\theta_4(\tau)\right)^{1/2}, \\
    \eta(2\tau) =&~ \left(\frac{\theta_2(\tau)\eta(\tau)}{2}\right)^{1/2} = \eta(\tau)^2\left(\theta_3(\tau)\theta_4(\tau)\right)^{-1/2}.
\end{aligned}
\end{align}
In addition, we have
\begin{align}
\begin{aligned}
    \theta_2\left(\frac{\tau}{2}\right) =&~\left(2\theta_2(\tau)\theta_3(\tau)\right)^{1/2}, \\
    \theta_3\left(\frac{\tau}{2}\right) =&~ \left(\theta_3(\tau)^2+\theta_2(\tau)^2\right)^{1/2}, \\
    \theta_4\left(\frac{\tau}{2}\right) =&~ \left(\theta_3(\tau)^2-\theta_2(\tau)^2\right)^{1/2}, \\
    \eta\left(\frac{\tau}{2}\right) =&~ \left(\theta_4(\tau)\eta(\tau)\right)^{1/2} = \eta(\tau)^2\left(\frac{2}{\theta_2(\tau)\theta_3(\tau)}\right)^{1/2},
\end{aligned}
\end{align}
\begin{align}
\begin{aligned}
    \theta_2\left(\frac{\tau+1}{2}\right) =&~e^{\frac{i\pi}{8}}\left(2\theta_2(\tau)\theta_4(\tau)\right)^{1/2}, \\
    \theta_3\left(\frac{\tau+1}{2}\right) =&~ \left(\theta_4(\tau)^2+i\theta_2(\tau)^2\right)^{1/2}, \\
    \theta_4\left(\frac{\tau+1}{2}\right) =&~ \left(\theta_4(\tau)^2-i\theta_2(\tau)^2\right)^{1/2}, \\
    \eta\left(\frac{\tau+1}{2}\right) =&~ e^{\frac{i\pi}{24}}\left(\theta_3(\tau)\eta(\tau)\right)^{1/2} =e^{\frac{i\pi}{24}} \eta(\tau)^2\left(\frac{2}{\theta_2(\tau)\theta_4(\tau)}\right)^{1/2}.
\end{aligned}
\end{align}

\subsection{``$n$-tuple-angle'' formulas}
In contrast to the double version, there is no clear relation between $\theta_a(n\tau)$ and $\theta_a(\tau)$ generically. However, as we present in this subsection, we can rewrite ``$n$-tuple-angle'' thetas in terms of $\theta_a(z|\tau)$.
In order to derive the ``$n$-tuple-angle'' formulas, we use the following factorization formulas.
\begin{align}
&1-q^{n}=\prod_{k=0}^{n-1}\left(1-e^{2 \pi i \frac{k}{n}} q\right) \\
&1+q^{n}=\prod_{k=0}^{n-1}\left(1-e^{2 \pi i \frac{2 k+1}{2 n}} q\right)=\prod_{k=-\frac{n-1}{2}}^{\frac{n-1}{2}}\left(1+e^{2 \pi i \frac{k}{n}} q\right)
\end{align}
By applying these formulas to the product representations of theta functions, we obtain the following ``$n$-tuple-angle'' formulas:
\begin{align}
\begin{aligned}
\frac{\theta_{2}(n \tau)}{\eta(n \tau)}=&~
\prod_{k=-\frac{n-1}{2}}^{\frac{n-1}{2}} \frac{\theta_{2}\left(\frac{k}{n} \mid \tau \right)}{\eta(\tau)}, & \qquad
\frac{\theta_{2}\left(\frac{\tau}{n}\right)}{\eta\left(\frac{\tau}{n}\right)}=&~\prod_{k=0}^{n-1} e^{\pi i \left( \frac{k}{n}\right)^{2} \tau} \frac{\theta_{2}\left(\frac{k}{n} \tau \mid \tau\right)}{\eta(\tau)}, \\
\frac{\theta_{3}(n \tau)}{\eta(n \tau)}=&~\prod_{k=-\frac{n-1}{2}}^{\frac{n-1}{2}} \frac{\theta_{3}\left(\frac{k}{n} \mid \tau\right)}{\eta(\tau)}, &
\frac{\theta_{3}\left(\frac{\tau}{n}\right)}{\eta\left(\frac{\tau}{n}\right)}=&~\prod_{k=-\frac{n-1}{2}}^{\frac{n-1}{2}} e^{\pi i \left( \frac{k}{n}\right)^{2} \tau} \frac{\theta_{3}\left(\frac{k}{n} \tau \mid \tau\right)}{\eta(\tau)}, \\
\frac{\theta_{4}(n \tau)}{\eta(n \tau)}=&~\prod_{k=0}^{n-1} \frac{\theta_{4}\left(\frac{k}{n} \mid \tau\right)}{\eta(\tau)}, &
\frac{\theta_{4}\left(\frac{\tau}{n}\right)}{\eta\left(\frac{\tau}{n}\right)}=&~\prod_{k=-\frac{n-1}{2}}^{\frac{n-1}{2}} e^{\pi i \left( \frac{k}{n}\right)^{2} \tau} \frac{\theta_{4}\left(\frac{k}{n} \tau \mid \tau\right)}{\eta(\tau)}.
\end{aligned}
\end{align}
We can extend these formulas straightforwardly to the nonzero $z$ case as
\begin{align}
\begin{aligned}
\frac{\theta_{1}(\alpha |n \tau)}{\eta(n \tau)}=&~\prod_{k=0}^{n-1} \frac{\theta_{1}\left(\frac{\alpha}{n}+\frac{k}{n} \mid \tau\right)}{\eta(\tau)},&
-i\frac{\theta_{1}\left(\alpha |\frac{\tau}{n}\right)}{\eta\left(\frac{\tau}{n}\right)}=&~\prod_{k=0}^{n-1} (-i)e^{2\pi i  \frac{k}{n}\alpha }e^{\pi i \left( \frac{k}{n}\right)^{2} \tau} \frac{\theta_{1}\left(\alpha+\frac{k}{n} \tau \mid \tau\right)}{\eta(\tau)}, \\
\frac{\theta_{2}(\alpha |n \tau)}{\eta(n \tau)}=&~\prod_{k=-\frac{n-1}{2}}^{\frac{n-1}{2}} \frac{\theta_{2}\left(\frac{\alpha}{n}+\frac{k}{n} \mid \tau \right)}{\eta(\tau)},&
\frac{\theta_{2}\left(\alpha |\frac{\tau}{n}\right)}{\eta\left(\frac{\tau}{n}\right)}=&~\prod_{k=0}^{n-1} e^{2\pi i  \frac{k}{n}\alpha }e^{\pi i \left( \frac{k}{n}\right)^{2} \tau} \frac{\theta_{2}\left(\alpha+\frac{k}{n} \tau \mid \tau\right)}{\eta(\tau)}, \\
\frac{\theta_{3}(\alpha |n \tau)}{\eta(n \tau)}=&~\prod_{k=-\frac{n-1}{2}}^{\frac{n-1}{2}} \frac{\theta_{3}\left(\frac{\alpha}{n}+\frac{k}{n} \mid \tau\right)}{\eta(\tau)},&
\frac{\theta_{3}\left(\alpha |\frac{\tau}{n}\right)}{\eta\left(\frac{\tau}{n}\right)}=&~\prod_{k=-\frac{n-1}{2}}^{\frac{n-1}{2}} e^{2\pi i  \frac{k}{n}\alpha }e^{\pi i \left( \frac{k}{n}\right)^{2} \tau} \frac{\theta_{3}\left(\alpha+\frac{k}{n} \tau \mid \tau\right)}{\eta(\tau)}, \\
\frac{\theta_{4}(\alpha |n \tau)}{\eta(n \tau)}=&~\prod_{k=0}^{n-1} \frac{\theta_{4}\left(\frac{\alpha}{n}+\frac{k}{n} \mid \tau\right)}{\eta(\tau)},&
\frac{\theta_{4}\left(\alpha |\frac{\tau}{n}\right)}{\eta\left(\frac{\tau}{n}\right)}=&~\prod_{k=-\frac{n-1}{2}}^{\frac{n-1}{2}} e^{2\pi i  \frac{k}{n}\alpha }e^{\pi i \left( \frac{k}{n}\right)^{2} \tau} \frac{\theta_{4}\left(\alpha+\frac{k}{n} \tau \mid \tau\right)}{\eta(\tau)}.\label{formulatthe} 
\end{aligned}
\end{align}
The following enables us to change the range of $k$.
\begin{align}
\begin{alignedat}{2}
e^{2\pi i  \frac{k}{n}\alpha }e^{\pi i \left( \frac{k}{n}\right)^{2} \tau} \theta_{1}\left(\alpha+\frac{k}{n} \tau \mid \tau\right) =&~& - &~e^{2\pi i  \frac{k+n}{n}\alpha }e^{\pi i \left( \frac{k+n}{n}\right)^{2} \tau} \theta_{1}\left(\alpha+\frac{k+n}{n} \tau \mid \tau\right) \\
e^{2\pi i  \frac{k}{n}\alpha }e^{\pi i \left( \frac{k}{n}\right)^{2} \tau} \theta_{2}\left(\alpha+\frac{k}{n} \tau \mid \tau\right) =&~& &~e^{2\pi i  \frac{k+n}{n}\alpha }e^{\pi i \left( \frac{k+n}{n}\right)^{2} \tau} \theta_{2}\left(\alpha+\frac{k+n}{n} \tau \mid \tau\right) \\
e^{2\pi i  \frac{k}{n}\alpha }e^{\pi i \left( \frac{k}{n}\right)^{2} \tau} \theta_{3}\left(\alpha+\frac{k}{n} \tau \mid \tau\right) =&~& &~e^{2\pi i  \frac{k+n}{n}\alpha }e^{\pi i \left( \frac{k+n}{n}\right)^{2} \tau} \theta_{3}\left(\alpha+\frac{k+n}{n} \tau \mid \tau\right) \\
e^{2\pi i  \frac{k}{n}\alpha }e^{\pi i \left( \frac{k}{n}\right)^{2} \tau} \theta_{4}\left(\alpha+\frac{k}{n} \tau \mid \tau\right) =&~& - &~e^{2\pi i  \frac{k+n}{n}\alpha }e^{\pi i \left( \frac{k+n}{n}\right)^{2} \tau} \theta_{4}\left(\alpha+\frac{k+n}{n} \tau \mid \tau\right)
\end{alignedat}
\end{align}
We can derive the ordinary $n$-tuple-angle formulas by substituting the right-hand side of (\ref{formulatthe}) to the left-hand side,
\begin{align}
\label{fermionsquareidentity}
    \begin{alignedat}{2}
    \frac{\theta_{1}(\alpha | \tau)}{\eta( \tau)}=&~&i^{n-n^2}\prod_{k=0}^{n-1} \prod_{l=0}^{n-1} ~&e^{2\pi i  \frac{l}{n}\left(\frac{\alpha}{n}+\frac{k}{n}\right) }e^{\pi i \left( \frac{l}{n}\right)^{2} \tau} \frac{\theta_{1}\left(\frac{\alpha}{n}+\frac{k}{n}+\frac{l}{n} \tau \mid \tau\right)}{\eta(\tau)}, \\
    \frac{\theta_{2}(\alpha |\tau)}{\eta( \tau)}=&~&\prod_{k=-\frac{n-1}{2}}^{\frac{n-1}{2}} \prod_{l=0}^{n-1} ~&e^{2\pi i  \frac{l}{n}\left(\frac{\alpha}{n}+\frac{k}{n}\right) } e^{\pi i \left( \frac{l}{n}\right)^{2} \tau} \frac{\theta_{2}\left(\frac{\alpha}{n}+\frac{k}{n}+\frac{l}{n} \tau \mid \tau\right)}{\eta(\tau)}, \\
    \frac{\theta_{3}(\alpha |\tau)}{\eta( \tau)}=&~&\prod_{k=-\frac{n-1}{2}}^{\frac{n-1}{2}} \prod_{l=-\frac{n-1}{2}}^{\frac{n-1}{2}} &e^{2\pi i  \frac{l}{n}\left(\frac{\alpha}{n}+\frac{k}{n}\right) } e^{\pi i \left( \frac{l}{n}\right)^{2} \tau} \frac{\theta_{3}\left(\frac{\alpha}{n}+\frac{k}{n}+\frac{l}{n} \tau \mid \tau\right)}{\eta(\tau)}, \\
    \frac{\theta_{4}(\alpha |\tau)}{\eta( \tau)}=&~&\prod_{k=0}^{n-1} \prod_{l=-\frac{n-1}{2}}^{\frac{n-1}{2}} &e^{2\pi i  \frac{l}{n}\left(\frac{\alpha}{n}+\frac{k}{n}\right) } e^{\pi i \left( \frac{l}{n}\right)^{2} \tau} \frac{\theta_{4}\left(\frac{\alpha}{n}+\frac{k}{n}+\frac{l}{n} \tau \mid \tau\right)}{\eta(\tau)}, 
    \end{alignedat}
\end{align}
which are analogous to the following trigonometric function identities,
\begin{align}
    \begin{alignedat}{2}
    2\sin nz =&~& \prod_{k=0}^{n-1} ~&2\sin \left(z+\frac{k \pi}{n}\right), \\
    2\cos nz =&~& \prod_{k=-\frac{n-1}{2}}^{\frac{n-1}{2}} &2\cos \left(z+\frac{k\pi}{n}\right). \\
    \end{alignedat}
\end{align}

\section{Examples of Hecke operators and orbifold theory partition functions}
\label{ExamplesHeckePartitionFunction}

\subsection{Examples of Hecke operators}
We repeat the definition of Hecke operators:
\begin{equation*}
    T_k Z\left(\tau\right) \equiv \frac{1}{k}\sum_{i|k} \sum_{j=0}^{i-1} Z\left(\frac{k \tau}{i^{2}}+\frac{j}{i}\right).
\end{equation*}
For most simple case, we have
\begin{align*}
T_1 Z(\tau)=&Z(\tau), \\
T_2 Z(\tau)=&\frac{1}{2}\left(Z(2\tau)+Z\left(\frac{\tau}{2}\right)+Z\left(\frac{\tau+1}{2}\right)\right), \\
T_3 Z(\tau)=&\frac{1}{3}\left(Z(3\tau)+Z\left(\frac{\tau}{3}\right)+Z\left(\frac{\tau+1}{3}\right)+Z\left(\frac{\tau+2}{3}\right)\right) .
\end{align*}
For the case $k$ is composite number, we have
\begin{align*}
T_4 Z(\tau)=&\frac{1}{4}\biggl( Z(4\tau) \biggr.\\
&\left.+Z\left(\frac{2\tau}{2}\right)+Z\left(\frac{2\tau+1}{2}\right) \right.\\
&\left.+Z\left(\frac{\tau}{4}\right)+Z\left(\frac{\tau+1}{4}\right)+Z\left(\frac{\tau+2}{4}\right)+Z\left(\frac{\tau+3}{4}\right)\right), \\
T_6 Z(\tau)=&\frac{1}{6} \biggl( Z(6\tau) \biggr.\\
&\left.+Z\left(\frac{3 \tau}{2}\right)+Z\left(\frac{3 \tau+1}{2}\right) \right.\\
&\left.+Z\left(\frac{2 \tau}{3}\right)+Z\left(\frac{2 \tau+1}{3}\right)+Z\left(\frac{2 \tau+2}{3}\right) \right.\\
&\left.+Z\left(\frac{\tau}{6}\right)+Z\left(\frac{\tau+1}{6}\right)+Z\left(\frac{\tau+2}{6}\right) \right.\\
&\left.+Z\left(\frac{\tau+3}{6}\right)+Z\left(\frac{\tau+4}{6}\right)+Z\left(\frac{\tau+5}{6}\right)\right).
\end{align*}
We notice here that the 4-th Hecke operator includes $\frac{1}{4} Z(\frac{2\tau}{2}) = \frac{1}{4} T_1 Z(\tau)$. This is because 4 has a square-factor.
We can find similar situation in more complicated example,
\begin{align*}
T_{12} Z(\tau)=&\frac{1}{12} \left( Z(12\tau)+Z\left(\frac{6 \tau}{2}\right)+Z\left(\frac{6 \tau+1}{2}\right) \right.\\
&\left.+Z\left(\frac{4 \tau}{3}\right)+Z\left(\frac{4 \tau+1}{3}\right)+Z\left(\frac{4 \tau+2}{3}\right) \right.\\
&\left.+Z\left(\frac{3 \tau}{4}\right)+Z\left(\frac{3 \tau+1}{4}\right)+Z\left(\frac{3 \tau+2}{4}\right)+Z\left(\frac{3 \tau+3}{4}\right) \right.\\
&\left.+Z\left(\frac{2\tau}{6}\right)+Z\left(\frac{2\tau+1}{6}\right)+Z\left(\frac{2\tau+2}{6}\right) \right.\\
&\left.+Z\left(\frac{2\tau+3}{6}\right)+Z\left(\frac{2\tau+4}{6}\right)+Z\left(\frac{2\tau+5}{6}\right) \right.\\
&\left.+Z\left(\frac{\tau}{12}\right)+Z\left(\frac{\tau+1}{12}\right)+Z\left(\frac{\tau+2}{12}\right) \right.\\
&\left.+Z\left(\frac{\tau+3}{12}\right)+Z\left(\frac{\tau+4}{12}\right)+Z\left(\frac{\tau+5}{12}\right) \right.\\
&\left.+Z\left(\frac{\tau+6}{12}\right)+Z\left(\frac{\tau+7}{12}\right)+Z\left(\frac{\tau+8}{12}\right) \right.\\
&\left.+Z\left(\frac{\tau+9}{12}\right)+Z\left(\frac{\tau+10}{12}\right)+Z\left(\frac{\tau+11}{12}\right)\right),
\end{align*}
here we notice again that this 12-th Hecke operator includes $\frac{1}{4} T_3 Z(\tau)$,
\begin{align*}
    \frac{1}{4} T_3 Z(\tau) = \frac{1}{12} \left(
    Z\left(\frac{6 \tau}{2}\right)+Z\left(\frac{2\tau}{6}\right)+Z\left(\frac{2\tau+2}{6}\right)+Z\left(\frac{2\tau+4}{6}\right)
    \right).
\end{align*}

\subsubsection{Examples of square-free Hecke operators}
Recall that we introduced square-free Hecke operators as
\begin{align*}
T_k^{sf}Z(\tau) = T_k Z(\tau) - \sum_{
    \substack{a\in \mathbb{Z}>1 \\
    a^2 | k}
    }
\frac{1}{a^2}T_{\frac{k}{a^2}}^{sf}Z(\tau),
\end{align*}
to specify the minimally modular invariant part of the Hecke operators. As is obvious from the definition, $T_k$ and $T_k^{sf}$ coincide when $k$ has no square-factor.
\begin{align*}
T_1^{sf} Z(\tau)=&T_1 Z(\tau)=Z(\tau) \\
T_3^{sf} Z(\tau)=&T_3 Z(\tau) \\
T_4^{sf} Z(\tau)=&T_4 Z(\tau)- \frac{1}{4}T_1^{sf} Z(\tau)\\
=&T_4 Z(\tau)- \frac{1}{4}Z(\tau)\\
T_9^{sf} Z(\tau)=&T_9 Z(\tau)- \frac{1}{9}T_1^{sf} Z(\tau)\\
=&T_9 Z(\tau)- \frac{1}{9}Z(\tau)\\
T_{12}^{sf} Z(\tau)=&T_{12} Z(\tau)- \frac{1}{4}T_3^{sf} Z(\tau)\\
=&T_{12} Z(\tau)- \frac{1}{4}T_3 Z(\tau)\\
T_{36}^{sf} Z(\tau)=&T_{36} Z(\tau)- \frac{1}{4}T_9^{sf} Z(\tau)- \frac{1}{9}T_4^{sf} Z(\tau)- \frac{1}{36}T_1^{sf} Z(\tau)\\
=&T_{36} Z(\tau)- \frac{1}{4}T_9 Z(\tau)-\frac{1}{9}T_4 Z(\tau)+ \frac{1}{36}Z(\tau)
\end{align*}

\subsubsection{Number of terms in (square-free) Hecke operators}
For later convenience, we refer here the number of terms that composes $k$-th (square-free) Hecke operators.
We assume that $k$ is prime-factorized as 
\begin{equation}
    k = \prod_{i} p_i^{n_i},
\end{equation}
where $p$s are prime integers and $n$s are positive integers.
It is obvious from the definition that the number of terms that composes the $k$-th Hecke operator is the sum of all divisors of $k$,
\begin{equation}
\label{NumTermsT}
    \left(\mathrm{\#~terms~of}~T_k\right)=\prod_{i}\left(\sum_{l_i=0}^{n_i} p_i^{l_i}\right).
\end{equation}
In the square-free case, $T_k^{sf}$ is constructed via excluding terms from $T_k$ corresponding to square factors of $k$, thus the following holds;
\begin{equation}
\label{NumTermsTsf}
    \left(\mathrm{\#~terms~of}~T_k^{sf}\right)=\prod_{i}\left(p_i^{n_i}+p_i^{n_i -1}\right).
\end{equation}
One can explicitly confirm these statements in examples.

\subsection{Examples of orbifold CFT partition functions}
We present here some examples of orbifold CFT partition functions.
We can check that $Z_{2,S}=Z_{2,\mathbb{Z}}$ for example. (This is obvious because $S_2\simeq\mathbb{Z}_2$.)
\subsubsection{Symmetric orbifold CFT}
Recall that we have well-known formula
\begin{equation*}
    Z_{N,S}(\tau) = 
    \sum_{\mathrm{partition~of}~N}
    \prod_{k=1}^{N}\frac{1}{(N_k)!} \left( T_k Z(\tau) \right)^{N_k},
\end{equation*}
where the partition of $N$ runs over $\displaystyle (N_1,\dots,N_N) ~\mathrm{s.t.}~ \sum_{k=1}^{N} k N_k = N$.
\begin{align*}
Z_{1,S}(\tau) =&T_1 Z(\tau) =Z(\tau) \\
Z_{2,S}(\tau) =&\frac{1}{2 !}\left(T_1 Z(\tau)\right)^{2}+T_2 Z(\tau) \\
=&\frac{1}{2}\left((Z(\tau))^{2}+Z(2 \tau)+Z\left(\frac{\tau}{2}\right)+Z\left(\frac{\tau+1}{2}\right)\right) \\
Z_{3,S}(\tau) =&\frac{1}{3 !}\left(T_1 Z(\tau)\right)^{3}+T_1 Z(\tau)~T_2 Z(\tau)+T_3 Z(\tau) \\
=&\frac{1}{6}(Z(\tau))^{3}+Z(\tau)~\frac{1}{2}\left(Z(2 \tau)+Z\left(\frac{\tau}{2}\right)+Z\left(\frac{\tau+1}{2}\right)\right) \\
&+\frac{1}{3}\left(Z(3 z)+Z\left(\frac{\tau}{3}\right)+Z\left(\frac{\tau+1}{3}\right)+Z\left(\frac{\tau+2}{3}\right)\right) \\
Z_{4,S}(\tau) =&\frac{1}{4 !}\left(T_1 Z(\tau)\right)^{4}+T_1 Z(\tau)~T_3 Z(\tau) \\
&+\frac{1}{2 !}\left(T_1 Z(\tau)\right)^{2}~T_2 Z(\tau)+\frac{1}{2 !}\left(T_2 Z(\tau)\right)^{2}+T_4 Z(\tau)
\end{align*}
\subsubsection{Cyclic orbifold CFT}
We use the following formula that presented in (\ref{GeneralCyclicPF}),
\begin{align}
    Z_{N,\mathbb{Z}}(\tau) =& \sum_{d|N} \frac{\phi(N/d)}{d} T_{\frac{N}{d}}^{sf}\left(Z(\tau)^{d}\right),
\end{align}
where $\phi$ is the Euler function.
\begin{align*}
Z_{1,\mathbb{Z}}(\tau) =&~T_1^{sf} Z(\tau) =Z(\tau) \\
Z_{2,\mathbb{Z}}(\tau) =&~\frac{1}{2}T_1^{sf}\left(Z(\tau)^{2}\right)+T_2^{sf} \left(Z(\tau)\right) \\
=&~\frac{1}{2}\left((Z(\tau))^{2}+Z(2 \tau)+Z\left(\frac{\tau}{2}\right)+Z\left(\frac{\tau+1}{2}\right)\right) \\
Z_{3,\mathbb{Z}}(\tau) =&~\frac{1}{3}T_1^{sf}\left(Z(\tau)^{3}\right)+\frac{2}{1}T_3^{sf} \left(Z(\tau)\right) \\
=&~\frac{1}{3}\left((Z(\tau))^{3}+
2\left(Z(3 z)+Z\left(\frac{\tau}{3}\right)+Z\left(\frac{\tau+1}{3}\right)+Z\left(\frac{\tau+2}{3}\right)\right)\right) \\
Z_{4,\mathbb{Z}}(\tau) =&~\frac{1}{4} T_{1}^{sf}\left(Z(\tau)^4\right)+ \frac{1}{2} T_{2}^{sf}\left(Z(\tau)^2\right)+ \frac{2}{1} T_{4}^{sf}\left(Z(\tau)\right) \\
=&~ \frac{1}{4} \left(
Z(\tau)^4 + Z(2 \tau)^2+Z\left(\frac{\tau}{2}\right)^2+Z\left(\frac{\tau+1}{2}\right)^2\right. \\
&\Biggl.+2\left(
Z(4\tau)+Z\left(\frac{2\tau+1}{2}\right) +Z\left(\frac{\tau}{4}\right)+Z\left(\frac{\tau+1}{4}\right)+Z\left(\frac{\tau+2}{4}\right)+Z\left(\frac{\tau+3}{4}\right)
\right)\Biggr)
\end{align*}
As we can explicitly check in above examples, $Z_{N,\mathbb{Z}}$ consists of $N^2$ terms of $Z$s, multiplied by $1/N$.

\section{Construction of cyclic orbifold partition function}
\label{COPproof}
In this section, we prove the equation (\ref{GeneralCyclicPF}), which constructs the $N$-th cyclic orbifold partition function in terms of square-free Hecke operators.
The general construction we have presented is,
\begin{equation}
    Z_{N,\mathbb{Z}}(\tau) = \sum_{d|N} \frac{(\mathrm{\#}k~(1\leq k\leq N)~\mathrm{s.t.}~\gcd(N,k)=d)}{d} T_{\frac{N}{d}}^{sf}\left(Z(\tau)^{d}\right).
\end{equation}
The proof of this equation is the goal of this section.

In the first place, we need to calculate $G=\mathbb{Z}_N$ version of (\ref{GeneralOrbifoldPF}),
\begin{align*}
    Z_{N,\mathbb{Z}}(\tau) =&
    \sum_{l=0}^{N-1} \mathrm{Tr}_{(l)} \left(\frac{\sum_{l'=1}^{N}g^{l'}}{N}q^{H_L}\bar{q}^{H_R}\right) \\
    =& \frac{1}{N} \sum_{l,l'}\mathrm{Tr}_{(l)} \left(g^{l'} q^{H_L}\bar{q}^{H_R}\right),
\end{align*}
where $g$ is one of the generators of $\mathbb{Z}_N$ and $l$ labels the (un)twisted sectors. The last line shows that we need to determine $N^2$ terms that contributes to the partition function.

It is easy to determine $N$ terms in the untwisted sector. They have $\frac{N}{d}\tau$ periods in the $\tau$ direction if $k$ satisfies $\gcd(N,k)=d$. Thus the untwisted sector consists of
\begin{align*}
    &Z(N\tau) \times (\mathrm{\#}k~(1\leq k\leq N)~\mathrm{s.t.}~\gcd(N,k)=1),\\
    &\vdots \\
    &Z\left(\frac{N}{d}\tau\right)^{d} \times (\mathrm{\#}k~(1\leq k\leq N)~\mathrm{s.t.}~\gcd(N,k)=d), \\
    &\vdots \\
    &Z(\tau)^N ,
\end{align*}
($d$ runs over the divisors of $N$). Distinguishing each $Z\left(\frac{N}{d}\tau\right)^{d}$ from others, these $Z$s actually give $N$ terms.
Other than $Z(\tau)^N$, these $Z$s are NOT modular invariant itself. The minimally modular invariant function that includes $Z\left(\frac{N}{d}\tau\right)^{d}$ is what we introduced in \ref{sec:squarefreeheckeopIntro}, the square-free Hecke operator
\begin{equation*}
    \frac{N}{d}T_{\frac{N}{d}}^{sf}\left(Z(\tau)^{d}\right).
\end{equation*}
This implies that we need to include $\frac{N}{d}T_{\frac{N}{d}}^{sf}\left(Z(\tau)^{d}\right)$ for each $Z\left(\frac{N}{d}\tau\right)^{d}$.

Here, we have determined the necessary terms to construct the cyclic orbifold partition function. In other words, we found that $Z_{N,\mathbb{Z}}(\tau)$ includes terms of the form
\begin{equation*}
    \sum_{d|N} \frac{(\mathrm{\#}k~(1\leq k\leq N)~\mathrm{s.t.}~\gcd(N,k)=d)}{d} T_{\frac{N}{d}}^{sf}\left(Z(\tau)^{d}\right).
\end{equation*}
Now what remains to do is to prove that this summation is sufficient. As we will see below, this summation actually consists of totally $N^2$ terms, thus is sufficient.

The key to the proof is
\begin{itemize}
    \item $\mathrm{\#}k~(1\leq k\leq N)~\mathrm{s.t.}~\gcd(N,k)=d$,
    \item $\mathrm{\#~terms~in}~\frac{N}{d}T_{\frac{N}{d}}^{sf}\left(Z(\tau)^{d}\right)$.
\end{itemize}
We get the total number of terms by summing their product over all $d$ that divides $N$.
Before proceeding, we set the prime factorization of $N$ and $N$'s divisor $d$ as follows:
\begin{align}
    N =& \prod_{i=1}^{n} p_i^{n_i} \\
    d =& \prod_{i=1}^{n} p_i^{k_i}
\end{align}
where $p$s are prime, $n$s are positive, and $k$s are non-negative integers. Of course, $k$s are in the range of $0\leq k_i \leq n_i$.
Firstly, let us evaluate $\mathrm{\#}k~(1\leq k\leq N)~\mathrm{s.t.}~\gcd(N,k)=d$. One can check that this quantity is $\phi(\frac{N}{d})$ where $\phi$ is the Euler function:
\begin{align}
    \left(\mathrm{\#}k~(1\leq k\leq N)~\mathrm{s.t.}~\gcd(N,k)=d\right) =&~ \phi\left(\frac{N}{d}\right) \nonumber\\
    =&~ \prod_{i=1}^{n} \phi\left(p_i^{n_i - k_i}\right) \nonumber\\
    =&~ \prod_{i=1}^{n} p_i^{n_i - k_i}\left(1-\frac{1-\delta_{n_i,k_i}}{p_i}\right).
\end{align}
Next, we evaluate $\mathrm{\#~terms~in}~\frac{N}{d}T_{\frac{N}{d}}^{sf}\left(Z(\tau)^{d}\right)$. Recall the formula (\ref{NumTermsTsf}), we find that 
\begin{align}
    \left(\mathrm{\#~terms~in}~\frac{N}{d}T_{\frac{N}{d}}^{sf}\left(Z(\tau)^{d}\right)\right) =&~ \prod_{i=1}^{n} \left(p_i^{n_i - k_i}+p_i^{n_i - k_i-1}(1-\delta_{n_i,k_i})\right) \nonumber\\
    =&~ \prod_{i=1}^{n} p_i^{n_i - k_i}\left(1+\frac{1-\delta_{n_i,k_i}}{p_i}\right).
\end{align}
The total number of terms is
\begin{align}
    \left(\#~\mathrm{total~terms}\right)
    =&\sum_{d|N}\left(\mathrm{\#}k~(1\leq k\leq N)~\mathrm{s.t.}~\gcd(N,k)=d\right) \nonumber\\
    &\times \left(\mathrm{\#~terms~in}~\frac{N}{d}T_{\frac{N}{d}}^{sf}\left(Z(\tau)^{d}\right)\right) \nonumber\\
    =&\sum_{d|N} \prod_{i=1}^{n} p_i^{n_i - k_i}\left(1-\frac{1-\delta_{n_i,k_i}}{p_i}\right)p_i^{n_i - k_i}\left(1+\frac{1-\delta_{n_i,k_i}}{p_i}\right) \nonumber\\
    =&\prod_{i=1}^{n}\sum_{k_i=0}^{n_i} p_i^{2\left(n_i - k_i\right)}\left(1-\frac{1-\delta_{n_i,k_i}}{p_i}\right)\left(1+\frac{1-\delta_{n_i,k_i}}{p_i}\right) \nonumber\\
    =&\prod_{i=1}^{n}\left(\left(p_i^{2n_i}-p_i^{2n_i - 2}\right)+\left(p_i^{2n_i - 2}-p_i^{2n_i - 4}\right)+\dots+\left(p_i^{2}-1\right)+1\right) \nonumber\\
    =&\prod_{i=1}^{n}{p_i}^{2n_i},
\end{align}
this is exactly $N^2$. This result proves that the formula (\ref{GeneralCyclicPF}) sufficiently gives $N^2$ terms, thus is the $N$-th cyclic orbifold partition function.

\bibliographystyle{JHEP.bst}
\bibliography{Article}

\providecommand{\href}[2]{#2}\begingroup\raggedright\begin{thebibliography}{10}

\bibitem{Strominger:1996sh}
A.~Strominger and C.~Vafa, {\it {Microscopic origin of the Bekenstein-Hawking
  entropy}},  {\em Phys. Lett. B} {\bf 379} (1996) 99--104,
  [\href{http://arxiv.org/abs/hep-th/9601029}{{\tt hep-th/9601029}}].

\bibitem{Vafa:1995bm}
C.~Vafa, {\it {Instantons on D-branes}},  {\em Nucl. Phys. B} {\bf 463} (1996)
  435--442, [\href{http://arxiv.org/abs/hep-th/9512078}{{\tt hep-th/9512078}}].

\bibitem{Vafa:1995zh}
C.~Vafa, {\it {Gas of d-branes and Hagedorn density of BPS states}},  {\em
  Nucl. Phys. B} {\bf 463} (1996) 415--419,
  [\href{http://arxiv.org/abs/hep-th/9511088}{{\tt hep-th/9511088}}].

\bibitem{Douglas:1995bn}
M.~R. Douglas, {\it {Branes within branes}},  {\em NATO Sci. Ser. C} {\bf 520}
  (1999) 267--275, [\href{http://arxiv.org/abs/hep-th/9512077}{{\tt
  hep-th/9512077}}].

\bibitem{Dijkgraaf:1998gf}
R.~Dijkgraaf, {\it {Instanton strings and hyperKahler geometry}},  {\em Nucl.
  Phys. B} {\bf 543} (1999) 545--571,
  [\href{http://arxiv.org/abs/hep-th/9810210}{{\tt hep-th/9810210}}].

\bibitem{Aspinwall:1995zi}
P.~S. Aspinwall, {\it {Enhanced gauge symmetries and K3 surfaces}},  {\em Phys.
  Lett. B} {\bf 357} (1995) 329--334,
  [\href{http://arxiv.org/abs/hep-th/9507012}{{\tt hep-th/9507012}}].

\bibitem{Maldacena:1997re}
J.~M. Maldacena, {\it {The Large N limit of superconformal field theories and
  supergravity}},  {\em Adv. Theor. Math. Phys.} {\bf 2} (1998) 231--252,
  [\href{http://arxiv.org/abs/hep-th/9711200}{{\tt hep-th/9711200}}].

\bibitem{Larsen:1999uk}
F.~Larsen and E.~J. Martinec, {\it {U(1) charges and moduli in the D1 - D5
  system}},  {\em JHEP} {\bf 06} (1999) 019,
  [\href{http://arxiv.org/abs/hep-th/9905064}{{\tt hep-th/9905064}}].

\bibitem{Avery:2010er}
S.~G. Avery, B.~D. Chowdhury, and S.~D. Mathur, {\it {Deforming the D1D5 CFT
  away from the orbifold point}},  {\em JHEP} {\bf 06} (2010) 031,
  [\href{http://arxiv.org/abs/1002.3132}{{\tt arXiv:1002.3132}}].

\bibitem{Carson:2014ena}
Z.~Carson, S.~Hampton, S.~D. Mathur, and D.~Turton, {\it {Effect of the
  deformation operator in the D1D5 CFT}},  {\em JHEP} {\bf 01} (2015) 071,
  [\href{http://arxiv.org/abs/1410.4543}{{\tt arXiv:1410.4543}}].

\bibitem{deBoer:1998us}
J.~de~Boer, {\it {Large N elliptic genus and AdS / CFT correspondence}},  {\em
  JHEP} {\bf 05} (1999) 017, [\href{http://arxiv.org/abs/hep-th/9812240}{{\tt
  hep-th/9812240}}].

\bibitem{Heemskerk:2009pn}
I.~Heemskerk, J.~Penedones, J.~Polchinski, and J.~Sully, {\it {Holography from
  Conformal Field Theory}},  {\em JHEP} {\bf 10} (2009) 079,
  [\href{http://arxiv.org/abs/0907.0151}{{\tt arXiv:0907.0151}}].

\bibitem{El-Showk:2011yvt}
S.~El-Showk and K.~Papadodimas, {\it {Emergent Spacetime and Holographic
  CFTs}},  {\em JHEP} {\bf 10} (2012) 106,
  [\href{http://arxiv.org/abs/1101.4163}{{\tt arXiv:1101.4163}}].

\bibitem{Keller:2011xi}
C.~A. Keller, {\it {Phase transitions in symmetric orbifold CFTs and
  universality}},  {\em JHEP} {\bf 03} (2011) 114,
  [\href{http://arxiv.org/abs/1101.4937}{{\tt arXiv:1101.4937}}].

\bibitem{Hartman:2014oaa}
T.~Hartman, C.~A. Keller, and B.~Stoica, {\it {Universal Spectrum of 2d
  Conformal Field Theory in the Large c Limit}},  {\em JHEP} {\bf 09} (2014)
  118, [\href{http://arxiv.org/abs/1405.5137}{{\tt arXiv:1405.5137}}].

\bibitem{Pakman:2009zz}
A.~Pakman, L.~Rastelli, and S.~S. Razamat, {\it {Diagrams for Symmetric Product
  Orbifolds}},  {\em JHEP} {\bf 10} (2009) 034,
  [\href{http://arxiv.org/abs/0905.3448}{{\tt arXiv:0905.3448}}].

\bibitem{Belin:2015hwa}
A.~Belin, C.~A. Keller, and A.~Maloney, {\it {Permutation Orbifolds in the
  large N Limit}},  \href{http://arxiv.org/abs/1509.01256}{{\tt
  arXiv:1509.01256}}.

\bibitem{Perlmutter:2016pkf}
E.~Perlmutter, {\it {Bounding the Space of Holographic CFTs with Chaos}},  {\em
  JHEP} {\bf 10} (2016) 069, [\href{http://arxiv.org/abs/1602.08272}{{\tt
  arXiv:1602.08272}}].

\bibitem{Caputa:2017tju}
P.~Caputa, Y.~Kusuki, T.~Takayanagi, and K.~Watanabe, {\it {Evolution of
  Entanglement Entropy in Orbifold CFTs}},  {\em J. Phys. A} {\bf 50} (2017),
  no.~24 244001, [\href{http://arxiv.org/abs/1701.03110}{{\tt
  arXiv:1701.03110}}].

\bibitem{Giusto:2014aba}
S.~Giusto and R.~Russo, {\it {Entanglement Entropy and D1-D5 geometries}},
  {\em Phys. Rev. D} {\bf 90} (2014), no.~6 066004,
  [\href{http://arxiv.org/abs/1405.6185}{{\tt arXiv:1405.6185}}].

\bibitem{Balasubramanian:2014sra}
V.~Balasubramanian, B.~D. Chowdhury, B.~Czech, and J.~de~Boer, {\it
  {Entwinement and the emergence of spacetime}},  {\em JHEP} {\bf 01} (2015)
  048, [\href{http://arxiv.org/abs/1406.5859}{{\tt arXiv:1406.5859}}].

\bibitem{Balasubramanian:2016ids}
V.~Balasubramanian, B.~Craps, B.~Czech, and G.~S\'arosi, {\it {Echoes of chaos
  from string theory black holes}},  {\em JHEP} {\bf 03} (2017) 154,
  [\href{http://arxiv.org/abs/1612.04334}{{\tt arXiv:1612.04334}}].

\bibitem{Balasubramanian:2018ajb}
V.~Balasubramanian, B.~Craps, T.~De~Jonckheere, and G.~S\'arosi, {\it
  {Entanglement versus entwinement in symmetric product orbifolds}},  {\em
  JHEP} {\bf 01} (2019) 190, [\href{http://arxiv.org/abs/1806.02871}{{\tt
  arXiv:1806.02871}}].

\bibitem{Apolo:2022fya}
L.~Apolo, A.~Belin, S.~Bintanja, A.~Castro, and C.~A. Keller, {\it {Deforming
  Symmetric Product Orbifolds: A tale of moduli and higher spin currents}},
  \href{http://arxiv.org/abs/2204.07590}{{\tt arXiv:2204.07590}}.

\bibitem{Belin:2020nmp}
A.~Belin, N.~Benjamin, A.~Castro, S.~M. Harrison, and C.~A. Keller, {\it
  {$\mathcal{N}=2$ Minimal Models: A Holographic Needle in a Symmetric Orbifold
  Haystack}},  {\em SciPost Phys.} {\bf 8} (2020), no.~6 084,
  [\href{http://arxiv.org/abs/2002.07819}{{\tt arXiv:2002.07819}}].

\bibitem{Belin:2021nck}
A.~Belin, S.~Biswas, and J.~Sully, {\it {The spectrum of boundary states in
  symmetric orbifolds}},  {\em JHEP} {\bf 01} (2022) 123,
  [\href{http://arxiv.org/abs/2110.05491}{{\tt arXiv:2110.05491}}].

\bibitem{Gaberdiel:2014cha}
M.~R. Gaberdiel and R.~Gopakumar, {\it {Higher Spins \& Strings}},  {\em JHEP}
  {\bf 11} (2014) 044, [\href{http://arxiv.org/abs/1406.6103}{{\tt
  arXiv:1406.6103}}].

\bibitem{Giribet:2018ada}
G.~Giribet, C.~Hull, M.~Kleban, M.~Porrati, and E.~Rabinovici, {\it
  {Superstrings on AdS$_{3}$ at $\mathcal{k} =$ 1}},  {\em JHEP} {\bf 08}
  (2018) 204, [\href{http://arxiv.org/abs/1803.04420}{{\tt arXiv:1803.04420}}].

\bibitem{Gaberdiel:2018rqv}
M.~R. Gaberdiel and R.~Gopakumar, {\it {Tensionless string spectra on
  AdS$_{3}$}},  {\em JHEP} {\bf 05} (2018) 085,
  [\href{http://arxiv.org/abs/1803.04423}{{\tt arXiv:1803.04423}}].

\bibitem{Eberhardt:2018ouy}
L.~Eberhardt, M.~R. Gaberdiel, and R.~Gopakumar, {\it {The Worldsheet Dual of
  the Symmetric Product CFT}},  {\em JHEP} {\bf 04} (2019) 103,
  [\href{http://arxiv.org/abs/1812.01007}{{\tt arXiv:1812.01007}}].

\bibitem{Eberhardt:2019ywk}
L.~Eberhardt, M.~R. Gaberdiel, and R.~Gopakumar, {\it {Deriving the
  AdS$_{3}$/CFT$_{2}$ correspondence}},  {\em JHEP} {\bf 02} (2020) 136,
  [\href{http://arxiv.org/abs/1911.00378}{{\tt arXiv:1911.00378}}].

\bibitem{Eberhardt:2021jvj}
L.~Eberhardt, {\it {Summing over Geometries in String Theory}},  {\em JHEP}
  {\bf 05} (2021) 233, [\href{http://arxiv.org/abs/2102.12355}{{\tt
  arXiv:2102.12355}}].

\bibitem{Holzhey:1994we}
C.~Holzhey, F.~Larsen, and F.~Wilczek, {\it {Geometric and renormalized entropy
  in conformal field theory}},  {\em Nucl. Phys. B} {\bf 424} (1994) 443--467,
  [\href{http://arxiv.org/abs/hep-th/9403108}{{\tt hep-th/9403108}}].

\bibitem{Calabrese:2004eu}
P.~Calabrese and J.~L. Cardy, {\it {Entanglement entropy and quantum field
  theory}},  {\em J. Stat. Mech.} {\bf 0406} (2004) P06002,
  [\href{http://arxiv.org/abs/hep-th/0405152}{{\tt hep-th/0405152}}].

\bibitem{Furukawa_2009}
S.~Furukawa, V.~Pasquier, and J.~Shiraishi, {\it Mutual information and boson
  radius in a c=1 critical system in one dimension},  {\em Physical Review
  Letters} {\bf 102} (apr, 2009).

\bibitem{Headrick:2010zt}
M.~Headrick, {\it {Entanglement Renyi entropies in holographic theories}},
  {\em Phys. Rev. D} {\bf 82} (2010) 126010,
  [\href{http://arxiv.org/abs/1006.0047}{{\tt arXiv:1006.0047}}].

\bibitem{Azeyanagi:2007bj}
T.~Azeyanagi, T.~Nishioka, and T.~Takayanagi, {\it {Near Extremal Black Hole
  Entropy as Entanglement Entropy via AdS(2)/CFT(1)}},  {\em Phys. Rev. D} {\bf
  77} (2008) 064005, [\href{http://arxiv.org/abs/0710.2956}{{\tt
  arXiv:0710.2956}}].

\bibitem{Ogawa:2011bz}
N.~Ogawa, T.~Takayanagi, and T.~Ugajin, {\it {Holographic Fermi Surfaces and
  Entanglement Entropy}},  {\em JHEP} {\bf 01} (2012) 125,
  [\href{http://arxiv.org/abs/1111.1023}{{\tt arXiv:1111.1023}}].

\bibitem{Mukhi:2017rex}
S.~Mukhi, S.~Murthy, and J.-Q. Wu, {\it {Entanglement, Replicas, and Thetas}},
  {\em JHEP} {\bf 01} (2018) 005, [\href{http://arxiv.org/abs/1706.09426}{{\tt
  arXiv:1706.09426}}].

\bibitem{Takayanagi:2010wp}
T.~Takayanagi and T.~Ugajin, {\it {Measuring Black Hole Formations by
  Entanglement Entropy via Coarse-Graining}},  {\em JHEP} {\bf 11} (2010) 054,
  [\href{http://arxiv.org/abs/1008.3439}{{\tt arXiv:1008.3439}}].

\bibitem{Calabrese:2005in}
P.~Calabrese and J.~L. Cardy, {\it {Evolution of entanglement entropy in
  one-dimensional systems}},  {\em J. Stat. Mech.} {\bf 0504} (2005) P04010,
  [\href{http://arxiv.org/abs/cond-mat/0503393}{{\tt cond-mat/0503393}}].

\bibitem{Abajo-Arrastia:2010ajo}
J.~Abajo-Arrastia, J.~Aparicio, and E.~Lopez, {\it {Holographic Evolution of
  Entanglement Entropy}},  {\em JHEP} {\bf 11} (2010) 149,
  [\href{http://arxiv.org/abs/1006.4090}{{\tt arXiv:1006.4090}}].

\bibitem{Hartman:2013qma}
T.~Hartman and J.~Maldacena, {\it {Time Evolution of Entanglement Entropy from
  Black Hole Interiors}},  {\em JHEP} {\bf 05} (2013) 014,
  [\href{http://arxiv.org/abs/1303.1080}{{\tt arXiv:1303.1080}}].

\bibitem{Ryu:2006bv}
S.~Ryu and T.~Takayanagi, {\it {Holographic derivation of entanglement entropy
  from AdS/CFT}},  {\em Phys. Rev. Lett.} {\bf 96} (2006) 181602,
  [\href{http://arxiv.org/abs/hep-th/0603001}{{\tt hep-th/0603001}}].

\bibitem{Ryu:2006ef}
S.~Ryu and T.~Takayanagi, {\it {Aspects of Holographic Entanglement Entropy}},
  {\em JHEP} {\bf 08} (2006) 045,
  [\href{http://arxiv.org/abs/hep-th/0605073}{{\tt hep-th/0605073}}].

\bibitem{Hubeny:2007xt}
V.~E. Hubeny, M.~Rangamani, and T.~Takayanagi, {\it {A Covariant holographic
  entanglement entropy proposal}},  {\em JHEP} {\bf 07} (2007) 062,
  [\href{http://arxiv.org/abs/0705.0016}{{\tt arXiv:0705.0016}}].

\bibitem{Takayanagi:2011zk}
T.~Takayanagi, {\it {Holographic Dual of BCFT}},  {\em Phys. Rev. Lett.} {\bf
  107} (2011) 101602, [\href{http://arxiv.org/abs/1105.5165}{{\tt
  arXiv:1105.5165}}].

\bibitem{Fujita:2011fp}
M.~Fujita, T.~Takayanagi, and E.~Tonni, {\it {Aspects of AdS/BCFT}},  {\em
  JHEP} {\bf 11} (2011) 043, [\href{http://arxiv.org/abs/1108.5152}{{\tt
  arXiv:1108.5152}}].

\bibitem{Dixon:1985jw}
L.~J. Dixon, J.~A. Harvey, C.~Vafa, and E.~Witten, {\it {Strings on
  Orbifolds}},  {\em Nucl. Phys. B} {\bf 261} (1985) 678--686.

\bibitem{Dijkgraaf:1996xw}
R.~Dijkgraaf, G.~W. Moore, E.~P. Verlinde, and H.~L. Verlinde, {\it {Elliptic
  genera of symmetric products and second quantized strings}},  {\em Commun.
  Math. Phys.} {\bf 185} (1997) 197--209,
  [\href{http://arxiv.org/abs/hep-th/9608096}{{\tt hep-th/9608096}}].

\bibitem{Haehl:2014yla}
F.~M. Haehl and M.~Rangamani, {\it {Permutation orbifolds and holography}},
  {\em JHEP} {\bf 03} (2015) 163, [\href{http://arxiv.org/abs/1412.2759}{{\tt
  arXiv:1412.2759}}].

\bibitem{Klemm:1990df}
A.~Klemm and M.~G. Schmidt, {\it {Orbifolds by Cyclic Permutations of Tensor
  Product Conformal Field Theories}},  {\em Phys. Lett. B} {\bf 245} (1990)
  53--58.

\bibitem{Casini:2005rm}
H.~Casini, C.~D. Fosco, and M.~Huerta, {\it {Entanglement and alpha entropies
  for a massive Dirac field in two dimensions}},  {\em J. Stat. Mech.} {\bf
  0507} (2005) P07007, [\href{http://arxiv.org/abs/cond-mat/0505563}{{\tt
  cond-mat/0505563}}].

\bibitem{Lunin:2000yv}
O.~Lunin and S.~D. Mathur, {\it {Correlation functions for M**N / S(N)
  orbifolds}},  {\em Commun. Math. Phys.} {\bf 219} (2001) 399--442,
  [\href{http://arxiv.org/abs/hep-th/0006196}{{\tt hep-th/0006196}}].

\bibitem{Chen:2014ehg}
B.~Chen and J.-q. Wu, {\it {Universal relation between thermal entropy and
  entanglement entropy in conformal field theories}},  {\em Phys. Rev. D} {\bf
  91} (2015), no.~8 086012, [\href{http://arxiv.org/abs/1412.0761}{{\tt
  arXiv:1412.0761}}].

\bibitem{Yellow}
P.~Di~Francesco, P.~Mathieu, and D.~Senechal, {\em {Conformal Field Theory}}.
\newblock Graduate Texts in Contemporary Physics. Springer-Verlag, New York,
  1997.

\bibitem{Diaconescu:1997br}
D.-E. Diaconescu, M.~R. Douglas, and J.~Gomis, {\it {Fractional branes and
  wrapped branes}},  {\em JHEP} {\bf 02} (1998) 013,
  [\href{http://arxiv.org/abs/hep-th/9712230}{{\tt hep-th/9712230}}].

\bibitem{Billo:1999nf}
M.~Billo, B.~Craps, and F.~Roose, {\it {On D-branes in type 0 string theory}},
  {\em Phys. Lett. B} {\bf 457} (1999) 61--69,
  [\href{http://arxiv.org/abs/hep-th/9902196}{{\tt hep-th/9902196}}].

\bibitem{Diaconescu:1999dt}
D.-E. Diaconescu and J.~Gomis, {\it {Fractional branes and boundary states in
  orbifold theories}},  {\em JHEP} {\bf 10} (2000) 001,
  [\href{http://arxiv.org/abs/hep-th/9906242}{{\tt hep-th/9906242}}].

\bibitem{Billo:2000yb}
M.~Billo, B.~Craps, and F.~Roose, {\it {Orbifold boundary states from Cardy's
  condition}},  {\em JHEP} {\bf 01} (2001) 038,
  [\href{http://arxiv.org/abs/hep-th/0011060}{{\tt hep-th/0011060}}].

\bibitem{Takayanagi:2001aj}
T.~Takayanagi and T.~Uesugi, {\it {D-branes in Melvin background}},  {\em JHEP}
  {\bf 11} (2001) 036, [\href{http://arxiv.org/abs/hep-th/0110200}{{\tt
  hep-th/0110200}}].

\bibitem{Maldacena:1996ds}
J.~M. Maldacena and L.~Susskind, {\it {D-branes and fat black holes}},  {\em
  Nucl. Phys. B} {\bf 475} (1996) 679--690,
  [\href{http://arxiv.org/abs/hep-th/9604042}{{\tt hep-th/9604042}}].

\bibitem{Maldacena:2001kr}
J.~M. Maldacena, {\it {Eternal black holes in anti-de Sitter}},  {\em JHEP}
  {\bf 04} (2003) 021, [\href{http://arxiv.org/abs/hep-th/0106112}{{\tt
  hep-th/0106112}}].

\bibitem{tenenbaum2015introduction}
G.~Tenenbaum, {\em Introduction to Analytic and Probabilistic Number Theory}.
\newblock Graduate Studies in Mathematics. American Mathematical Society, 2015.

\end{thebibliography}\endgroup

\end{document}